\documentclass{jfm}
\pdfoutput=1
\usepackage{gensymb}
\usepackage{siunitx}

\usepackage{graphicx}
\usepackage{natbib}
\usepackage{amsmath}
\usepackage{color}
\usepackage[normalem,normalbf]{ulem}
\usepackage{multirow}
\usepackage{xcolor}
\usepackage{soul}
\usepackage{subcaption} 
\usepackage{booktabs} 
\usepackage[outdir=./]{epstopdf}

\graphicspath{{figures/}}

\newcommand{\vect}[1]{\boldsymbol{#1}}

\definecolor{brick}{rgb}{0.75,0,0}
\definecolor{rltgreen}{rgb}{0,0.5,0}

\newcommand{\CR}        [1] {{\color{red}{#1}}}

\title[Dynamics of afterbody vortices]{Meandering dynamics of streamwise vortex pairs in afterbody wakes}

\author[Ranjan \textit{et al.}]
{R. Ranjan$^{1}$\thanks{Email address for correspondence: ranjan.25@osu.edu},
J.-C. Robinet$^2$,
D. Gaitonde$^1$
}
\affiliation{$^1$Mechanical and Aerospace Engineering, The Ohio State University, Columbus, OH 43210, USA\\[\affilskip]
$^2$DynFluid Lab. - Arts \& M\'etiers Paris - 151, Bd. de l'H\^opital - 75013 - Paris - France\\[\affilskip]
}

\pubyear{2020}
\volume{xxx}
\pagerange{xxx--xxx}
\date{2020}

\begin{document}

\maketitle

\begin{abstract}
Wakes of upswept afterbodies are often characterized by a  counter-rotating streamwise vortex pair.
The unsteady dynamics of these vortices are examined with a spatio-temporally resolved Large-Eddy Simulation (LES) dataset on a representative configuration consisting of a cylinder with upswept basal surface.
Emphasis is placed on understanding the meandering motion of the vortices in the pair, including vortex core displacement, spectral content, stability mechanisms and overall rank-behavior.
The first two energy-ranked modes obtained through Proper orthogonal decomposition (POD) of the time-resolved vorticity field reveals a pair of vortex dipoles aligned relatively perpendicularly to each other.
The dynamics is successfully mapped to a matched Batchelor vortex pair whose spatial and temporal stability analyses indicate similar dipole structures associated with  an $|m|=1$ elliptic mode pair. 
This short-wave elliptic instability dominates the meandering motion, with strain due to axial velocity playing a key role in breakdown.
The low frequency of the unstable mode (Strouhal number $St_D \simeq 0.3$ based on cylinder diameter) is consistent with spectral analysis of meandering in the LES.
The wake is examined for its rank behavior; the number of modes required to reproduce the flow to given degree of accuracy diminishes rapidly outside of the immediate vicinity of the base. 
Beyond two diameters downstream, only two leading POD modes are required to reconstruct the dominant meandering motion and spatial structure in the LES data with $<15\%$ performance loss, while ten modes nearly completely recover the flow field.
This low-rank behavior may hold promise in constructing a reduced-order model for control purposes.

\end{abstract}

\section{Introduction}
\begin{figure}
\centering
\includegraphics[width=0.95\textwidth, trim={0.1cm 5cm 0.1cm 5cm},clip]{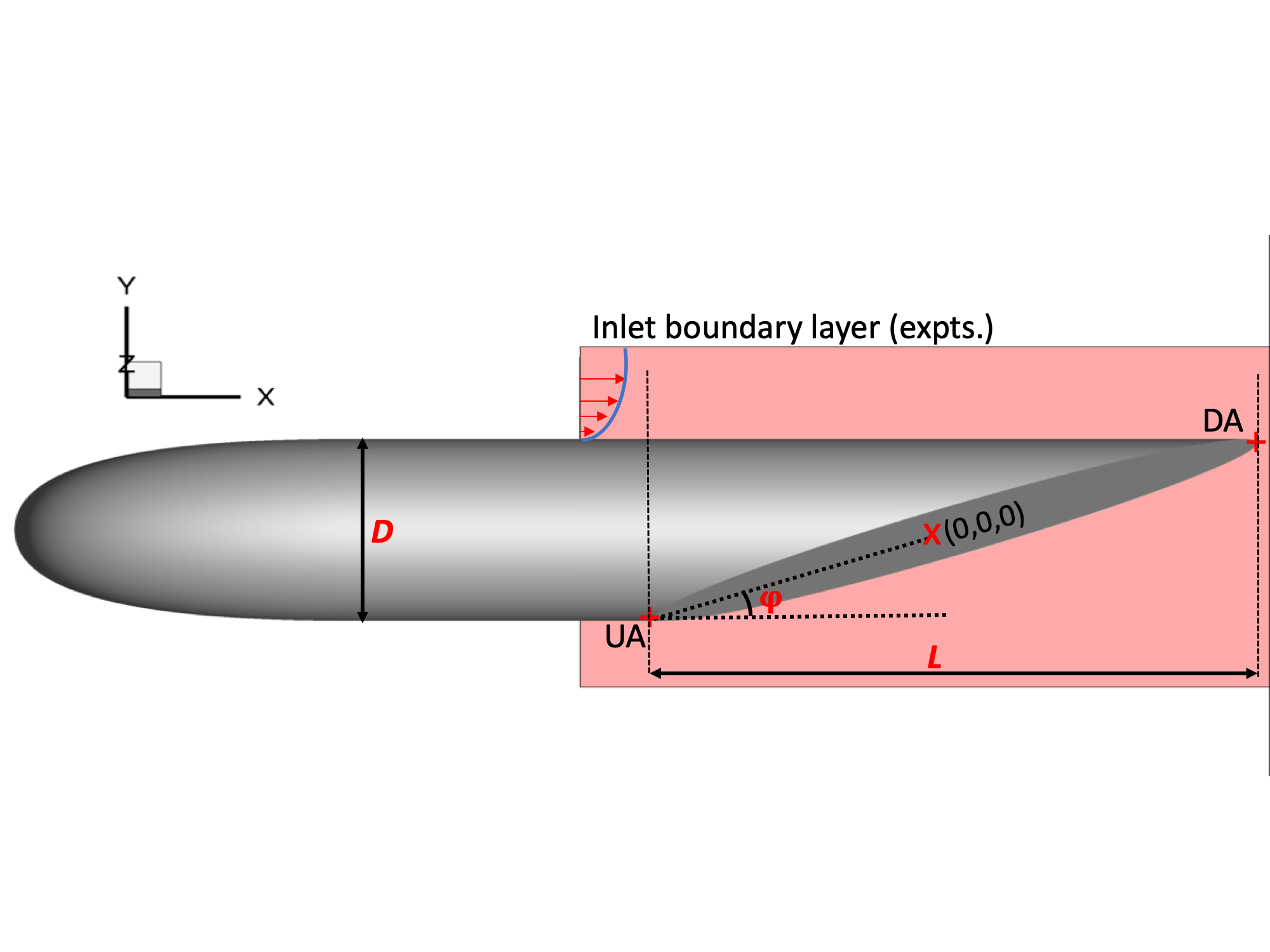}
\caption{Canonical configuration employed to reproduce the streamwise-oriented wake vortex pair.  
Cylinder diameter is $D$ and upsweep angle is $\phi=\ang{20}$.
The coordinate axis origin is located at the center of the upswept surface, $(0,0,0)$, (red cross: \CR{$\times$}) and the axial slant length is $L = D~\text{cot}(\phi)$. 
The shaded colored region encompasses the near-computational domain. 
The upstream apex (UA) and the downstream apex (DA) are located at $x/D=-\frac{\text{cot}(\phi)}{2},y/D=-0.5,z/D=0$ and $x/D=\frac{\text{cot}(\phi)}{2},y/D=0.5,z/D=0$ respectively. 
To describe the flow around the upswept base, the transformation $X=x+\frac{D}{2}\text{cot}(\phi)$ is used such that UA and DA are at $X/L=0$ and $X/L=1$ respectively. 
At the inlet, simulated boundary layer data matching the experiments~\citep{Zigunov2020} are used.}  
\label{fig:sketch}
\end{figure}

The complexity of turbulent afterbody flows is associated with presence of separation bubbles, rollup and flapping of shear layer, vortex-shear layer interactions and vortex-vortex interactions.
The present work considers the situation where the afterbody is  characterized by a cylinder, with axis aligned with the freestream,  terminated by a planar upswept base.
The configuration, shown in figure~\ref{fig:sketch}, is representative of the aft fuselage of a typical cargo aircraft~\citep{morel1978effect, bulathsinghala2017afterbody}, but the flowfield is relevant to many applications including automobiles \citep{lienhart2002flow, rossitto2016influence} and high-speed trains \citep{bell2016dynamics}.
The configuration consists of a nose section (spherical or ogive forebody) followed by a main cylindrical section of diameter $D$.
The trailing end of the cylinder is cut by a slanted base, whose angle with the horizontal is $\phi$, giving rise to an upstream apex (UA) and a downstream apex (DA) as the extreme endpoints on the base.
The range of upsweep angles typically employed in aircraft ranges from  $\ang{7} \le \phi \le \ang{28}$; we select $\ang{20}$ as a representative value that generates the phenomena of interest.

\begin{figure}
\centering
  \begin{subfigure}[b]{0.45\textwidth}
    \includegraphics[width=\textwidth, trim={0.5cm 4cm 1cm 1cm},clip]{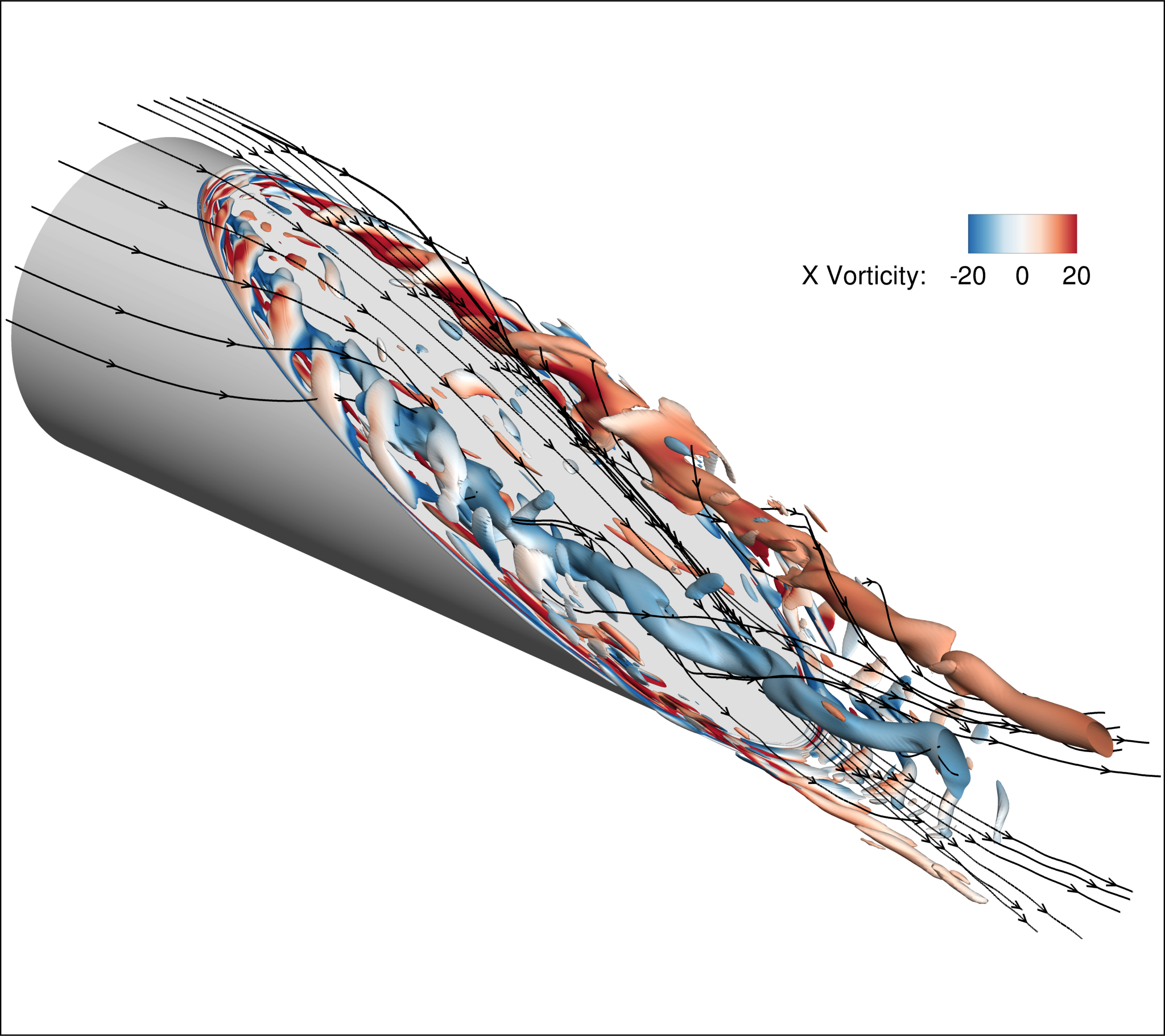}
    \caption{Instantaneous flow}
  \end{subfigure}
  \begin{subfigure}[b]{0.45\textwidth}
        \includegraphics[width=\textwidth, trim={2.2cm 2cm 3cm 2cm},clip]{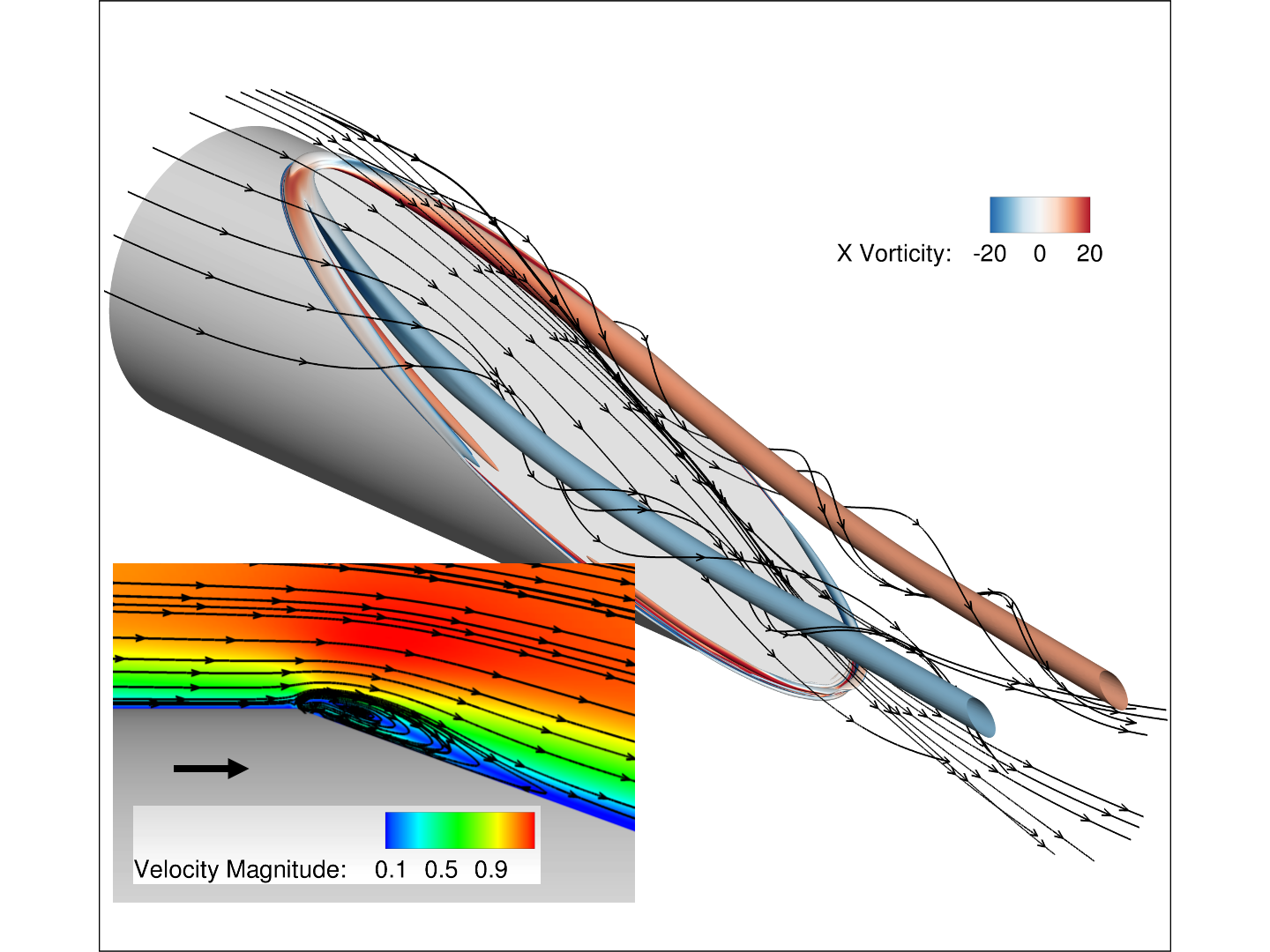}
    \caption{Mean flow}
  \end{subfigure}
      \caption{Longitudinal vortices behind the slanted body. An iso-surface of $Q=30$ is plotted, colored by streamwise vorticity $\omega_x$. The mean separation bubble near the upstream apex is shown in inset (b). The configuration is inverted with positive $y$-axis pointing downward for display purposes.} 
            \label{fig:vortmean}
\end{figure}

The overall features of the flow have been described in numerous works; the dominant feature arising from flow separation around the periphery of the base region is a streamwise-oriented vortex pair downstream \citep{bulathsinghala2017afterbody,garmann2019high2,Zigunov2020, ranjan2020mean}. 
For illustration and future reference, visualizations from the current Large Eddy Simulations (LES) are shown in figure~\ref{fig:vortmean}.
In figure~\ref{fig:vortmean}(a), a 
Q-criterion \citep{haller2005objective} iso-level ($Q=30$) is depicted, colored by streamwise vorticity, $\omega_x$, together with select superposed streamlines to provide an indication of the swirling motion in the instantaneous flow.
The counter-rotating streamwise vortex pair as well as the twisting motion of each vortex in the pair, are clearly evident.
Figure ~\ref{fig:vortmean}(b) displays the mean flowfield using the same flow variables.
In the mean sense, each vortex in the pair gradually assumes an axisymmetric form in the region outside of the immediate vicinity of the upstream apex.
\S~\ref{sec:flowfield} and \S~\ref{sec:meanvortex} provide further insights into the mean structure of the flowfield and lay the foundation to examine the spatio-temporal features of the vortex pair as influenced by a breakdown of the separated shear layer,  fluid entrainment, potential vortex-vortex interactions and meandering.

Depending on the Reynolds number, this structure persists in a range of $\phi \leq \ang{45}$;  descriptions at other upsweep angles in $\ang{24} \leq \phi \leq \ang{45}$ may be found in \citet{bulathsinghala2017afterbody, garmann2019high2, Zigunov2020} and \citet{ranjan2020mean}.
At even higher upsweep angles ($\phi > \ang{45}$) however, the flow changes to a turbulent wake regime typical of bluff-body wakes~\citep{morel1978effect, ranjan2020mean}.
Hysteresis or bi-stable states are possible, typically at higher angles than that considered here~\citep{morel1978effect,Zigunov2020, ranjan2020hysteresis} but these phenomena are outside the scope of the current effort.

From a practical point of view, the presence of the vortex pair results in higher drag and significant unsteadiness for long distances downstream due to the meandering (or wandering) of vortices, both of which are major concerns for designers.
For example, the unsteadiness and meandering motion can affect payload accuracy and paratrooper safety, as well as impose constraints on distance from trailing aircraft and airport operations~\citep{holzapfel2019assessment}. 
The overall goal of the present work is to: 1) describe the phenomenology of streamwise-oriented vortex pair dynamics, including  meandering, 2)  place this phenomenology in the context of stability  mechanisms, and 3) examine the low-rank behavior of the complex flowfield for potential model-order reduction.

The overwhelming number of studies on axial vortex pairs representative of the aircraft fuselages have been experimental in nature.  
Most early studies \citep{morel1978effect, britcher1991interference, epstein1994experimental, bury2013experimental} focus on more practical aspects such as drag due to upsweep angle.
Recently, however, details of the unsteady motion have become clearer through experiments using advanced diagnostics and high-fidelity simulations.
\citet{bulathsinghala2017afterbody} performed extensive experimental studies on slanted cylindrical bodies, similar to that of figure~\ref{fig:sketch} for five upsweep angles between \ang{24} and \ang{32}, and Reynolds numbers of $Re_D = 2\times 10^4$ and $2\times 10^5$ based on cylinder diameter $D$ and free-stream velocity $U_{\infty}$. 
Proper Orthogonal Decomposition (POD) of flowfields from high-resolution Particle Image Velocimetry (PIV) measurements reveal clear evidence of vortex meandering, similar to that observed in wing-tip~\citep{chow1997mean} and delta-wing~\citep{zhang2016interaction} vortices. 
Subsequently, \citet{jackson2019controlJ} used control techniques such as continuous and pulsed jets on the same configuration to  reduce drag; however, an undesirable side effect was an increase in vortex meandering.
Recently, \citet{Zigunov2020} also consider a similar topology at three upsweep angles $\phi = 20\degree, 32\degree$ and 45$\degree$, and a range of Reynolds numbers  between $Re_D = 2.5\times10^4$ and $5\times10^5$.
Based on a comparison of vortex strengths at different $\phi$, they observed that the core becomes weaker with the upsweep angle, and that may thus result in increased  meandering.
Experimental observations are generally limited to regions near the base; as such, data on the downstream evolution of vortex pair, or the dynamics of the turbulence field, is not accessible in regions where meandering can be prominent.

Fully resolved simulations have aided in filling this gap.
Large Eddy Simulations (LES) were performed by \citet{garmann2019high2} on one of the configurations of \citet{bulathsinghala2017afterbody}, specifically  at $Re_D=2\times10^5$ and $\phi = \ang{28}$.
They characterized the meandering of vortices downstream of the body in previously unachievable detail by simulating the entire configuration, including the upstream nose region.
The boundary layer transitioned in the mid-section of the cylinder with a Klebanoff-type (K-type) breakdown and eventually became turbulent near the downstream apex (see figure~\ref{fig:sketch} for definition). 
They found that the meandering amplitude is maximum below the upswept base where there is a strong interaction of the vortex with the surface, and its value decreases with the streamwise distance.
The phenomenology  and dynamics of meandering in the current problem are discussed in detail in \S~\ref{sec:phenomenology}, using a reliable vortex center detection technique.
The motion of each vortex about its corresponding mean location is characterized, as are correlations between the motions.
The evolution of energetic structures in the turbulent field with streamwise distance is also examined using POD in \S~\ref{sec:pod}.


The motions associated with vortex meandering are considerably broader than those observed specifically in the configuration of figure~\ref{fig:vortmean} and include, for example,  wingtip~\citep{del2011dynamics} and delta-wing vortices~\citep{ma2017symmetry}.
A consensus regarding underlying mechanisms in all these flows that sustain the displacement of vortex cores during meandering remains elusive.
For example, some early efforts~\citep{corsiglia1973rapid} ascribe meandering phenomena in wind tunnels to the disturbance environment \textit{i.e.,} free-stream turbulence effects in the test environment.
This attribution was motivated by the fact that some experiments on wingtip vortices~\citep{baker1974laser} observed coherent side-to-side motions, similar to those associated with wandering at zero angle of attack \textit{i.e.,} when no vortices are formed.
More recent experimental efforts \citep{devenport1996structure, beresh2010meander} have observed that wandering amplitude increases with downstream distance, which is indicative of the possible existence of an instability mechanism. 

Simplified situations with theoretical vortex models have greatly aided in developing a better understanding of the instability mechanism.
\citet{antkowiak2004transient} deduced the transient amplification of a disturbance to Lamb-Oseen vortex as being associated with the Orr mechanism as well as vortex induction.
Other efforts have incorporated complications associated with external strain and the presence of a wake on the stability of the flow.
The Batchelor vortex model that includes the effects of axial velocity, for example, has been fruitfully employed in some recent studies.  
At high swirl and Reynolds numbers, \citet{parras2007spatial} connected the stability of these vortices to the dynamics of actual large-aircraft vortices.
Through rigorous spatial stability analysis, they showed that at very high Reynolds number, absolute instabilities are present only in the presence of an axial velocity. 
\citet{edstrand2016mechanism, edstrand2018parallel} used temporal and spatial stability analyses of a model Batchelor vortex resembling a tip vortex behind a NACA0012 wing. 
A key finding was the presence of marginal elliptic stability \citep{kerswell2002elliptical} as a possible mechanism for meandering. 
This was also confirmed by \cite{cheng2019quantitative}, who examined the problem at a range of Reynolds numbers and angles-of-attack in water tunnel experiments and performed corresponding linear stability analyses of a fitted Batchelor vortex model.
Optimal perturbation analysis by \citet{navrose2019transient} for a trailing vortex system concluded that random perturbations near the wing surface, with relatively low initial energy, can non-linearly trigger vortex displacement.

While illustrative, studies on wake vortices modeled on isolated constructs of vortex models, may not fully represent the dynamics of vortex pairs, which could be either co- or counter-rotating~\citep{fabre2002optimal, leweke2016dynamics}. 
The temporal stability results of \citet{hein2004instability} on a pair of Batchelor vortices exhibiting long-wavelength (Crow) instability, found stronger modification of the instability characteristics for the vortex pair system relative to the isolated vortex. 
For wingtip vortices, the vorticity dynamics of an isolated vortex, especially those occurring at short wavelengths, can be a good representative of counter-rotating vortex pair, because the individual vortices are well separated, typically by the wingspan. 
On the other hand, the afterbody counter-rotating vortices, such as those under consideration, are much closer to each other as they both are formed on the same slanted cylinder surface. 
Hence, it is imperative to consider both the vortices in stability studies.
Such efforts, which consider both vortices in the pair, are scarce due to much higher computational cost than a single vortex, although some exceptions have been reported~\citep{jugier2020linear}. 
For the current afterbody configuration, interactions between the afterbody vortices in the pair were postulated to be related to low-frequency ($St_D \sim 0.4$, where $St_D$ is the Strouhal number based on diameter) oscillations observed in experiments~\citep{zigunov2020dynamics}. 
However, a rigorous stability analysis to confirm this conjecture is not available. 
The stability analysis in the present work, \S~\ref{sec:stability}, 
is therefore conducted by considering both vortices in the pair, each modeled as a Batchelor vortex. 
The analysis is performed at a downstream location, where a good description of the mean vortex can be obtained without much interference of the shear layer.



Although the causal dynamics is predicated on the underlying stability mechanisms and the corresponding stability modes, considerable insight may be obtained by post-processing the fully turbulent fluctuation field.
Several models, as described above, are available to characterize the behavior of mean vortex fields, but those that capture the primary building block coherent structures underpinning the dynamics of meandering as well as shear layer are relatively scarce. 
A final contribution of the current work is therefore to propose a low-rank, large-scale fluctuating flowfield representation that describes the vortex meandering phenomenon.
Such representations can be readily used to construct a reduced-order model (ROM) using Galerkin-projections~\citep{rowley2004model}.
Potential low-rank representation, and therefore the ROM of a complex turbulent flowfield, depends on the behavior of coherent structures, which are the next dominant flow features after the mean. 
These structures can be obtained from a suitable decomposition of the flowfield depending on the primary objective of the reduction. 
Among the flows of interest, \citet{iungo2015data} used selected dynamic mode decomposition (DMD) modes to construct a reduced-order model (ROM) for the downstream evolution and dynamics of wind turbine wakes. 
Similarly, \citet{gupta2019low} used local linear stability modes of the time-averaged mean flow to develop a low-order model of wake meandering.  
Their model successfully predicts the increase in meandering amplitude and the advancement of the onset of meandering closer to the turbine.


In the current work, \S~\ref{sec:lods}, we employ the POD modes extracted in \S~\ref{sec:pod} as a prelude to stability studies, to propose a low-rank representation for meandering.
POD is an effective way to facilitate the construction of a physics-based low-rank approximation to meandering phenomena since it provides optimum convergence in extracting spatial modes and time-coefficients in terms of the time-mean energy of the flow~\citep{aubry1991hidden}. 
The current low-rank approximation approach is similar to that used in some related flows such as \citet{mula2015study} for a spiraling vortex filament, or \citet{karami2019coherent} for tornado-like vortices.
However, in the present study, we adopt a more rigorous validation by not only considering the accuracy of the spatial structures due to these models but also their ability to recreate the actual meandering motion.  
This study can thus be directly relevant to the control efforts directed at reducing meandering. 
The models are examined at several streamwise locations downstream of the body to compare their performance.

\section{Numerical Procedure}\label{sec:num}
The configuration considered, shown earlier in figure~\ref{fig:sketch} is a three-dimensional axisymmetric cylinder with a trailing sectional cut based on the upsweep angle $\phi=\ang{20}$.
The chosen Reynolds number $Re_D=2.5 \times 10^4$ is one of those based on the experimental campaign of \citet{Zigunov2020}.
Although this $Re$ is low compared to actual flight conditions, several experimentalists~\citep{epstein1994experimental,bury2013experimental, bulathsinghala2019modified} note that the primary features of the flow behind an afterbody at a given upsweep angle of the base are largely independent of $Re$. 
High-fidelity simulations have independently confirmed this observation \citep{ranjan2020mean}.

As shown in figure~\ref{fig:sketch}, $x,y,z$  denote streamwise, vertical and spanwise  directions respectively in the chosen coordinate system. 
All geometrical parameters are non-dimensionalized based on cylinder diameter ($D$) and free-stream velocity $U_{\infty}$. 
Axial distances are designated in terms of $L = D~\text{cot}(\phi)$, the distance between the upstream and downstream apexes, which are marked in figure~\ref{fig:sketch}.
The vertical plane at $z=0$ will be referred to as a mid-plane or symmetry plane. 
The compressible Navier-Stokes equations are solved in a curvilinear $(\xi,\eta,\zeta)$-coordinate system:
\begin{equation} 
\frac{\partial}{\partial \tau} \biggl(\frac{Q}{J}\biggl) = -\biggl[\biggl(\frac{\partial F_i}
{\partial \xi} + \frac{\partial G_i}{\partial \eta} + \frac{\partial H_i}
{\partial \zeta}\biggl) + 
\frac{1}{Re} \biggl(\frac{\partial F_v}{\partial \xi} + \frac{\partial G_v}{\partial \eta} 
+ \frac{\partial H_v}{\partial \zeta}\biggl) \biggl]                
\label{NSE}
\end{equation}      
\noindent where, $Q=[\rho,\rho u,\rho v, \rho w, \rho E]^T$ denotes the solution vector, defined in terms of the fluid density $\rho$, Cartesian velocity components $(u,v,w)$ and total  specific energy $E={T}/{(\gamma-1)M^2}+(u^2+v^2+w^2)/2$. 
Here, $M$ is the Mach number of the flow, which is set to 0.1. 
$\gamma$ is the ratio of the specific heats and $T$ is the fluid temperature.
$J=\partial{(\xi,\eta,\zeta,\tau)}/\partial{(x,y,z,t)}$ denotes the Jacobian of the transformation from Cartesian $(x,y,z)$  to curvilinear $(\xi,\eta,\zeta)$-coordinate system. 
The above governing equations are complemented by the ideal gas law, written in non-dimensional variables as $p=\rho T/{\gamma M^2}$. 
Sutherland's law is used to express fluid dynamic viscosity $\mu$ as a function of temperature $T$.  

\begin{figure}
\centering
\includegraphics[width=0.95\textwidth, trim={0.1cm 0.1cm 0.1cm 0.1cm},clip]{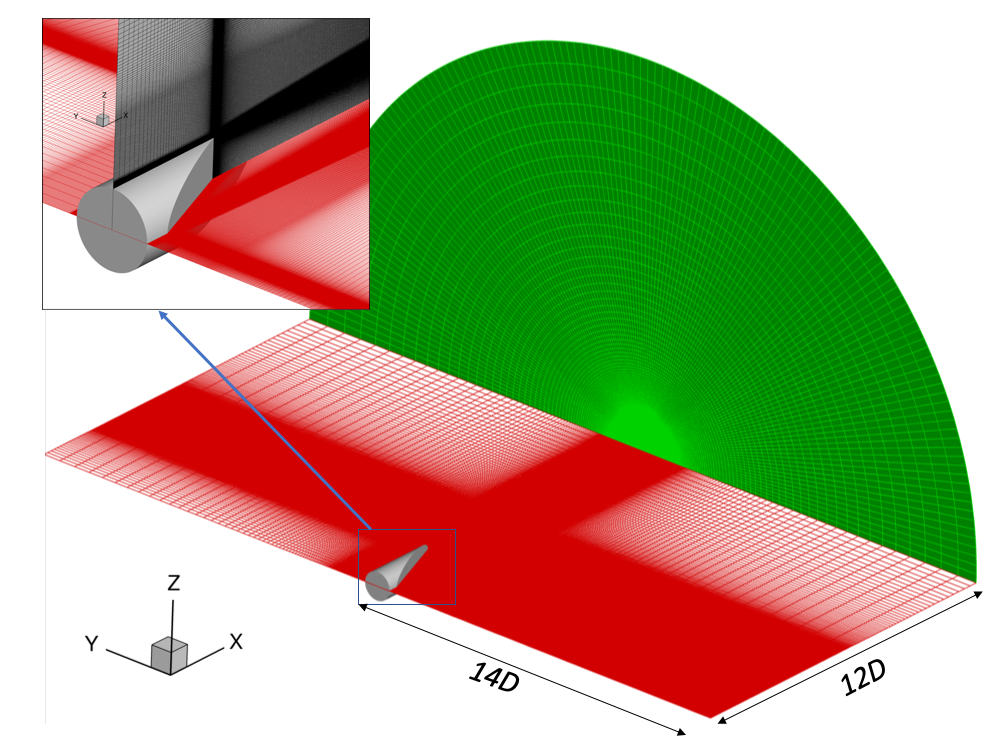}
\caption{Computational domain and grid employed for current afterbody simulations. Grids on the center $xy$-plane of the base, and at downstream farfield location are shown to illustrate the grid topology. Cylindrical grid with origin at the center of the base and periodicity in the azimuthal direction is used. Domain extends $12D$ and $14D$ in  downstream (green) and radial farfield (red) directions respectively. Bottom half of the computational domain is not shown. }  
\label{fig:mesh}
\end{figure}

The computational domain as well as grid topology employed are shown in figure~\ref{fig:mesh}.
The nose section of the body, shown in figure~\ref{fig:sketch}, is not simulated; instead, the inflow to the domain is placed upstream of the base at a location where experimental profiles are available.
At this location, results from precursor simulations are performed on an axisymmetric body to match the experimental conditions used in \citet{Zigunov2020}. 
Viscous no-slip and zero normal pressure gradient boundary conditions are used on the cylinder surface, whereas freestream conditions were applied at the radial farfield boundary which is placed approximately $14D$ away from all surfaces. 
At the downstream farfield boundary,  Neumann boundary condition with zero-gradient is specified for all flow variables. 

A structured cylindrical grid is considered with a total of $485$, $418$ and $285$ points in the longitudinal, radial and azimuthal  directions respectively.
The local mesh density is comparable to that employed by \citet{garmann2019high2} for a much higher Reynolds number. 
A $p$-refinement study using fourth- and sixth-order compact difference schemes with sixth-order and eighth-order implicit filters, demonstrates mesh independence as discussed in \citet{ranjan2020mean}.
The filter serves both to ensure numerical stability, as well as to provide an implicit subgrid closure mechanism \citep{MathewPhysFluids2003}. 
Time-stepping is performed using the implicit Beam-Warming scheme \citep{beam1976implicit} with the diagonalization of \citet{pulliam1981diagonal}. 
The simulations are performed with a constant non-dimensional time-step of $\Delta t = 2.5\times10^{-4}$, which is sufficient to ensure temporal accuracy.
Further details of the numerical algorithm used for the simulations can be found in \citet{gaitonde1998high,visbal2002use}.  



\section{Overview of Flowfield}\label{sec:flowfield}
The representative three-dimensional instantaneous and mean flow structures were shown earlier in figure~\ref{fig:vortmean}.
The latter was obtained by computing the time average over $700{,}000$ snapshots encompassing total non-dimensional time of $T_c \equiv tU_\infty/D=175$. 
The turbulent flowfield arises as a result of the strong interaction between the separating shear layer and vortices in the region adjacent to the upswept base.
Looking downstream, the left and right vortices have dominantly positive and negative streamwise vorticity components respectively.
In the instantaneous flow, figure ~\ref{fig:vortmean}(a), there are smaller convecting structures that are washed out in the time-averaged sense (figure ~\ref{fig:vortmean}(b)). 

The formation of the vortical structures may be summarized as follows.
The boundary layer approaching the edge separates around the entire periphery to form a shear layer.  
The segment separating near the upstream apex rolls up to form a bubble structure in the symmetry plane (shown in the inset of figure~\ref{fig:vortmean}(b)).
The flow separating around the periphery is entrained into each vortical structure.
The continuity of the vortex on the symmetry plane permits an alternative description of the vortex pair as leg components of a single  horseshoe-like vortical structure.  
Each vortex leg then lifts away from the upswept base, at about the midway point to form the streamwise oriented pair.
After the vortices orient away from the surface, the free shear layer arising from separation at the  downstream apex continues to entrain fluid into the vortices for some distance behind the body. 
This interaction, although not crucial in the formation phase of the vortices, affects the motion downstream.

\begin{figure}
\centering
  \begin{subfigure}[b]{0.6\textwidth}
        \includegraphics[width=\textwidth, trim={0.1cm 2.8cm 0.1cm 2.5cm},clip]{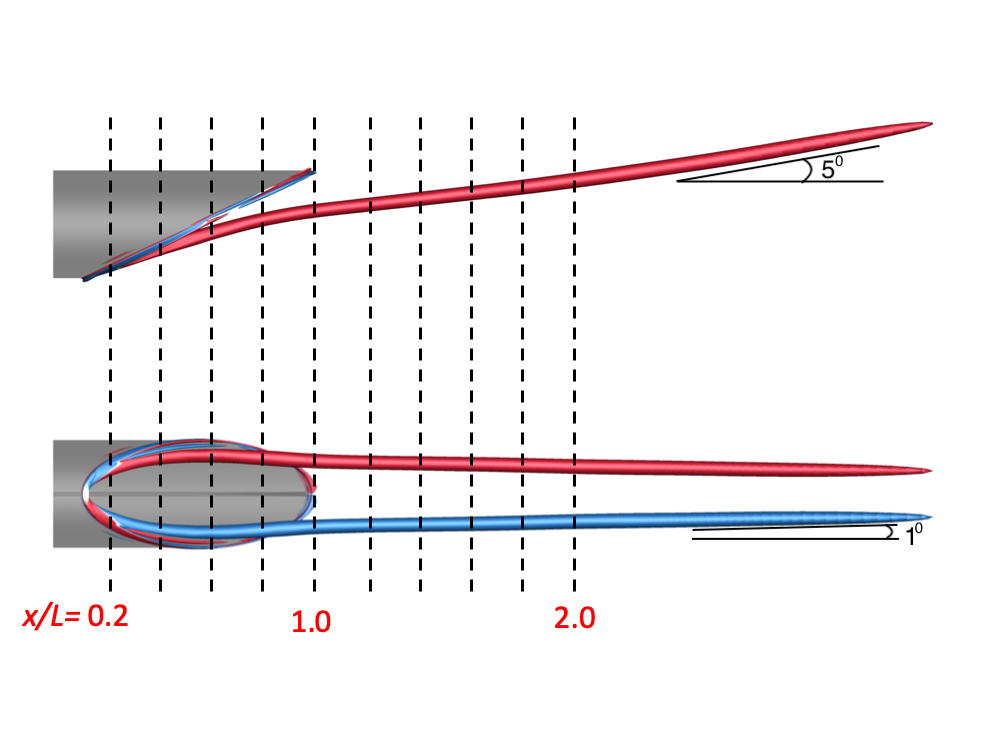}
        \caption{Streamwise vortices using $Q$-criterion.}
    \end{subfigure}\\
  \begin{subfigure}[b]{0.48\textwidth}
        \includegraphics[width=\textwidth]{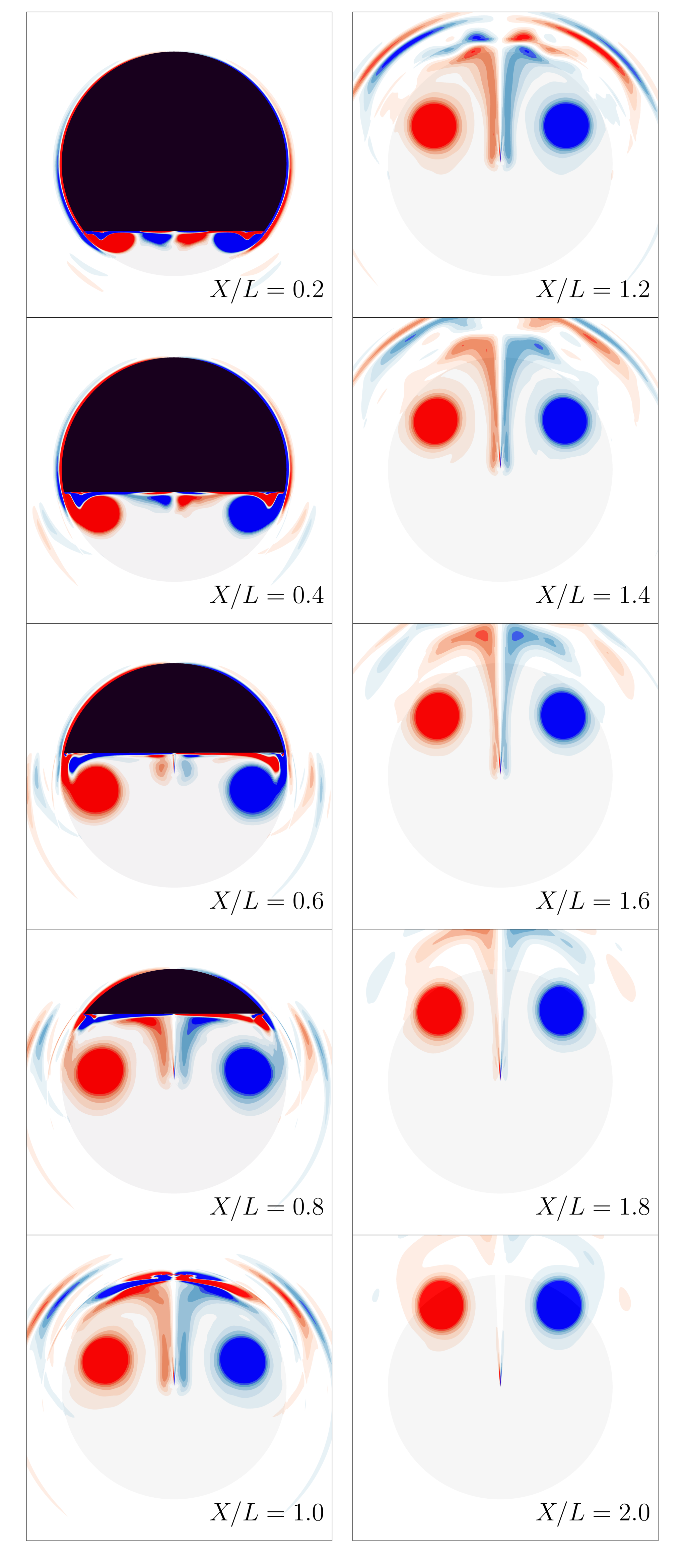}
    \caption{Mean}
  \end{subfigure}\quad
    \begin{subfigure}[b]{0.48\textwidth}
          \includegraphics[width=\textwidth]{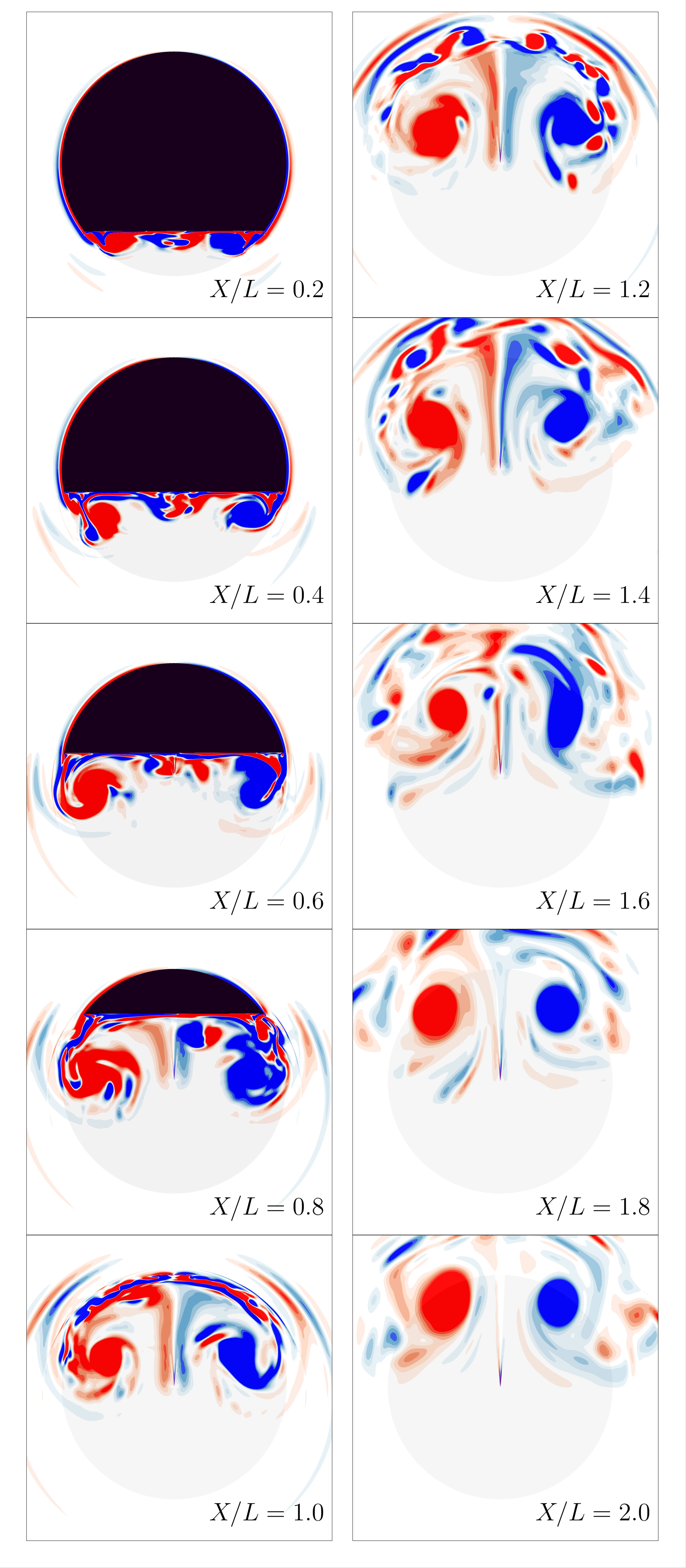}
    \caption{Instantaneous}
  \end{subfigure}
\caption{Evolution of streamwise vortex along the base of the upswept surface through $\omega_x$ contours at 11 levels between -3 and 3.   Crossflows in the intervals for 0.2 between $X/L=0.2$ and $2.0$ for the mean~(b) and a representative snapshot~(c) are shown. (a) shows the locations in front and top view of vortices.}
\label{fig:2dsections}
\end{figure}


A granular description of the formation of the vortex pair is now provided for reference.
The mean trajectory of the vortices is displayed using side and bottom views in figure~\ref{fig:2dsections}(a), again with the same isolevels as in figure~\ref{fig:vortmean}.
The side view indicates that after a streamwise turn immediately after lifting off from the base,  the vortex remains inclined to the freestream at an approximate angle of \ang{5} even far downstream.
The bottom view shows that the two mean vortices approach each other at a very small angle of approximately \ang{1} towards the symmetry plane.

%
Figures~\ref{fig:2dsections}(b) and (c) show the mean and instantaneous streamwise vorticity components, respectively, using cross-sectional contours at ten different locations between $0.2 \leq X/L \leq 2.0$, marked by dotted lines in figure~\ref{fig:2dsections}(a).
The mean vorticity field at $X/L=0.2$ displays a teardrop-shaped structure, with opposing signs in the dominant lobes, which are located symmetrically about the mid-plane, $z=0$.
For future reference, the two vortices in the pair are designated as \textbf{L} and \textbf{R} vortex based on whether, looking downstream, they are located to the left or right of the symmetry plane respectively.
Each vortex in the pair on the base is fed by the shear layer and gradually evolves into an axisymmetric structure at about $X/L \sim 0.6$.
As evident from (a), the vortex detaches from the surface; thus the connection with the feeding sheet is clearly weaker at $X/L=0.8$.

At the downstream apex of the upswept base ($X/L=1.0$), the vortices are completely detached from the surface and the direct effect of the base shear layer on the fully developed mean vortex pair gradually diminishes.
The process continues with streamwise distance with the vortices becoming essentially independent from the immediate effects of the base.
Secondary and tertiary vortices are also present beneath the primary vortex, as clearly evident in planes intersecting the base.
These have opposite and same signs, respectively, as the corresponding primary vortex in the pair and their influence far downstream of the base appears to be a higher-order effect.
Further details of these mean flow features for $\phi$ values of \ang{20}, \ang{32} and \ang{45}, together with various aspects of the flow on the base plane, such as surface oil flow and pressure patterns, as well as drag, have been discussed extensively in~\citet{ranjan2020mean}.

Instantaneous structures of the vortices, as well as their meandering downstream, are shown in figure~\ref{fig:2dsections}(c) on the same planes as in (b) for a representative snapshot.
Although the flow is unsteady and turbulent along the entire base region, the instantaneous flow displays the dominant signature of primary vortices, consistent with figure~\ref{fig:vortmean}(a).
However, the axisymmetric nature is not evident, and the meandering of vortices is discernible in the streamwise vorticity field from $X/L=0.8$ downstream as a result of interactions with the shear layer.
Animations indicate a swirling motion of the vortex center with time, as discussed further in \S~\ref{sec:meandering}.

Each vortex contains some regions of opposite-signed (secondary) vorticity field, 
these have some similarity to the ``vortical substructures" described in \citet{gordnier2005compact}, who note their presence around the primary cores of delta-wing vortices.
Their genesis is associated with the vortex-surface interaction and eruptions from the boundary layer on the upswept base surface.
Their influence on the unsteady motion is significant near the base of the bluff body, as they interact with the primary vortices \citep{yaniktepe2004flow, Gursul2005}.
Downstream of the base, these interactions diminish,  and between $X/L=1.2$ and $X/L = 2.0$, a perceptible decrease of intensity in the meandering is evident, as the effects of the near-surface interactions diminish.  
The entrainment of flow separating at the downstream apex into the central region between the vortex cores continues until further downstream.
These observations are generally consistent with those of \citet{garmann2019high2}, who performed LES of the entire configuration (including forebody) at a higher Reynolds number of $Re_D=2.0\times10^5$.

\section{Characterization of Mean Vortex Pair}\label{sec:meanvortex}

\begin{figure}
\centering
      \includegraphics[width=0.6\textwidth]{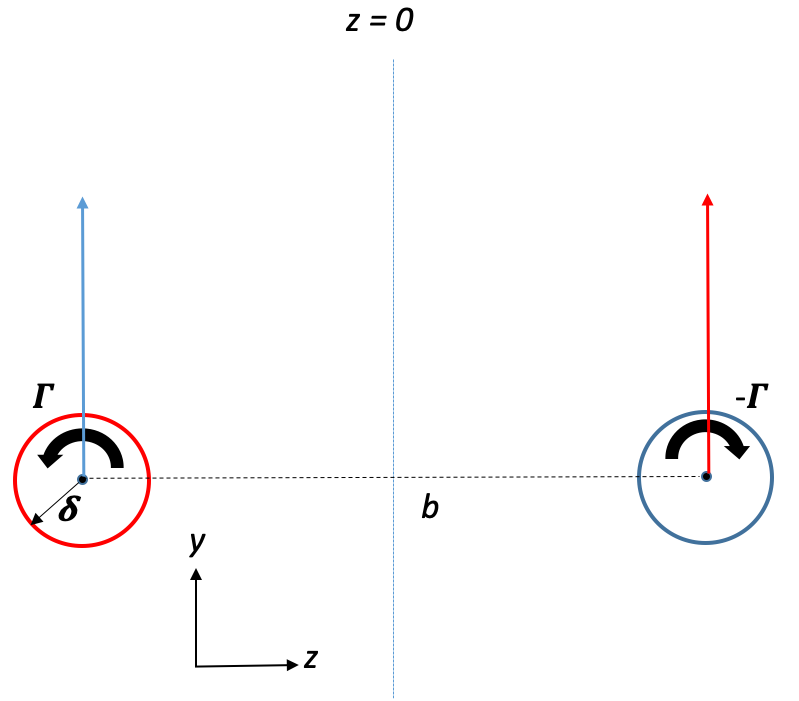}
 \caption{Pair of counter-rotating vortices at a streamwise location. $\delta$-radius and $|\Gamma|$-circulation of vortex. $b$-separation between vortices.} \label{fig:vortex_sketch}
\end{figure}

The mean vortex pair is discussed in a canonical sense to identify key parameters of interest; this also sets the stage for subsequent examination of underlying instability mechanisms. 
The afterbody flow on the slanted cylinder presents a classical counter-rotating vortex pair of equal strength, shown schematically in figure~\ref{fig:vortex_sketch}.
The main parameters of interest are the radius of the vortex core, $\delta$, the distance between the centroids of the vortices, $b$, and the circulation of the vortices, $\Gamma$. 
The \textbf{L} and \textbf{R} vortices thus lie at $X_L=-b/2$ and $X_R=b/2$ symmetrically about the midplane.  

The coordinates of the vortex centers are a primary quantity in the spatial evolution of the mean vortices~\citep{lombard2016implicit}.
In the literature on flow behind afterbodies, the maximum value of the $Q$-criterion \citep{bulathsinghala2017afterbody} or entropy \citep{garmann2019high2} is typically used to identify the core center.
In this work we use the $\Gamma_1$/$\Gamma_2$ approach proposed in \citet{graftieaux2001combining}, employed previously by \citet{bell2016dynamics} and concurrently by \citet{ranjan2020mean,Zigunov2020}. 
The method provides the center and boundary of a vortex by considering the local convection velocity field in the region of interest.
Once the vortex is identified, the circulation may be computed using the area integral of vorticity.

\begin{figure}
\centering
  \begin{subfigure}[b]{0.45\textwidth}
        \includegraphics[width=\textwidth, trim={0cm 0cm 0cm 0cm},clip]{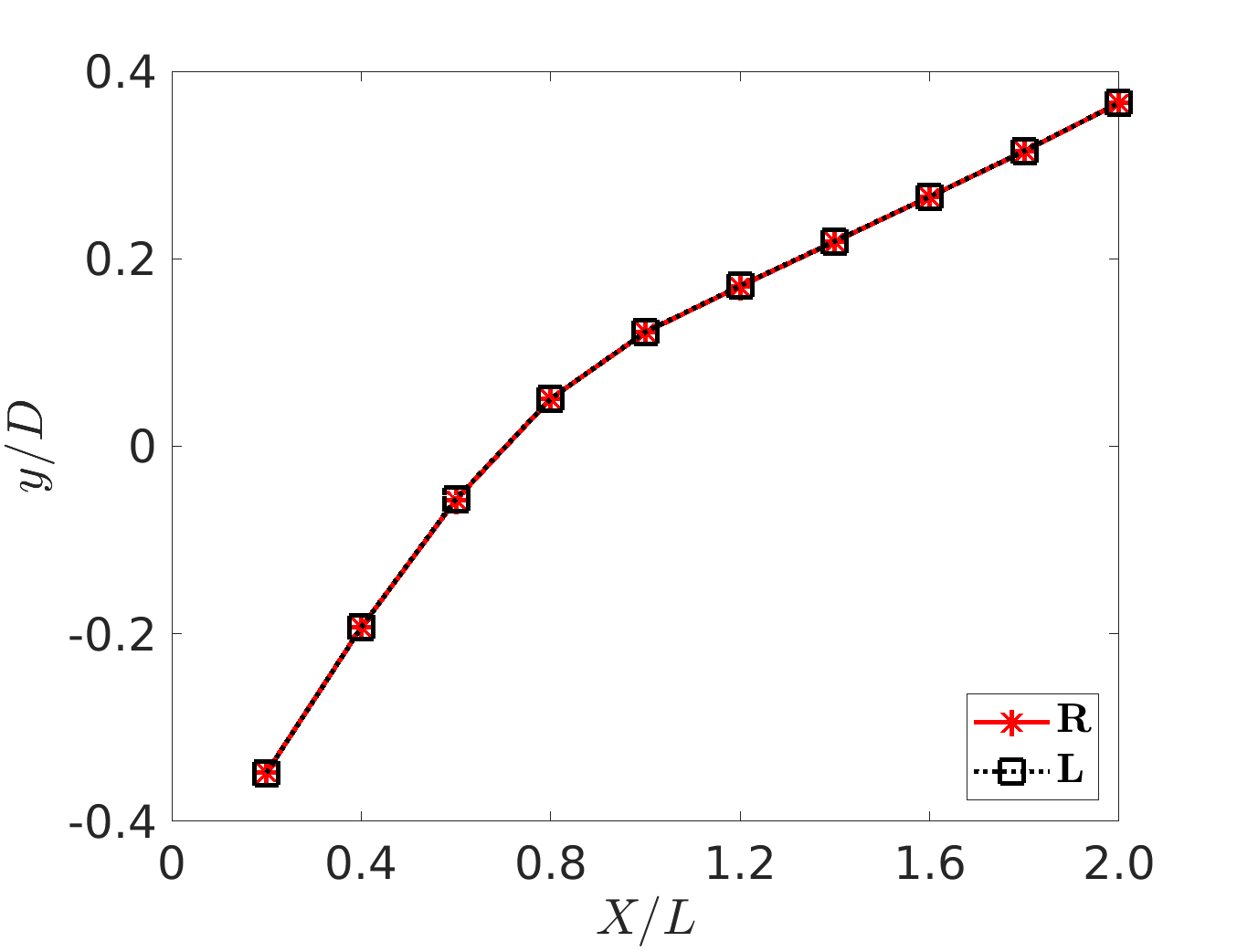}
        \caption{}
    \end{subfigure}\quad
  \begin{subfigure}[b]{0.45\textwidth}
        \includegraphics[width=\textwidth]{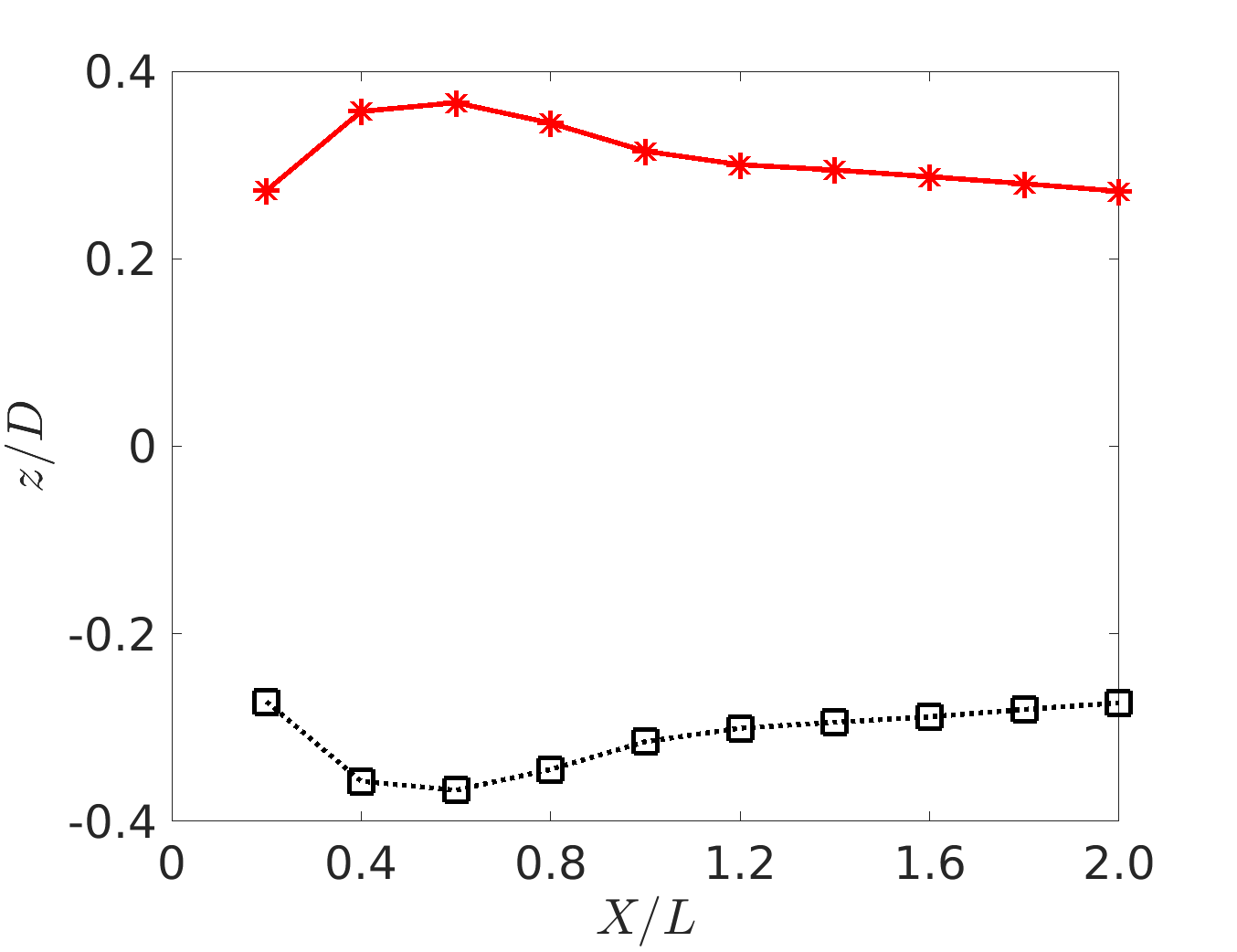}
    \caption{}
  \end{subfigure}\\
  \begin{subfigure}[b]{0.45\textwidth}
        \includegraphics[width=\textwidth, trim={0cm 0cm 0cm 0cm},clip]{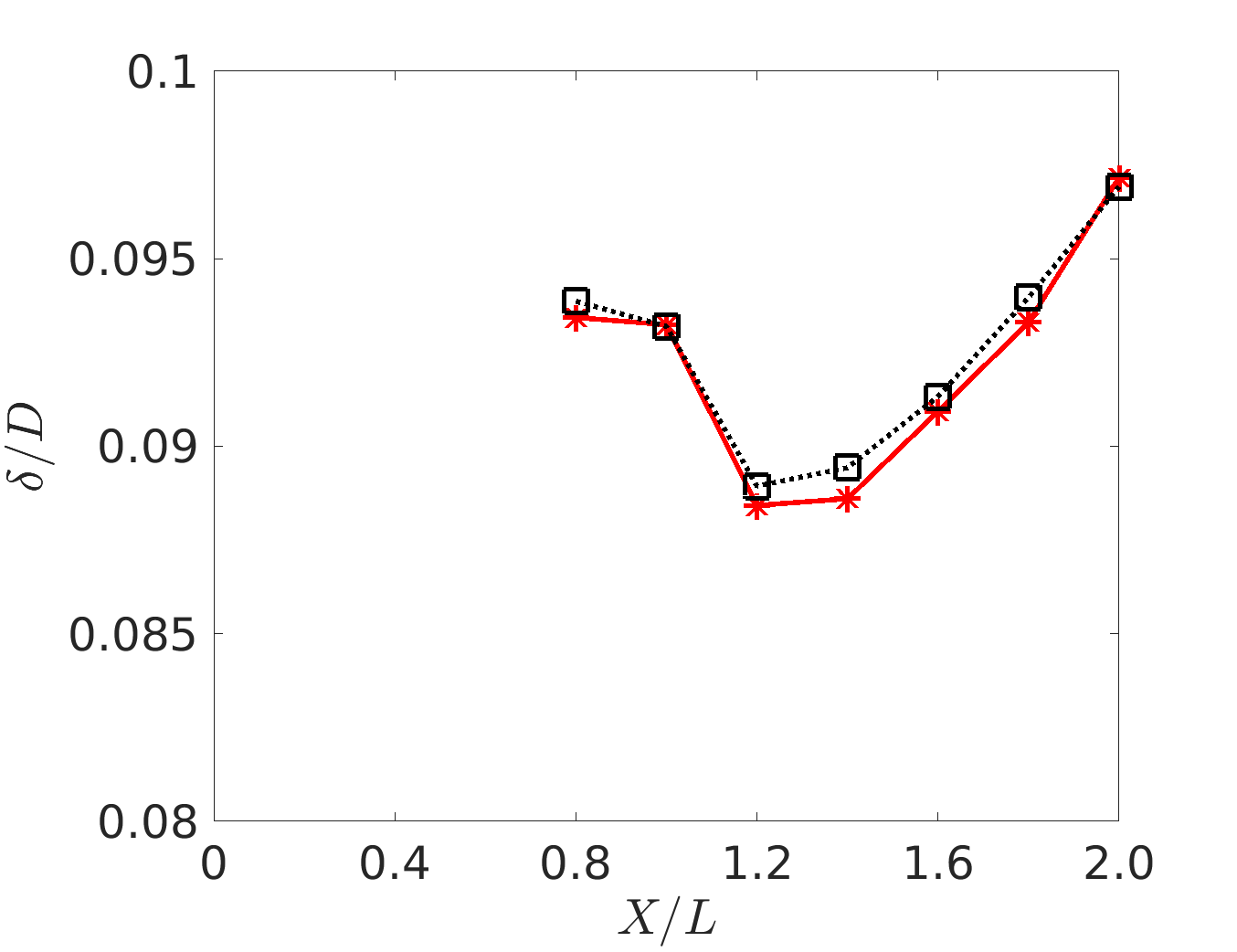}
        \caption{}
    \end{subfigure}\quad
  \begin{subfigure}[b]{0.45\textwidth}
        \includegraphics[width=\textwidth]{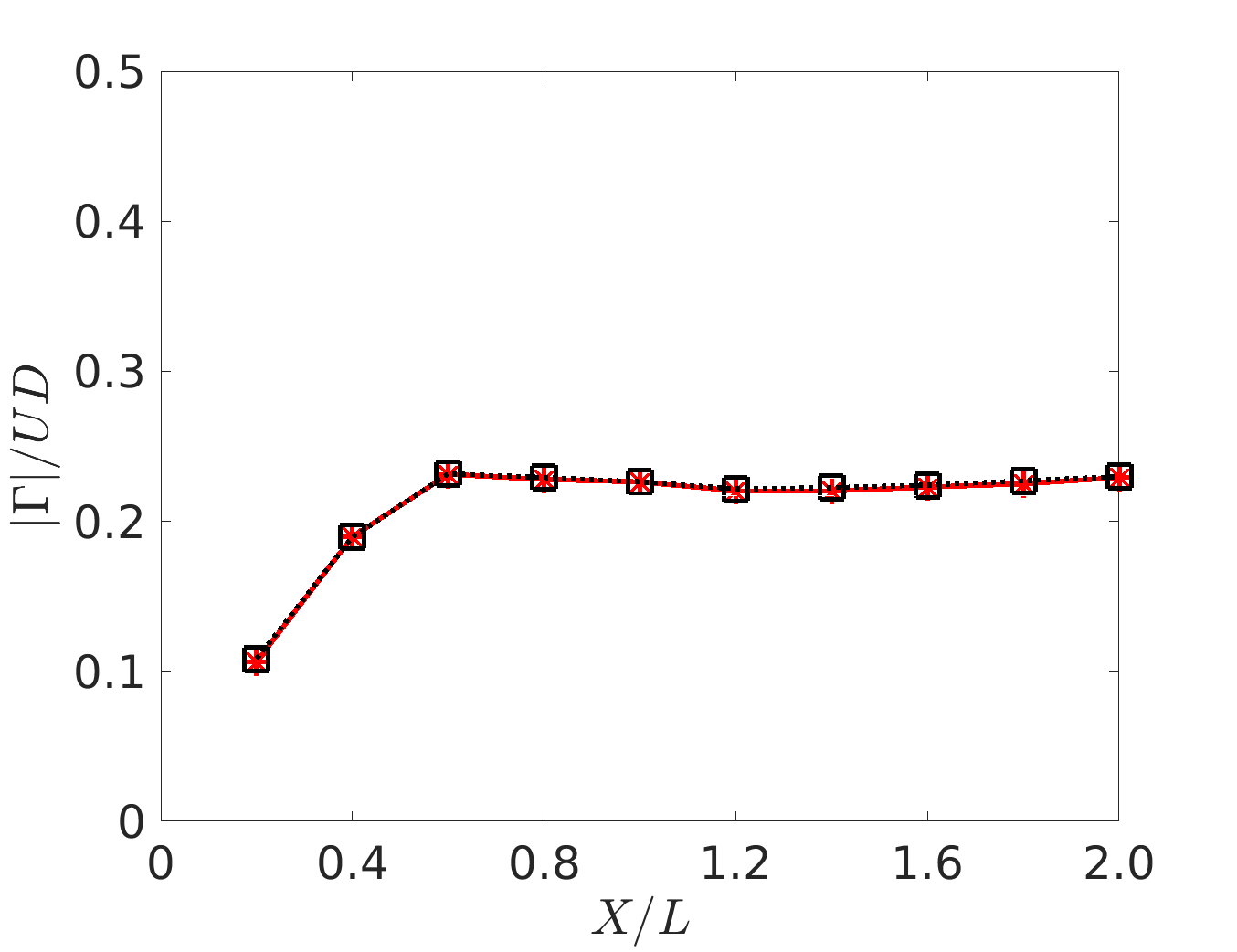}
    \caption{}
  \end{subfigure}
  \caption{Mean streamwise vortex location at different $X/L$. (a) Vertical location, (b) Spanwise location, (c) Radius and (d) Circulation of the core.} 
  \label{fig:meanvortex}
\end{figure}

Figure~\ref{fig:meanvortex}(a,b) respectively plot the vertical and spanwise locations of the \textbf{L} and \textbf{R} vortex cores. 
Figure~\ref{fig:meanvortex}(a) indicates that both vortices develop in the same manner along the streamwise distance, $x$, as anticipated, and in fact, Figure~\ref{fig:meanvortex}(b) displays symmetry about $z=0$.
A closer examination reveals that after the formation of the tear-drop shaped vortex at $X/L=0.2$, the distance between the vortices grows until $X/L\sim 0.6$.
This is consistent with the peripheral development of the 3D vortex structure shown in figure~\ref{fig:vortmean}(b).
Beyond $X/L\sim 0.6$, the centers approach each other until $X/L\sim 1.0$ as they conform to the elliptic shape of the base. 
Further downstream, beyond $X/L\sim 1.2$, the vortices achieve an asymptotic state inclined at about \ang{5} to the free-stream flow (see figure~\ref{fig:2dsections}(a)).
For future reference, the precise inter-vortical separation, $b$, is reported as a function of $x$ in Table~\ref{tab:vortex}, along with several other properties discussed below.
\begin{table}
 \begin{center}
\def~{\hphantom{0}}
 \begin{tabular}{lcccc}
 \toprule
          $\displaystyle \frac{X}{L}$  & $\displaystyle \frac{\delta}{D}$   &   $\displaystyle \frac{b}{D}$ & $\displaystyle \frac{\delta}{b}$  & $\displaystyle \frac{|\Gamma|}{UD}$  \\[3pt]
      \midrule
      0.2   & - & 0.5450  & - & 0.1070 \\
      0.4   & - & 0.7143 & - & 0.1895 \\
      0.6  & - & 0.7333 & - & 0.2315 \\
      0.8   & 0.0935  & 0.6893 & 0.1356  & 0.2288 \\
      1.0 & 0.0932 & 0.6297 & 0.1480 & 0.2262 \\
      1.2   & 0.0884  & 0.6010 & 0.1471 & 0.2210 \\
      1.4   & 0.0886 & 0.5890 & 0.1504 & 0.2218 \\
      1.6  & 0.0909 & 0.5761 & 0.1578 & 0.2234 \\
      1.8   & 0.0933 & 0.5607 & 0.1664 & 0.2255 \\
      2.0 & 0.0972 & 0.5456 & 0.1781 & 0.2295 \\
            \bottomrule
 \end{tabular}
 \caption{Properties of mean vortex at different streamwise locations. Note that there is less than 1\% difference between left and right vortices in computation of parameters such as $\delta$ and $|\Gamma|$. The reported quantities are thus averaged between the two. See sketch \protect\ref{fig:vortex_sketch} for definitions of parameters.}
 \label{tab:vortex}
 \end{center}
\end{table}


Figures~\ref{fig:meanvortex}(c) and~(d) depict the evolution of the radius and circulation of each vortex leg.
The radius of the vortex, $\delta$, is plotted only after it is fully developed and becomes axisymmetric at $X/L\sim 0.6$.
Figure~\ref{fig:meanvortex}(c,d), show that both \textbf{L} and \textbf{R} vortices nearly overlap, indicating acceptable convergence of the mean. 
Thus, both vortices are nearly identical except for the fact that they are counter-rotating in nature.
The results also indicate that both the area as well as circulation increase until $X/L\sim 0.6$, \textit{i.e.,} in the region where the vortex is fed through the shear layer (see figure~\ref{fig:2dsections}(a)). 
Once the vortex lifts away from the surface and gradually becomes axisymmetric, both radius and circulation decrease. 
The slight increase beyond $X/L=1.2$ is associated with the entrainment of separated fluid from the downstream apex.
For completeness, the radius and circulation of the mean vortex are tabulated in Table \ref{tab:vortex}.

\begin{figure}
\centering
  \includegraphics[width=0.95\textwidth, trim={0.2cm 0.2cm 0.2cm 0.2cm},clip]{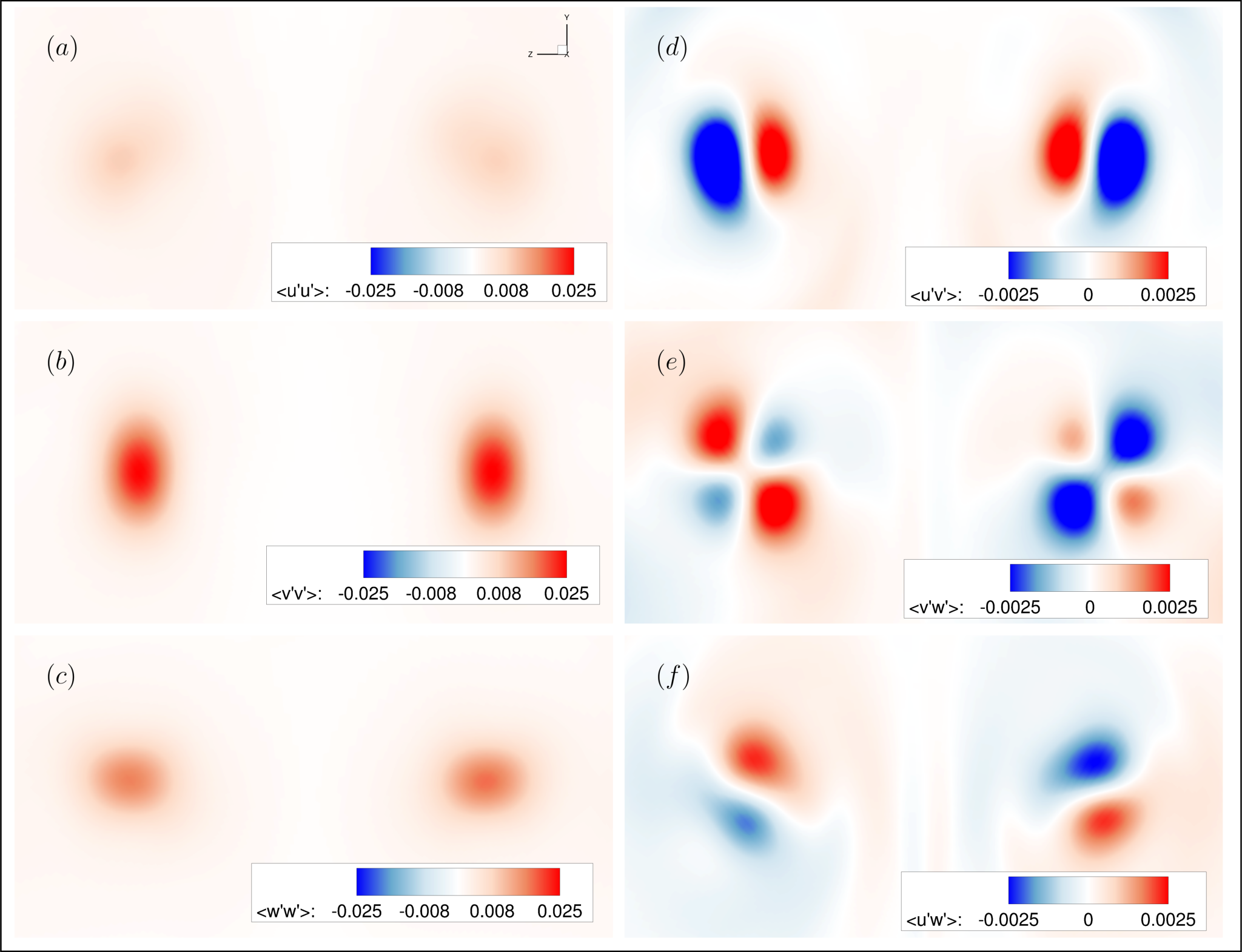}  
 \caption{isolines Reynolds stresses at $X/L=2.0$ in $yz$-plane. (a): $\langle u'u' \rangle$; (b): $\langle v'v' \rangle$; (c): $\langle w'w' \rangle$; (d): $\langle u'v' \rangle$; (e): $\langle v'w' \rangle$; (f): $\langle u'w' \rangle$. } 
  \label{fig:stresses}
\end{figure}

In addition to the kinematic nature of the vortex pair, examined through the velocity field and its variants, key insights into the structure emerge from an analysis of the Reynolds stress distribution.
For example,  in the context of wingtip vortices, \citet{zeman1995persistence} found that vortical structures suppress the Reynolds shear stress and therefore the vortex-core growth rate is controlled only by molecular viscosity; as such, vortex turbulence decays with distance.
Figure~\ref{fig:stresses} shows the distribution of normal and shear stresses on the $yz$-plane  near the vortex pair at the location $X/L=2.0$. 
The normal stresses, figure~\ref{fig:stresses}(a-c)  are symmetric about the mid-plane as expected.
$\overline{v'v'}$ and  $\overline{w'w'}$ are almost equal in magnitude (maximum non-dimensional value is 0.08) and considerably  larger than $\overline{u'u'}$.
Also, the distributions of $\overline{v'v'}$ and $\overline{w'w'}$ are elliptic in nature, with major axes oriented in the respective directions of the fluctuations. 

The shear stresses (figure~\ref{fig:stresses}(d-f)) are an order-of-magnitude smaller than the normal stresses and hence are plotted on a different scale.
Reynolds shear stress mediates the extraction of energy from the mean flow by turbulence, and therefore in all the stresses, the peak is off-center due to zero turbulence production at the centerline of the vortex \citep{zeman1995persistence}. 
The structure of in-plane shear stress, $\overline{v'w'}$, shows four lobes, similar to those obtained in \citet{zeman1995persistence} for a turbulent wing tip vortex in the farfield region, using both Reynolds stress closure model as well as through experiments.  
The structures of correlating in- and out-plane shear stresses, $\overline{u'v'}$ and $\overline{u'w'}$ have additional noteworthy features.  
These stresses indicate the strain in the field due to fluctuations in the axial velocity.
Briefly, these stresses show two-lobe structures similar to $|m|=1$ elliptic modes~\citep{kerswell2002elliptical}.
These manifestations are associated with displacements of vortex cores as discussed later.
The shear stresses are anti-symmetric with respect to the mid-plane due to the counter-rotating nature of vortices.



\section{Meandering of Streamwise Vortices}\label{sec:meandering}
\subsection{Phenomenology}\label{sec:phenomenology}
The motion of each vortex is now discussed by examining their behavior at location $X/L=2.0$, where the immediate effects of the base region have diminished. 
The definition of vortex meandering  follows \citet{green1991unsteady} who characterize it as the fluctuation of the vortex core  with time at a specific downstream location.
\begin{figure}
\centering
        \includegraphics[width=0.9\textwidth, trim={2.7cm 3cm 2.7cm 3cm},clip]{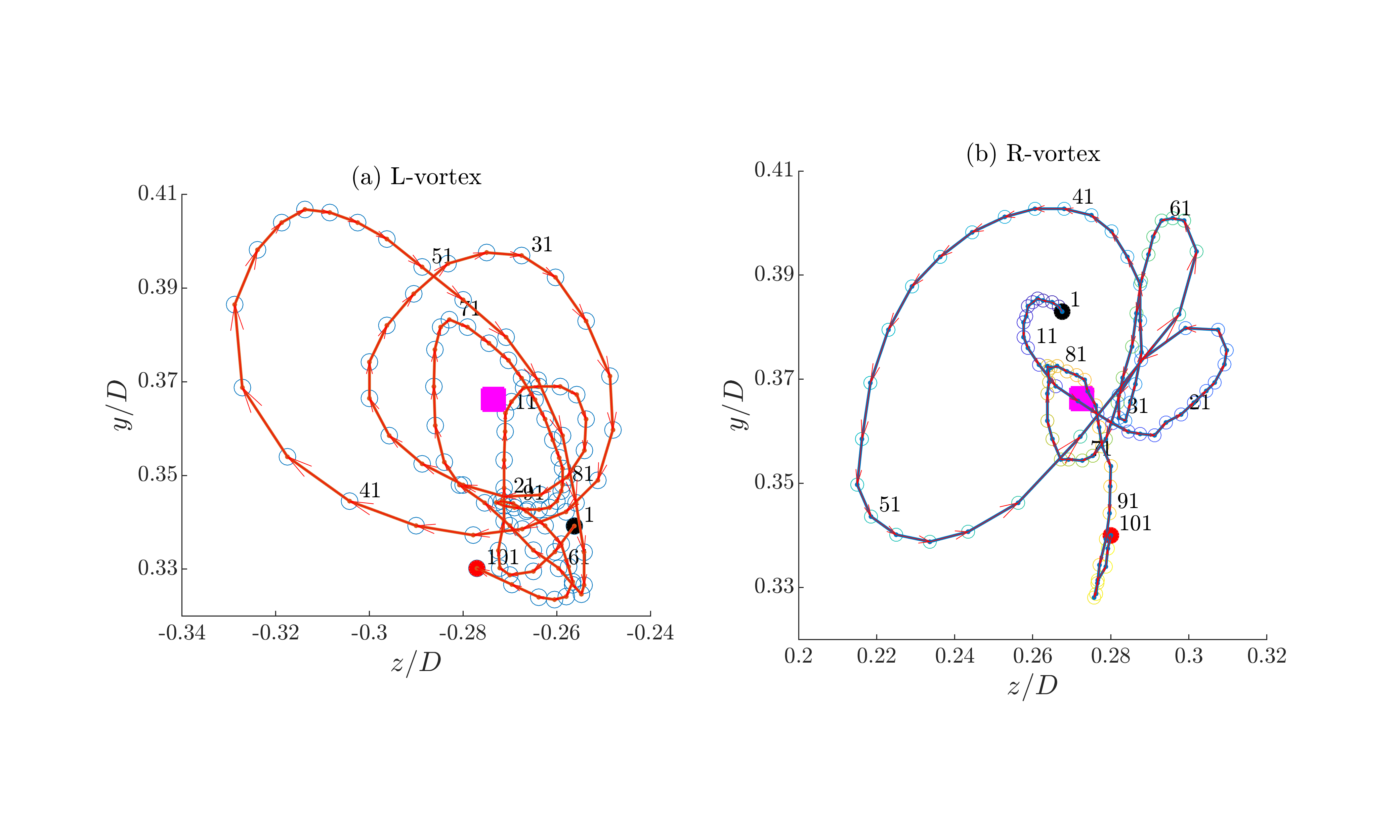}
 \caption{Phase-portrait of instantaneous vortices at $X/L=2.0$. The location of the mean vortex is shown as the large filled square. Black and red circles respectively indicate the beginning and end of the trajectory.} 
  \label{fig:meandering1}
\end{figure}
The locus of vortex cores in time is presented in the form of a phase portrait in figure~\ref{fig:meandering1}.
The trajectory of each core using $100$ snapshots, separated by a non-dimensional time $\Delta t = 0.025$, is shown on the crossflow ($yz$) plane.
The initial and final positions of each core are marked with black and red circles, and the position of the mean vortex core is shown as a large magenta square.
While the core of the \textbf{L}-vortex meanders in the clockwise direction, the meandering trajectory of the \textbf{R}-vortex is counter-clockwise. 
The sense of rotation is the same as the swirl of the vortex, similar to observations on delta wings \citep{gordnier2005compact}.
The instantaneous location of each vortex core deviates significantly from the mean, with $\Delta d_{max} / D = 0.05$, where $\Delta d_{max}$ is the maximum displacement of instantaneous core location from the mean. 
In order to quantify this phenomenon, the meandering amplitude $a_m$, defined as the average of distance between instantaneous core ($z_i,y_i$) and mean core locations ($z_c,y_c$), is measured:
\begin{eqnarray} \label{eq:amplitude}
 \displaystyle a_m = \sqrt{\frac{1}{N}\sum\limits_{i=1}^N (z_i - z_c)^2 + (y_i - y_c)^2} 
\end{eqnarray}
$a_m$ values are $0.0288$ and $0.0282$ for \textbf{L} and \textbf{R} vortex, respectively. 
Similar values of the amplitude for both the vortices indicate their statistically analogous behavior despite seemingly random motion. 



\begin{figure}
\centering
\begin{subfigure}[b]{0.5\textwidth}
 \includegraphics[width=\textwidth, trim={0.2cm 0.2cm 0.2cm 0.2cm},clip]{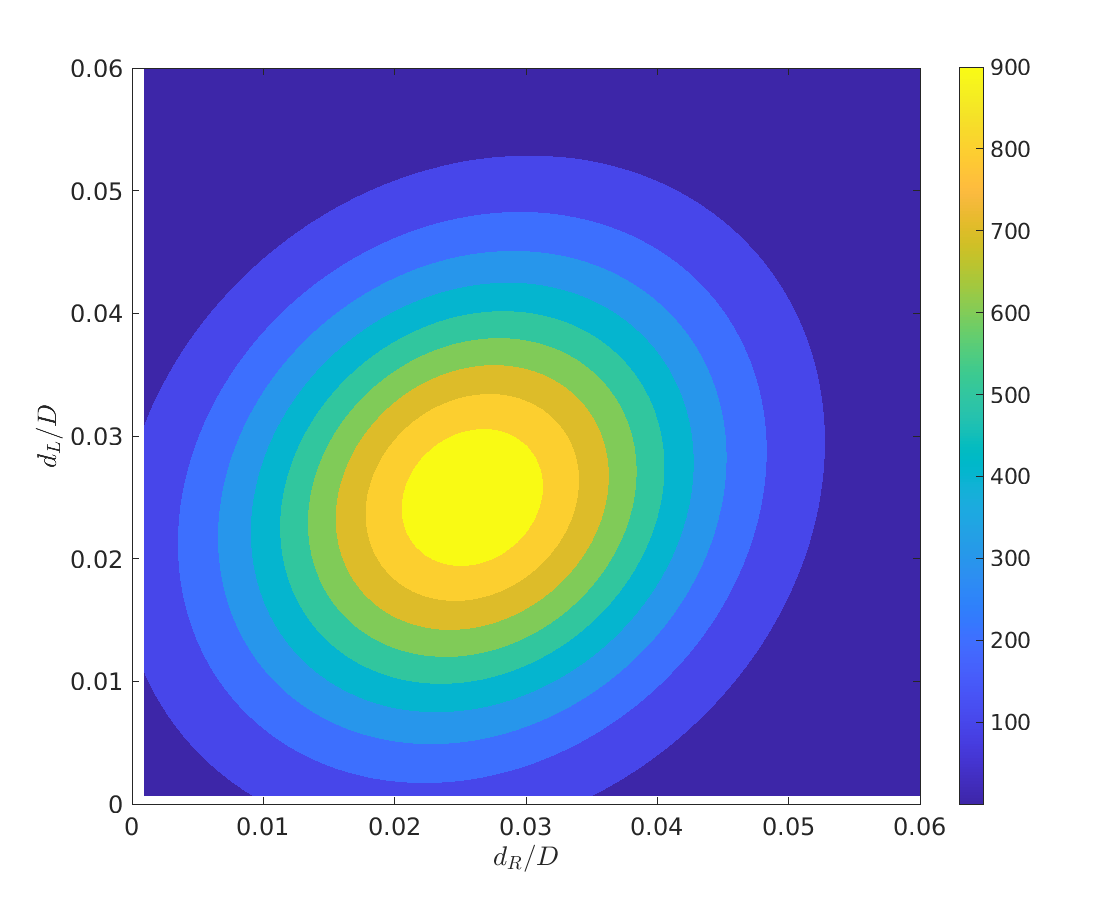}  \caption{}
\end{subfigure}\hspace{2em}
\begin{subfigure}[b]{0.43\textwidth} 
\includegraphics[width=\textwidth]{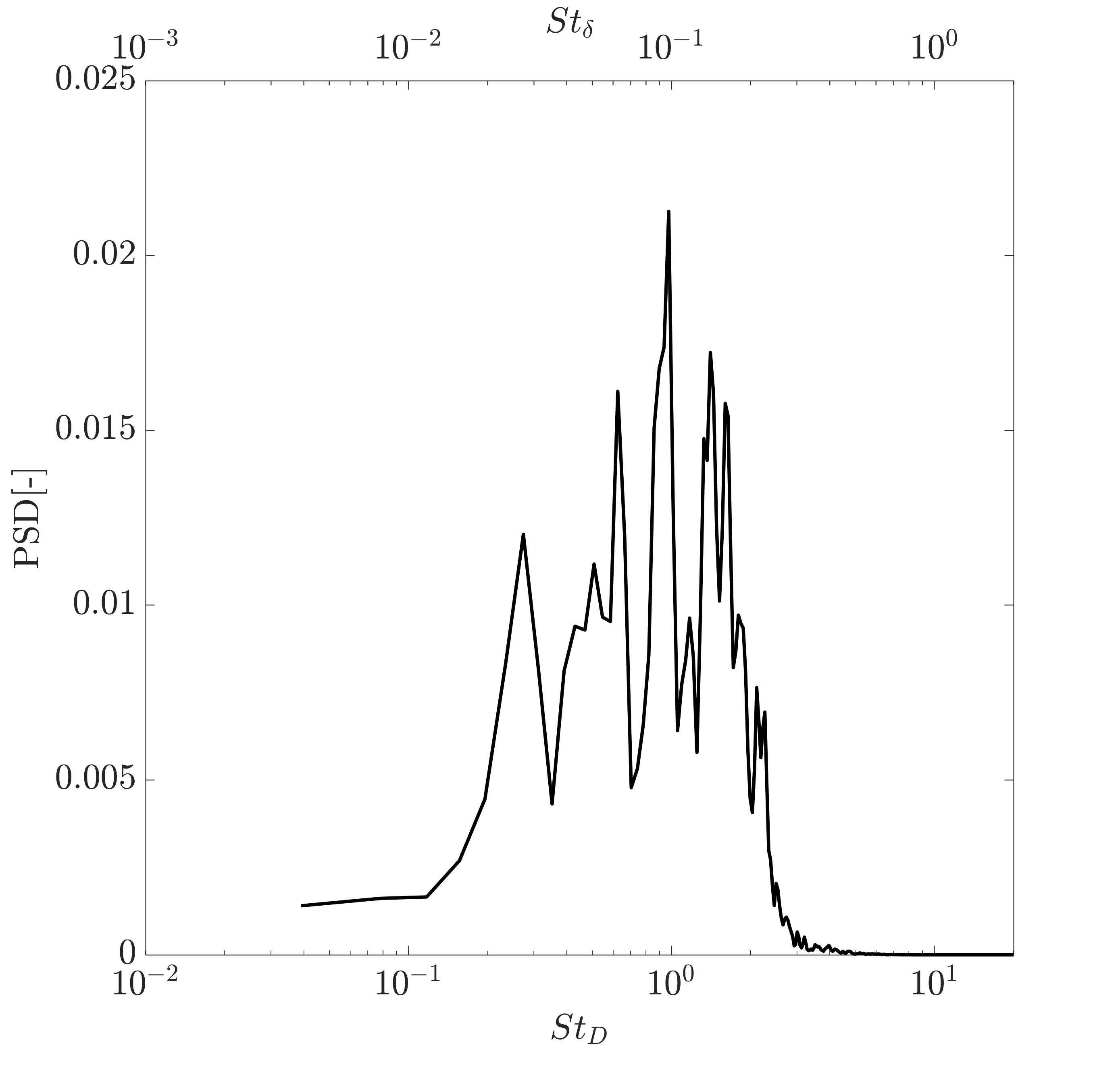}
\caption{}
\end{subfigure}
\caption{Characterization of streamwise vortices at $X/L=2.0$. (a) Gaussian joint PDF of instantaneous vortices. (b) Spectral content in vortex core fluctuations using Welch PSD for the \textbf{R}-vortex. Non-dimensional frequencies are shown based on the cylinder diameter, $St_{D} = \frac{U_{\infty}D}{\nu}$ as well as vortex core diameter, $St_{\delta} = \frac{U_{\infty}\delta}{\nu}$.}
  \label{fig:pdf}
\end{figure}

In order to estimate the correlation between meandering in the two vortices, we compute 
the correlation coefficient, $R$, between \textbf{L} and \textbf{R} vortex by using the displacements of instantaneous cores from their respective mean positions. 
Smaller values of this coefficient indicate lower influence of the vortices on one another. 
The correlation coefficient thus obtained by considering $3{,}000$ instantaneous vortices is $R=0.1622$.
This value is very near to that reported by \citet{jackson2015control}, but slightly higher than that of \citet{bulathsinghala2017afterbody}. 
According to \citet{jackson2015control}, these observations are rather independent of the streamwise location.
To further probe the nature of the correlation, the Gaussian joint probability distribution function (JPDF) is plotted in figure~\ref{fig:pdf}(a).
The maximum probability of the interaction between the two vortices is observed at $d_L/D = d_R/D \simeq 0.028$. 
Since the mean vortices are separated by $b/D \simeq 0.06$, a reasonable inference is that the interaction may only become important when the instantaneous vortices are closest to the symmetry plane. 
The JPDF diminishes as the vortices move away from the symmetry plane. 
Although the above analysis suggests the effects of vortices on one-another are weak, these calculations do not include the crucial axial velocity component that can play a significant role in determining the instability mechanism responsible for the meandering of these vortices as discussed in section~\ref{sec:stability}.  

The frequency of the meandering motion at this location is characterized using the spectral signature of the instantaneous vortices at the mean vortex core location.
Figure~\ref{fig:pdf}(b) shows the power spectral density (PSD) of fluctuations of $v$-velocity for the \textbf{R}-vortex using a long time-series signal, $tU_{\infty}/D > 60$. 
The signal shows a frequency band in Strouhal number based on diamater $St_D \equiv fD/U_\infty \in [0.3~2.0]$, with several low $(St_D \simeq 0.3, 0.6)$ and high ($St_D \simeq 1.0, 1.4, 1.6$) frequency peaks.
The maximum spectral energy is observed at $St_D \simeq 1.0$.
Though not shown for brevity, similar frequency content is also obtained for the \textbf{L}-vortex.
High-frequency peaks in the range $St_D \in [1.4~1.6]$ were also observed in the LES studies at high Reynolds number~\citep{garmann2019high2}, whereas ~\citet{zigunov2020dynamics} reported strong tones at low-frequency, $St_D \simeq 0.4$, in addition to the high-frequency peaks for a range of Reynolds numbers.  
\citet{zigunov2020dynamics} postulated the high-frequency content to be related to the shear layer downstream of the body, while the low-frequency fluctuations were conjectured to be related to interactions between the cores as observed in several vortex meandering studies~\citep{edstrand2016mechanism,devenport1996structure}.
Detailed stability analyses of vortex pair are performed in \S\ref{sec:stability} to investigate the vortex-vortex interactions, including effects on underlying frequency content as well as their correlation.

\subsection{Coherent Structures}\label{sec:pod}
The phenomenon of meandering is associated with the motion of coherent structures constituting the vortex \citep{devenport1996structure}. 
One approach to educe these is with proper orthogonal decomposition (POD) using the method of snapshots as introduced by \citet{sirovich1987turbulence}. 
For a given unsteady flowfield, the technique extracts modes with the largest energy content, which are often related to the coherence in the flow. 
The POD decomposition of time snapshots, represented by $\vect{q}(\vect{x},t)$, is given by:
\begin{eqnarray} \label{eq:pod}
  \displaystyle \vect{q}(\vect{x},t) = \vect{\bar{q}}(\vect{x}) + \sum\limits_{i=1}^\infty \sqrt{\Lambda_i} \vect{\Phi}_i (\vect{x}) \vect{a}_i(t) 
\end{eqnarray} 
where $\vect{\bar{q}}(\vect{x})$ is the mean flow. 
$\vect{\Phi}_i(\vect{x})$ and $\vect{a}_i(t)$ represent the $i^{th}$ orthonormal spatial mode and its time-coefficient respectively. 
$\Lambda_i$ is the energy associated with the $i^{th}$ POD mode.


The role of coherent structures during the development of streamwise vortices is analyzed first, as this provides a perspective of the meandering process downstream.  
This analysis is thus performed at different streamwise sections along the upswept base. 
$3{,}000$ snapshots, sampled  every $100$ iterations,  are collected, for an effective time-step $\Delta t_{\text{POD}}=0.025$.
Results performed for each  vortex in isolation indicate very similar spatial dynamics due to flow symmetry.
For brevity, results from only \textbf{R}-vortex are presented here. 

\begin{figure}
\centering
    \begin{subfigure}[b]{0.05\textwidth}\caption*{}\end{subfigure}
    \begin{subfigure}[b]{0.3\textwidth}\caption*{\scalebox{1.3}{$\displaystyle \vect{\bar{q}}(\vect{x})$}}\end{subfigure}
    \begin{subfigure}[b]{0.3\textwidth}\caption*{\scalebox{1.3}{$\vect{\Phi}_1 (\vect{x})$}}\end{subfigure}
    \begin{subfigure}[b]{0.3\textwidth}\caption*{\scalebox{1.3}{$\vect{\Phi}_2 (\vect{x})$}}\end{subfigure}
\\
    \begin{subfigure}[b]{0.05\textwidth}\caption*{\rotatebox{90}{(a) $~ X/L=0.2$}}\end{subfigure}
  \begin{subfigure}[b]{0.3\textwidth}\includegraphics[width=1.0\textwidth, trim={2cm 4cm 30cm 5cm},clip]{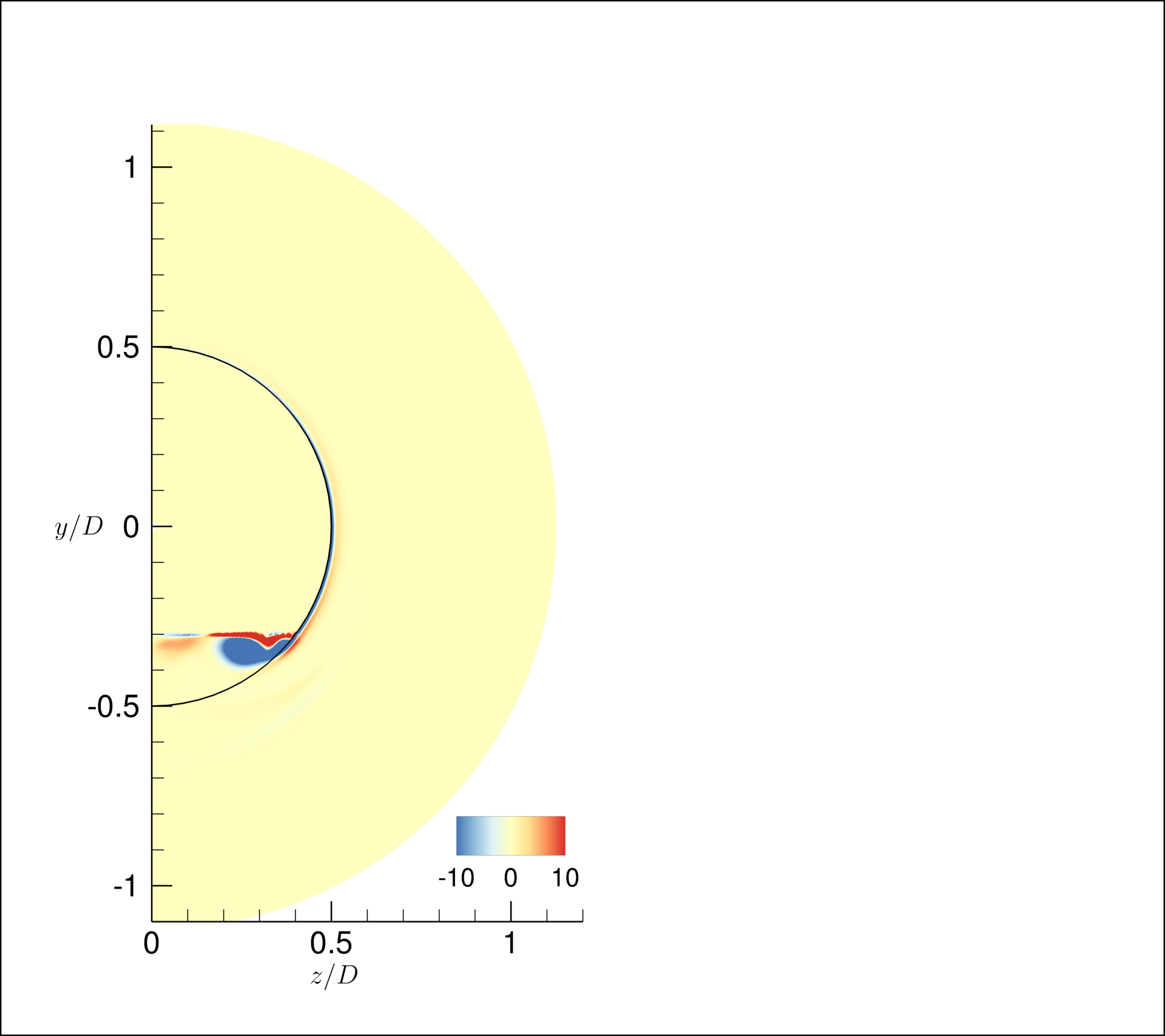}\end{subfigure}
  \begin{subfigure}[b]{0.3\textwidth}\includegraphics[width=1.0\textwidth, trim={2cm 4cm 30cm 5cm},clip]{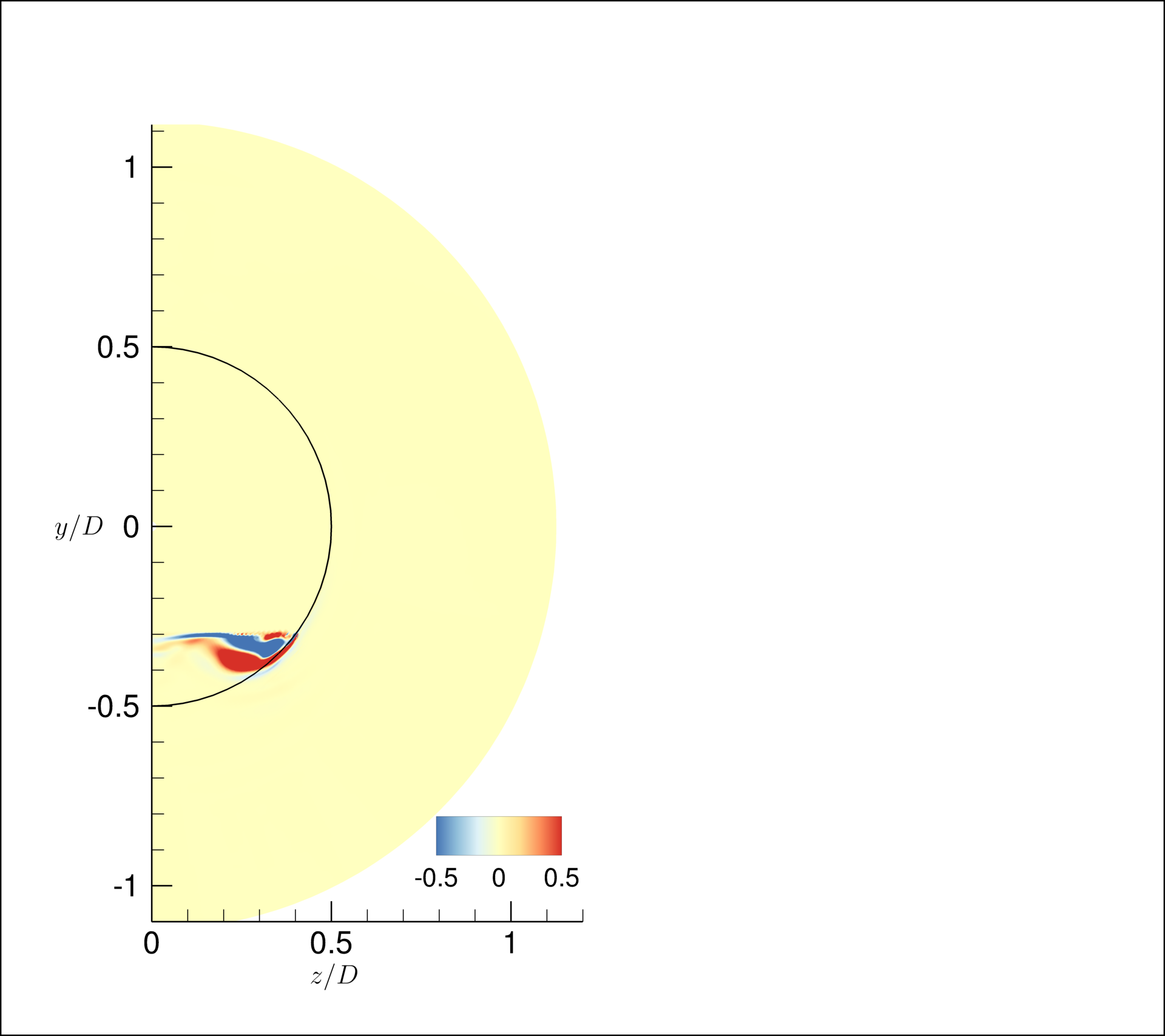}  \end{subfigure}
  \begin{subfigure}[b]{0.3\textwidth}\includegraphics[width=1.0\textwidth, trim={2cm 4cm 30cm 5cm},clip]{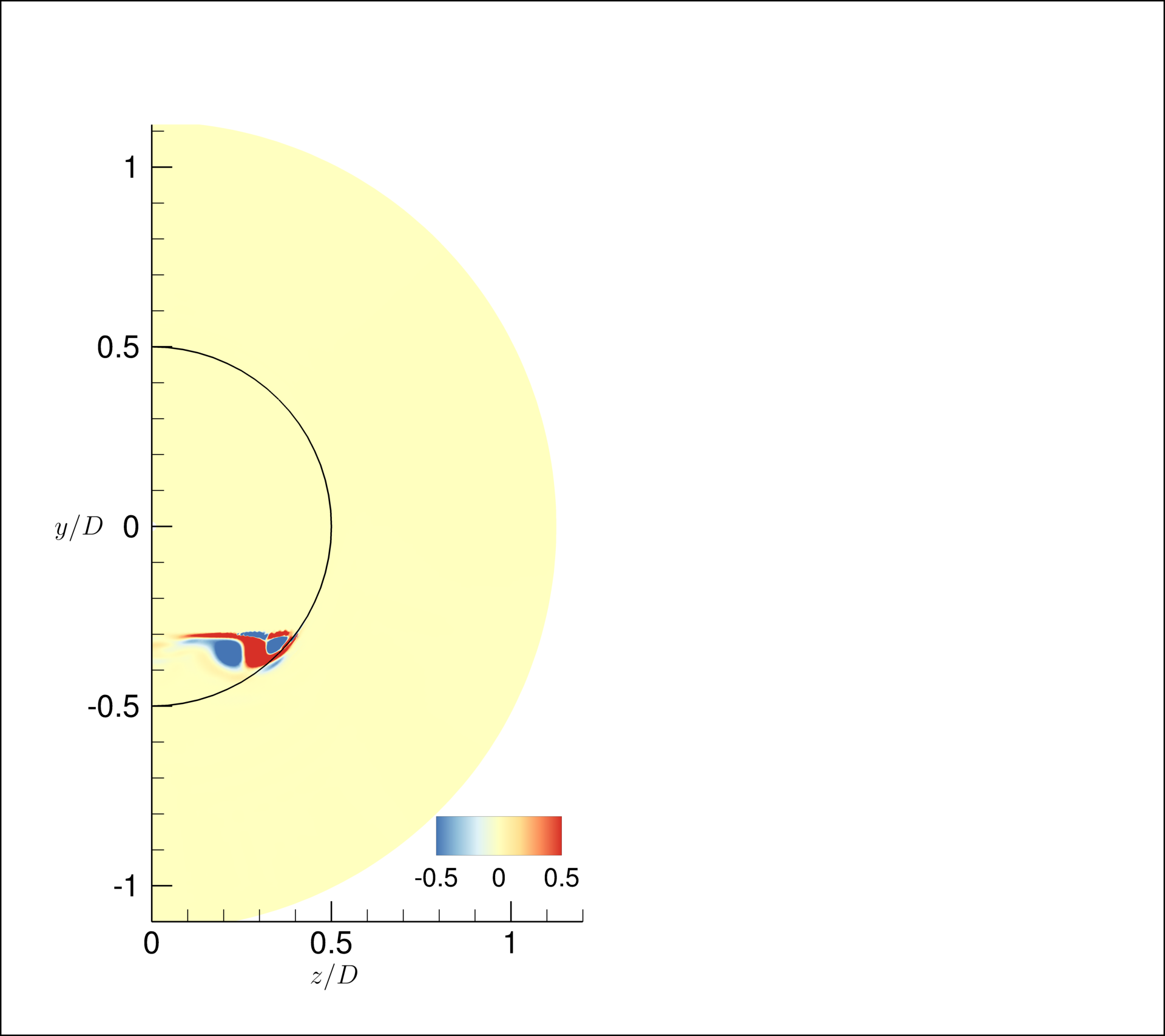}  \end{subfigure}
\\
    \begin{subfigure}[b]{0.05\textwidth}\caption*{\rotatebox{90}{(b) $~ X/L=0.6$}}\end{subfigure}
      \begin{subfigure}[b]{0.3\textwidth}\includegraphics[width=1.0\textwidth, trim={2cm 4cm 30cm 5cm},clip]{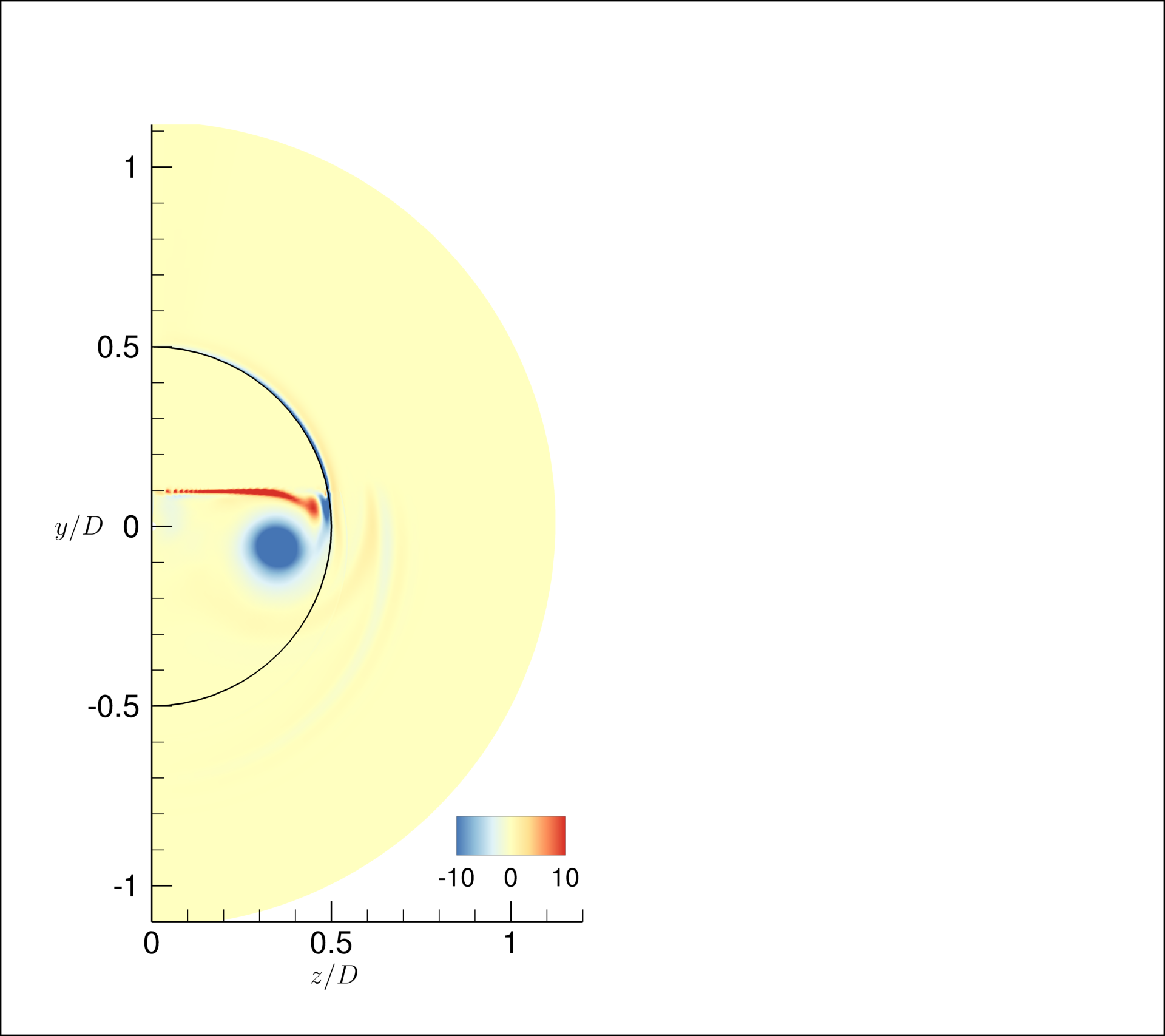}\end{subfigure}
  \begin{subfigure}[b]{0.3\textwidth}\includegraphics[width=1.0\textwidth, trim={2cm 4cm 30cm 5cm},clip]{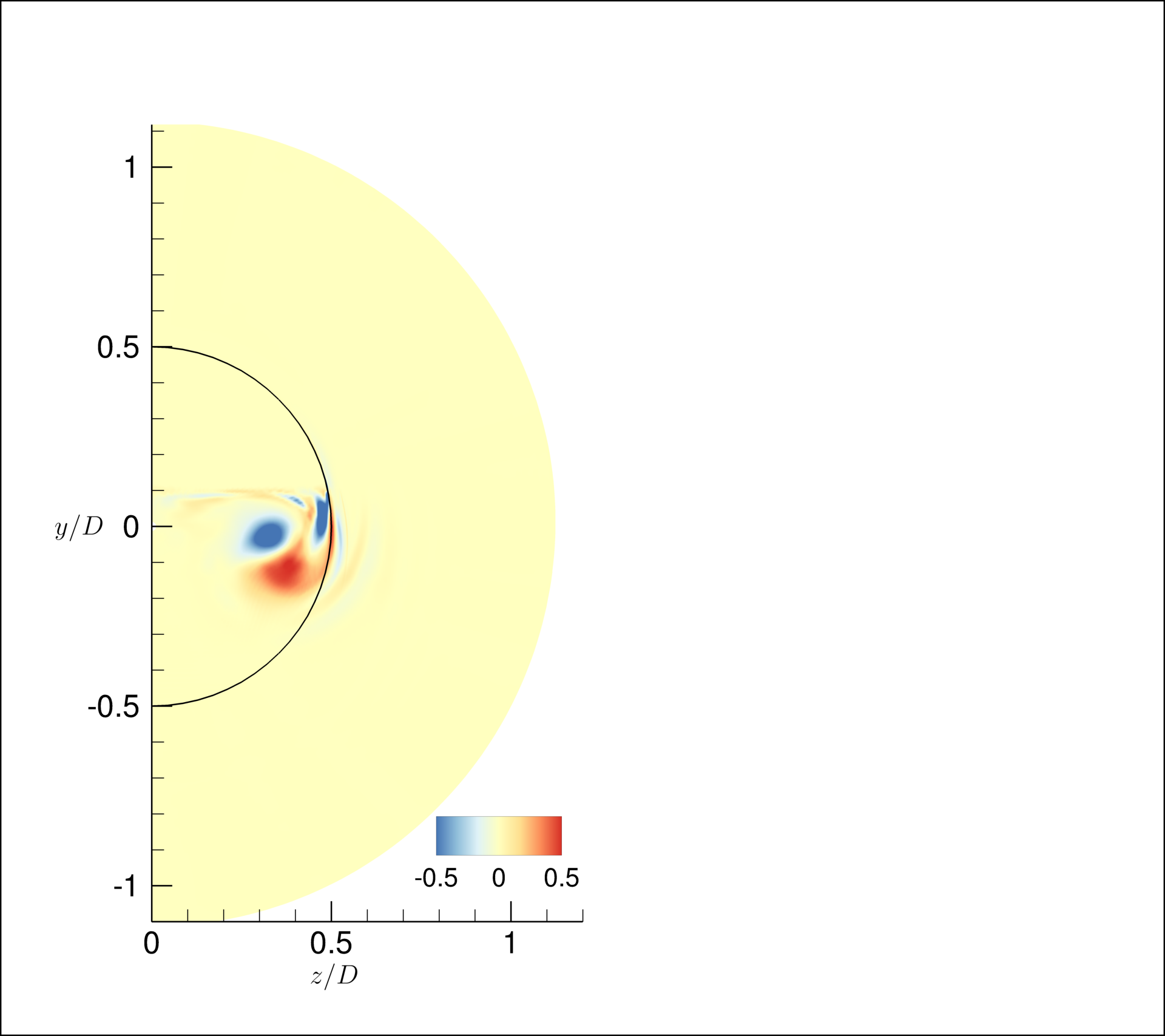}  \end{subfigure}
  \begin{subfigure}[b]{0.3\textwidth}\includegraphics[width=1.0\textwidth, trim={2cm 4cm 30cm 5cm},clip]{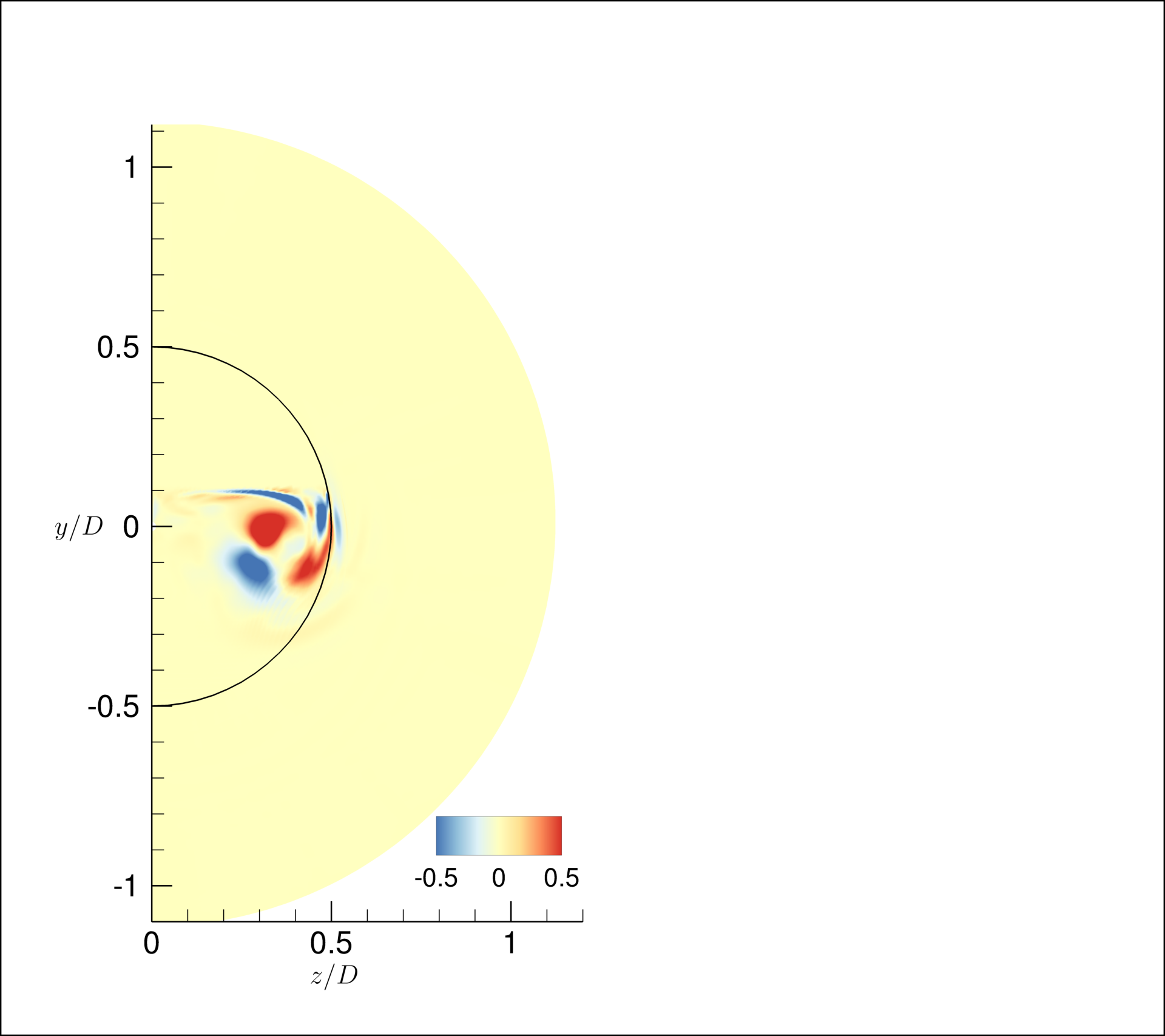}  \end{subfigure}
\\
    \begin{subfigure}[b]{0.05\textwidth}\caption*{\rotatebox{90}{(c)$~ X/L=1.0$}}\end{subfigure}
      \begin{subfigure}[b]{0.3\textwidth}\includegraphics[width=1.0\textwidth, trim={2cm 2cm 30cm 5cm},clip]{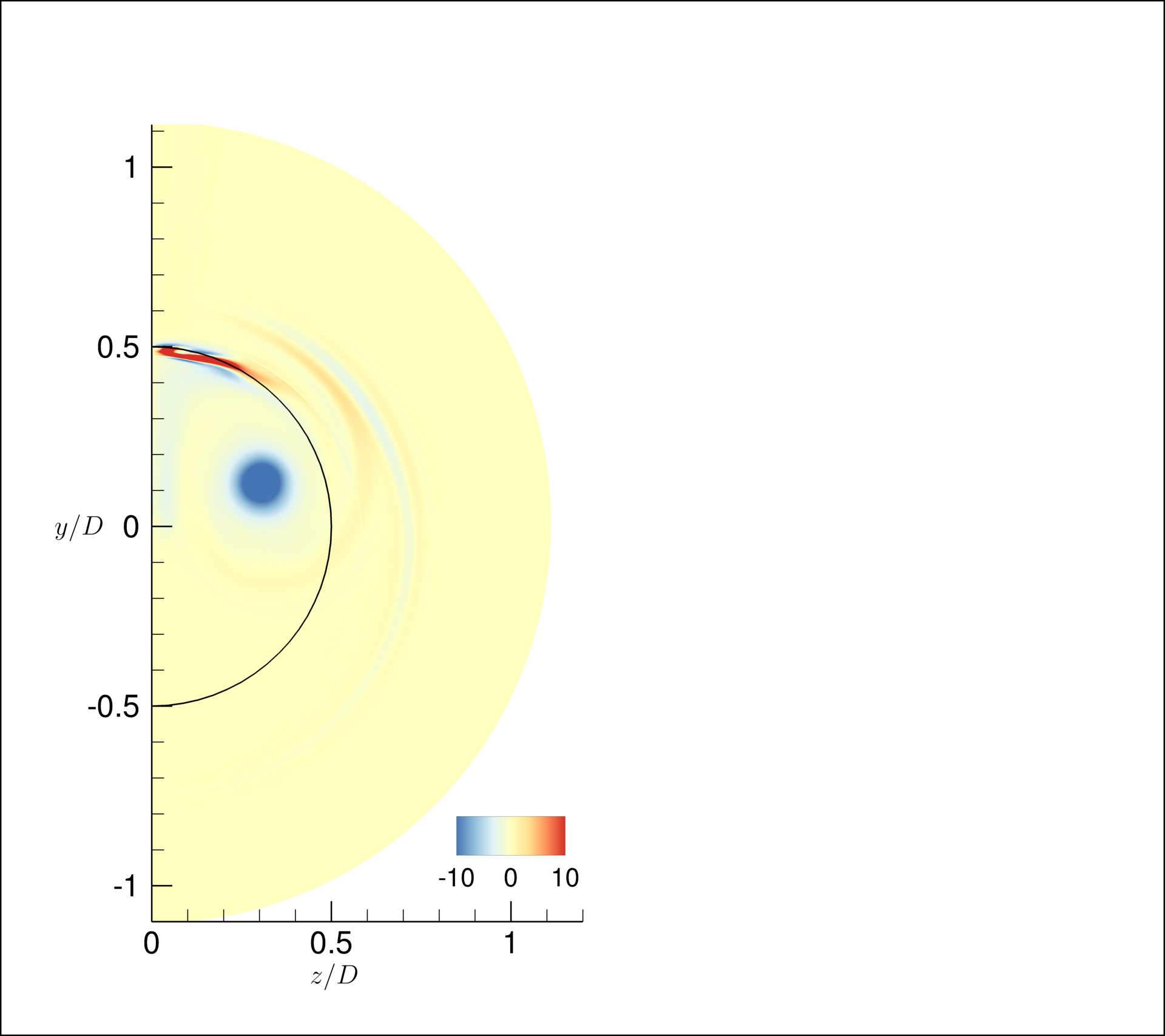}\end{subfigure}
  \begin{subfigure}[b]{0.3\textwidth}\includegraphics[width=1.0\textwidth, trim={2cm 2cm 30cm 5cm},clip]{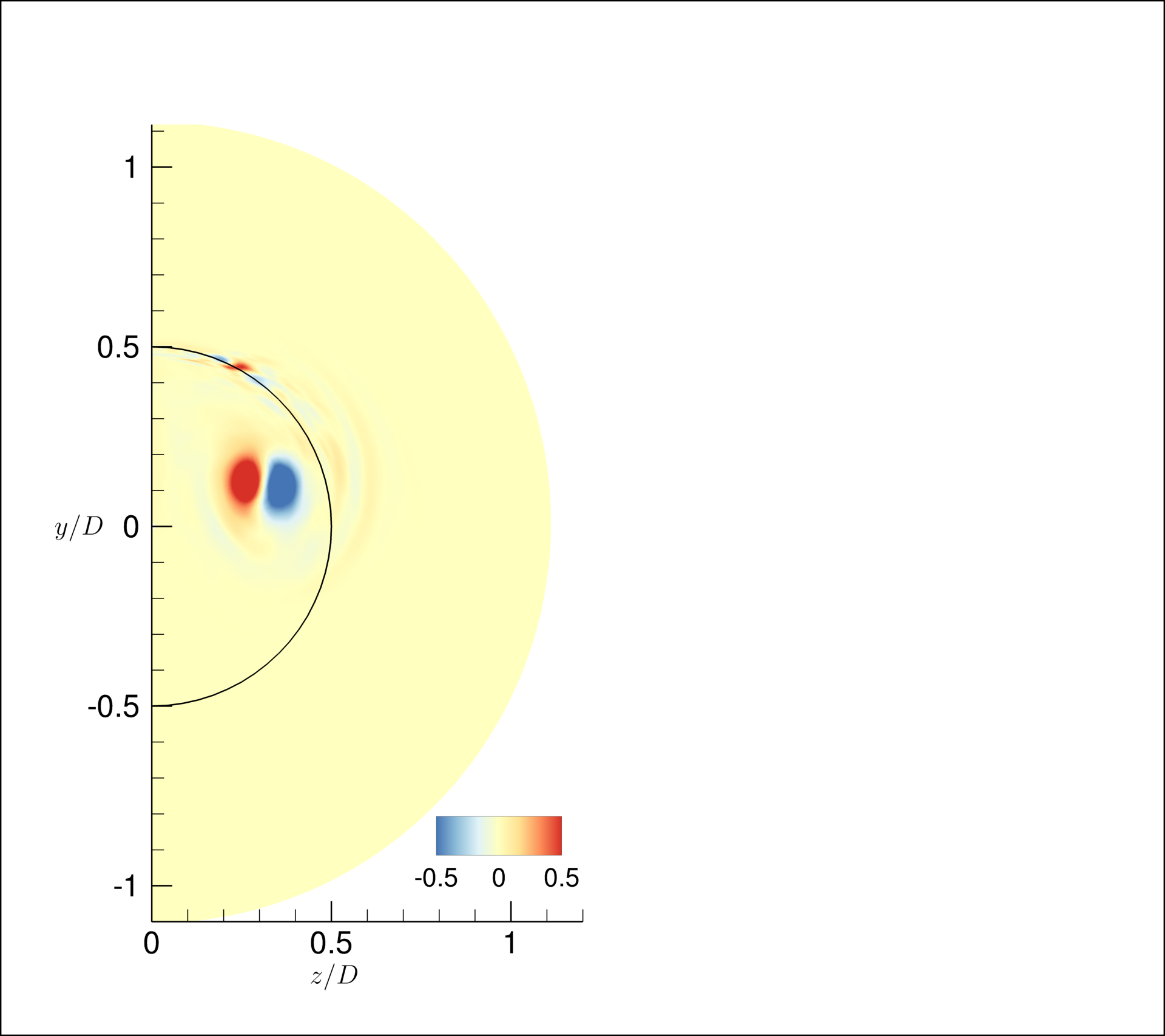}  \end{subfigure}
  \begin{subfigure}[b]{0.3\textwidth}\includegraphics[width=1.0\textwidth, trim={2cm 2cm 30cm 5cm},clip]{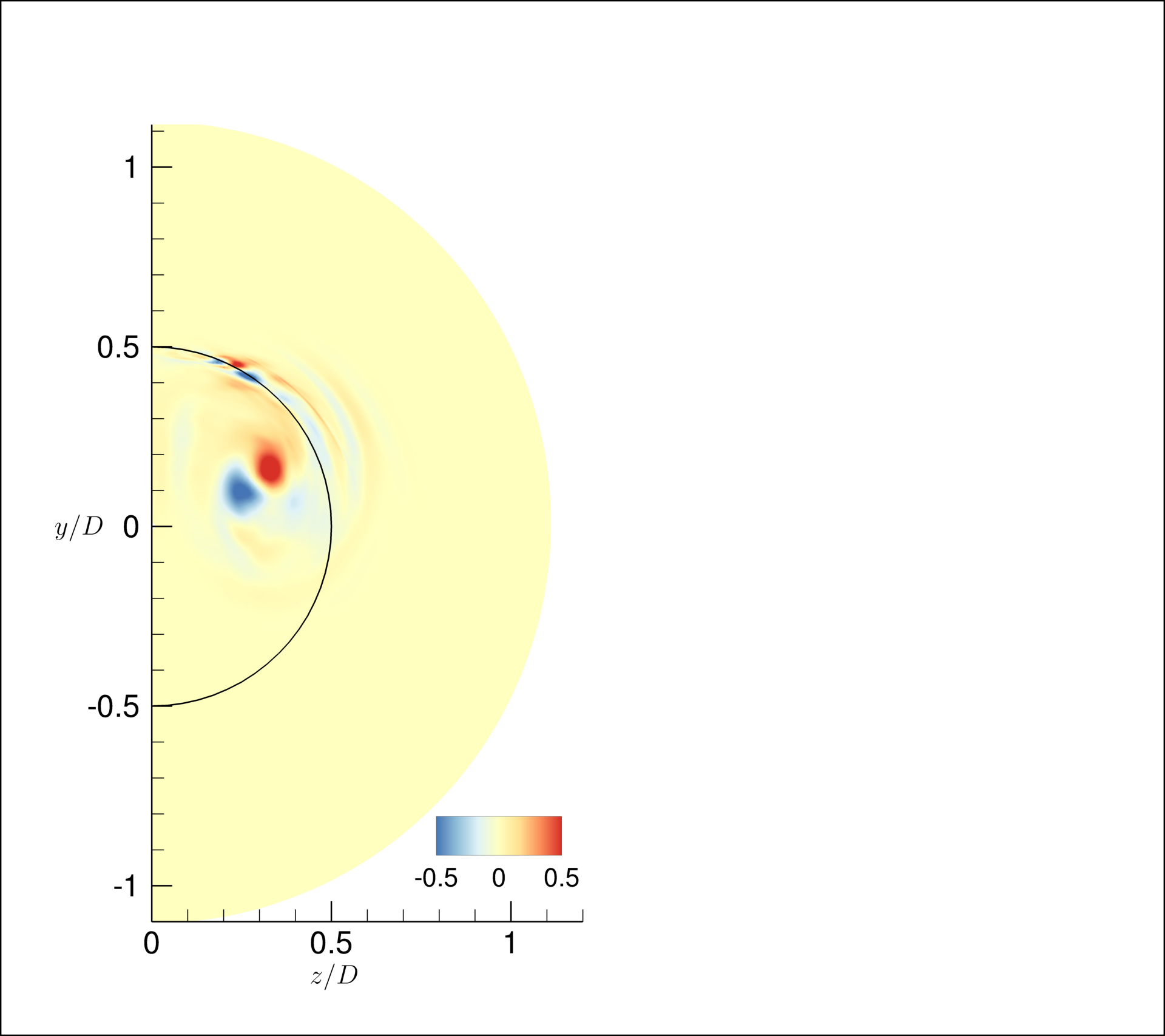}  \end{subfigure}
  \caption{From left to right, mean flow as well as first two most energetic POD modes are shown through streamwise vorticity at different locations within ramp. Semi-circle indicates the boundary of the cylinder body.} 
  \label{fig:pod20}
\end{figure}

The first two most energetic modes are shown along with the mean vortex in Figure~\ref{fig:pod20} at three different sections along the ramp, namely at $X/L=0.2$, $0.6$ and $1.0$. 
At $X/L=0.2$, figure~\ref{fig:pod20}(a), where the mean vortex is in a nascent stage and fed by the shear layer,  the first mode qualitatively resembles the mean field~\citep{bulathsinghala2017afterbody}.
This indicates the prominence of fluctuations in both the vortex as well as the shear layer.
Proceeding downstream, at $X/L=0.6$ in figure~\ref{fig:pod20}(b), where the mean vortices have become nearly axisymmetric, but are still attached to the surfaces, both POD modes display a vortex dipole-like structure around the time-averaged vortex.
However, the presence of a shear layer in the POD modes effectively diffuses these structures.
Further downstream, towards the edge of the base, at $X/L=1.0$ in figure~\ref{fig:pod20}(c), the vortex dipoles are clearly discernible near the vortex core for both POD modes.
Smaller structures near the periphery of the cylinder display the shear layer emanating from the top.
The influence of the shear layer on the vortex persists for some distance in the wake, complicating the flow dynamics.

\begin{figure}
\centering
  \begin{subfigure}[b]{0.43\textwidth}\includegraphics[width=1.0\textwidth, trim={8cm 4cm 12cm 5cm},clip]{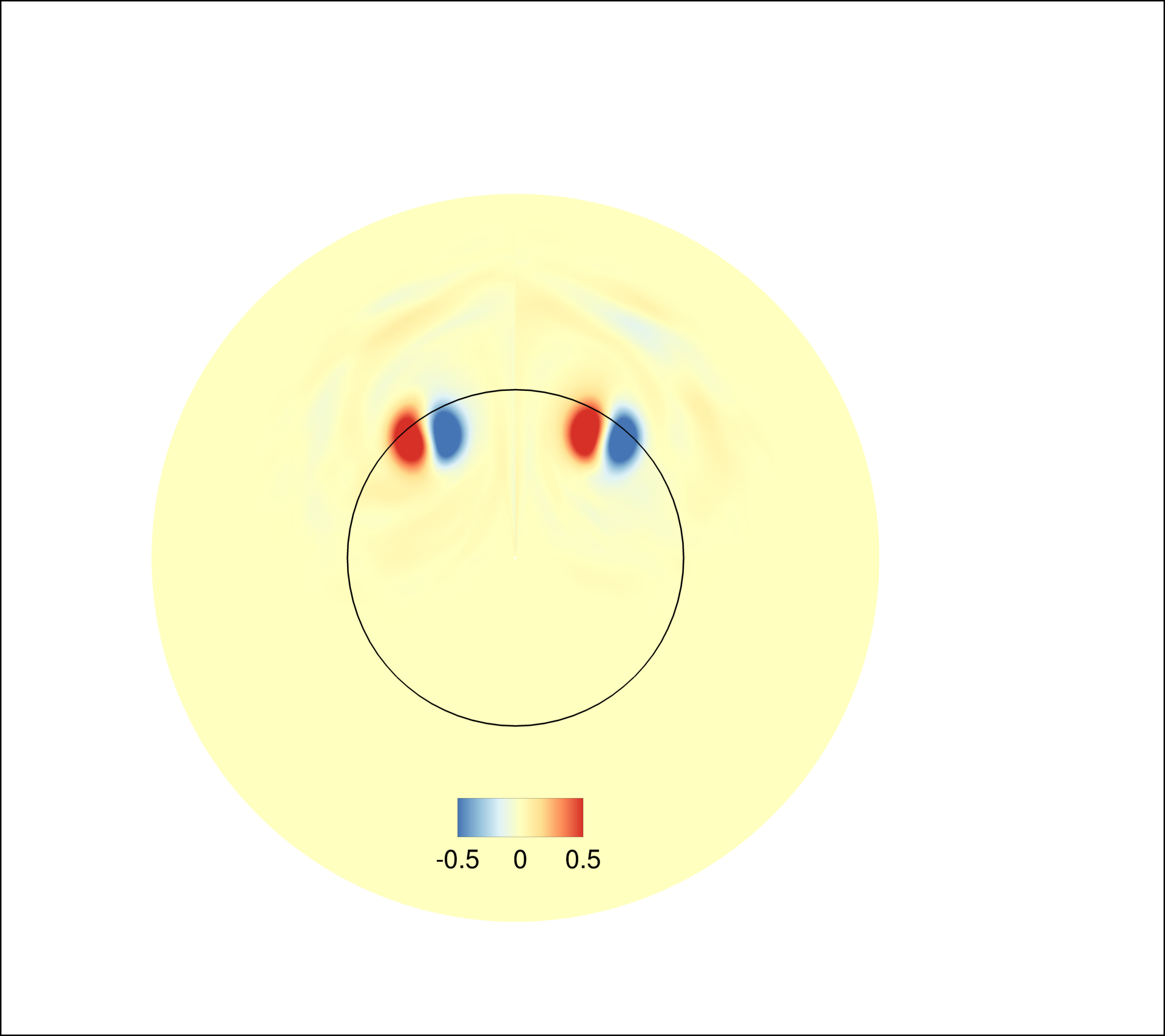} \caption{$\vect{\Phi}_1 (\vect{x})$} \end{subfigure}
  \begin{subfigure}[b]{0.43\textwidth}\includegraphics[width=1.0\textwidth, trim={8cm 4cm 12cm 5cm},clip]{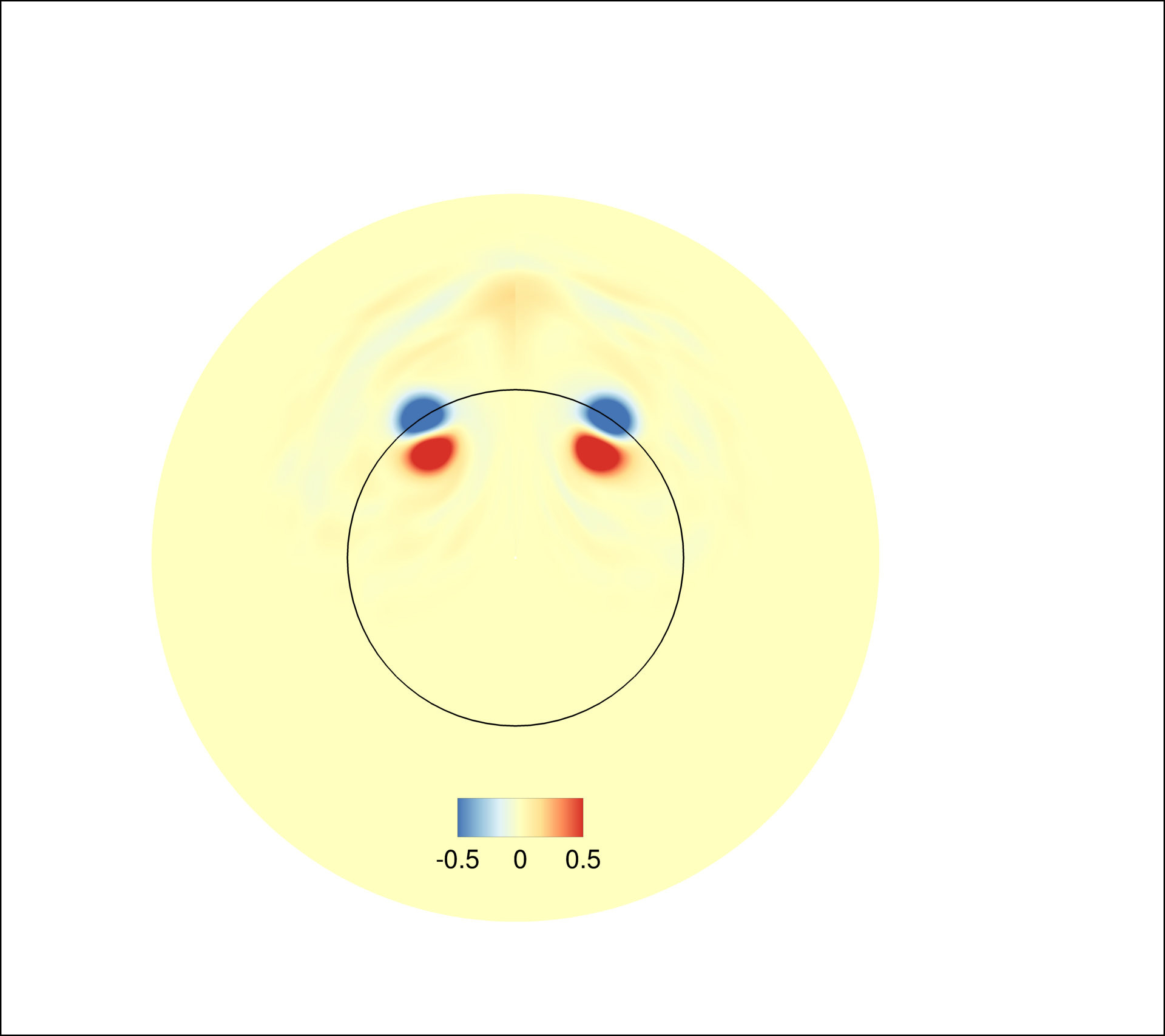} \caption{$\vect{\Phi}_2 (\vect{x})$} \end{subfigure}\\
        \begin{subfigure}[b]{0.46\textwidth}\includegraphics[width=1.0\textwidth, trim={0cm 0cm 0cm 0cm},clip]{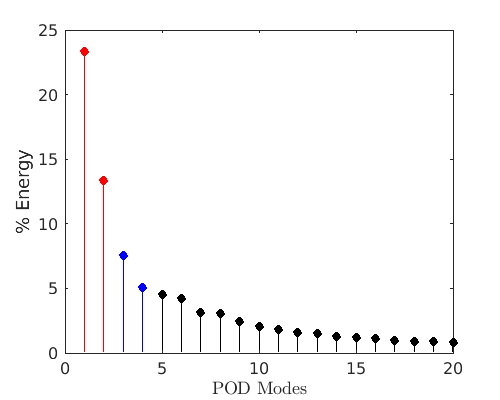}\caption{$\Lambda_i$(\textbf{L})}\end{subfigure}
         \begin{subfigure}[b]{0.46\textwidth}\includegraphics[width=1.0\textwidth, trim={0cm 0cm 0cm 0cm},clip]{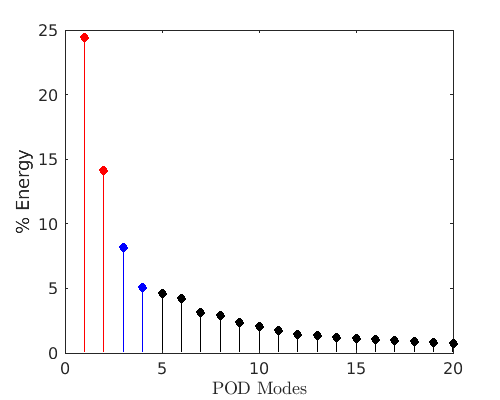}\caption{$\Lambda_i$(\textbf{R})}\end{subfigure}\\
        \begin{subfigure}[b]{0.41\textwidth}\includegraphics[width=1.0\textwidth, trim={0cm 0cm 0cm 0cm},clip]{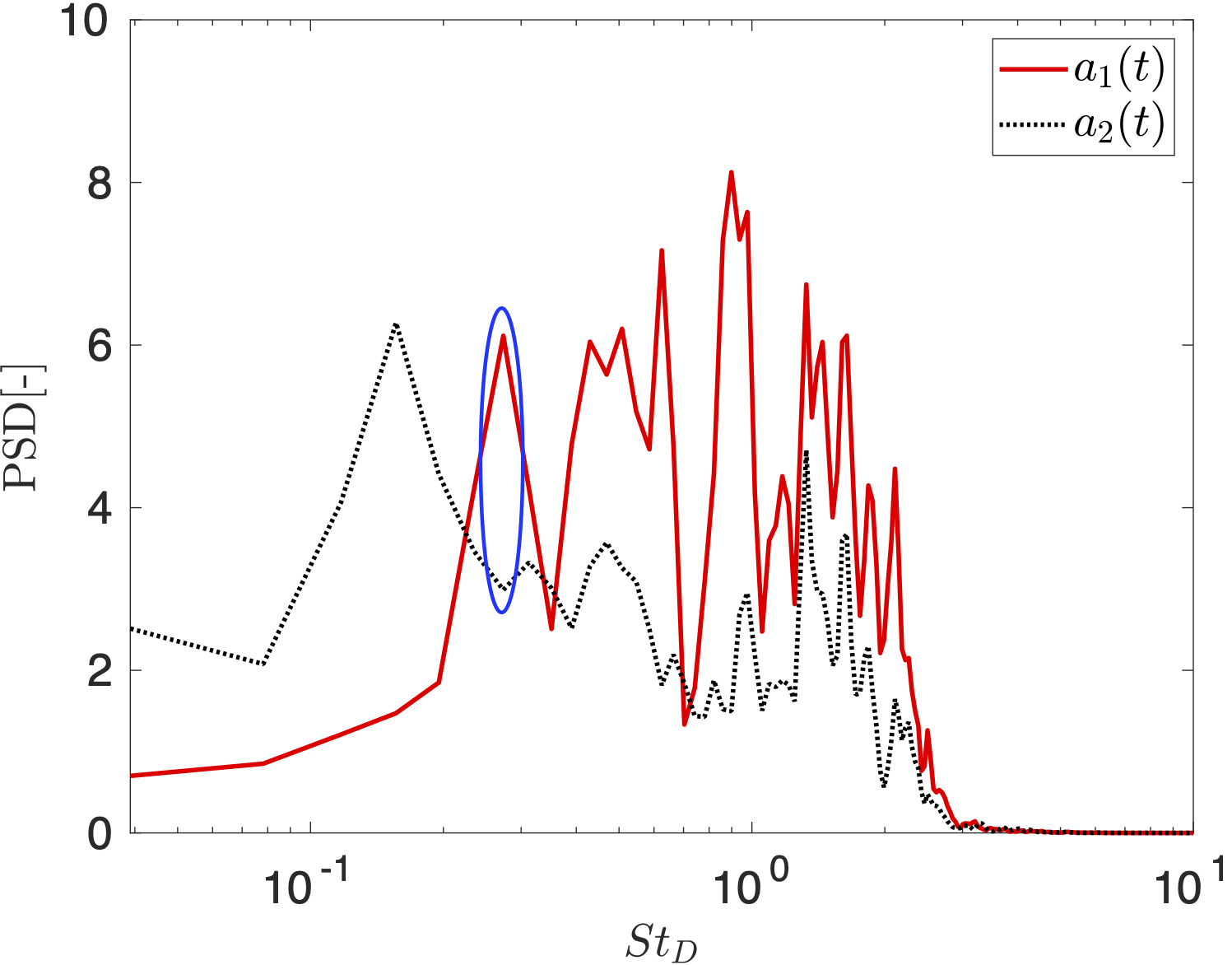} \caption{PSD(\textbf{L})} \end{subfigure}\quad
        \begin{subfigure}[b]{0.41\textwidth}\includegraphics[width=1.0\textwidth, trim={0cm 0cm 0cm 0cm},clip]{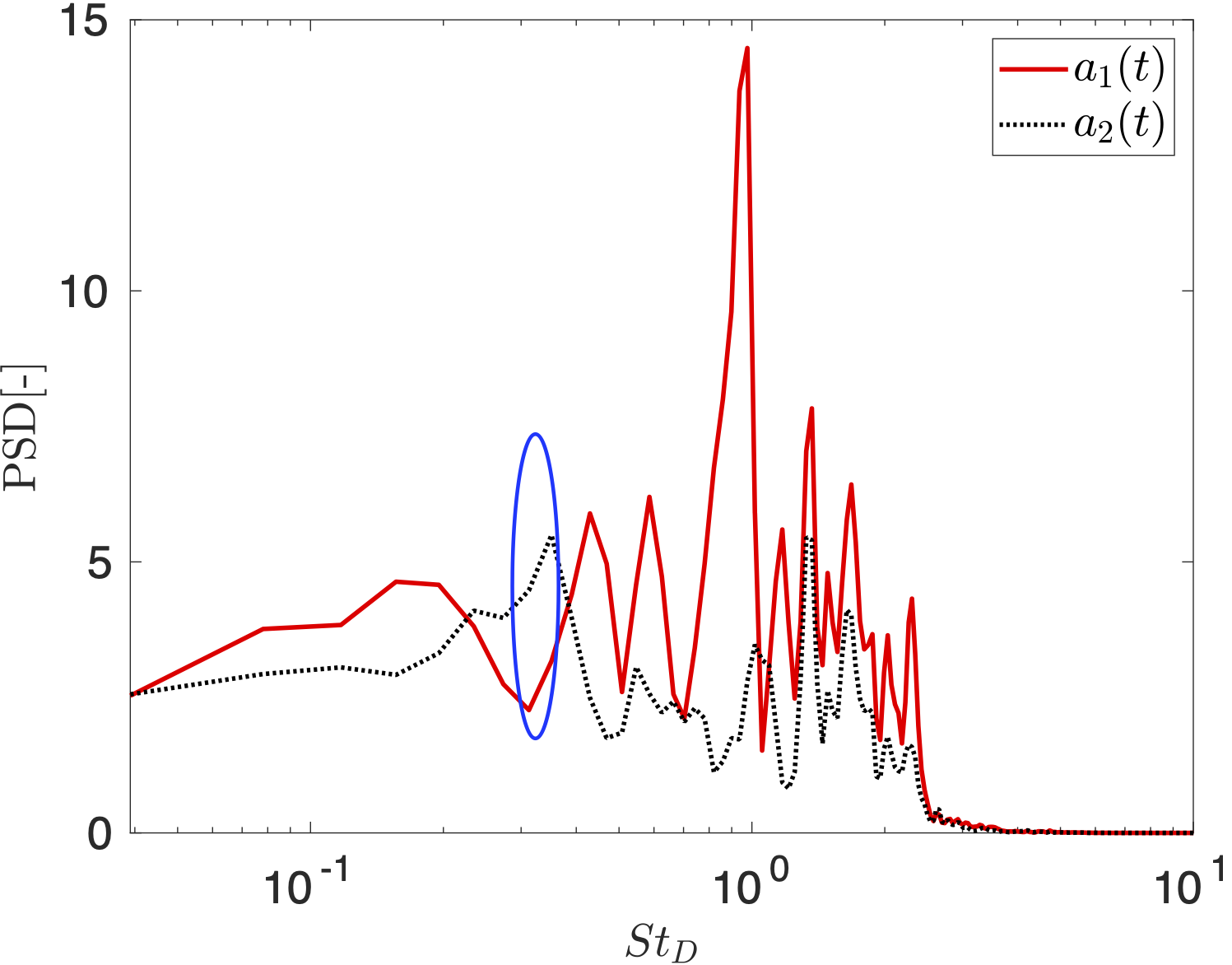} \caption{PSD(\textbf{R})} \end{subfigure}
 \caption{Dominant two POD modes at $X/L=2.0$. Structures of mode 1(a) and 2(b). Energy content in dominant modes for \textbf{L} and \textbf{R} vortices (c,d).  The spectral content of time-coefficients in first two modes for both the vortices (e,f). Blue ovals in (e,f) highlight the peaks around $St_D \simeq 0.3$, which are later obtained from stability analysis (section~\ref{sec:stab-analys-afterb}) as being associated with unstable modes.} 
  \label{fig:podmode}
\end{figure}

In order to isolate the vortex dynamics and the role of large-scale coherent structures in the overall meandering process, detailed analysis is performed at a  location relatively far downstream, $X/L=2.0$. 
Since the instantaneous meandering motions of the two vortices in the pair are very different, as shown in figure~\ref{fig:meandering1},  the vortices are considered separately for this component of the analysis.
Figure~\ref{fig:podmode}(a) and (b) show the spatial structures of the first two POD modes as obtained from a decomposition of the \textbf{L} and \textbf{R} vortices separately and then plotting them together.
The vortex dipoles are more distinctly defined compared to those observed at $X/L=1.0$ in figure \ref{fig:pod20}(c), while the shear layer component is very small.

The energies ($\Lambda_i$) of the first $20$ most energetic modes are shown in figure ~\ref{fig:podmode}(c,d) for the \textbf{L} and \textbf{R} vortex respectively. 
Both vortices show similar energy distribution among the modes. 
The first mode has nearly double the energy of the second, but together, these two modes contain almost $35\%$ of the total energy.  
Furthermore, 80\% of the flow field can be reconstructed by only considering the first 10 modes.
Although first two modes differ in energy content, they may together be considered as paired helical modes with azimuthal wave number ($|m|=1$), as in the Batchelor vortex results of \citet{edstrand2016mechanism}.

\begin{figure}
\centering
  \begin{subfigure}[b]{0.47\textwidth}\includegraphics[width=1.0\textwidth, trim={4.8cm 3cm 5cm 8cm},clip]{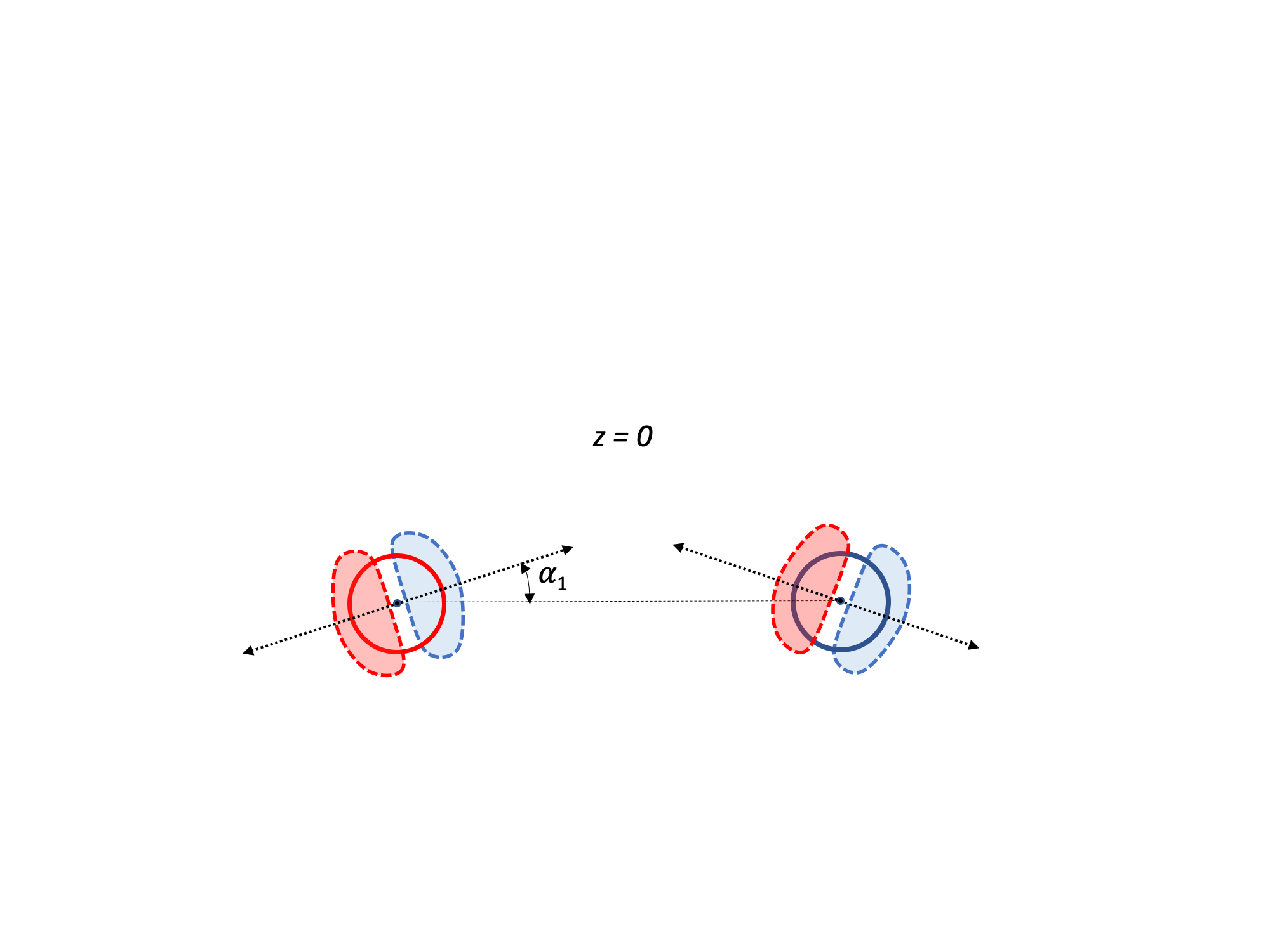}\caption{POD Mode1}\end{subfigure}
  \begin{subfigure}[b]{0.47\textwidth}\includegraphics[width=1.0\textwidth, trim={4.8cm 3cm 5cm 8cm},clip]{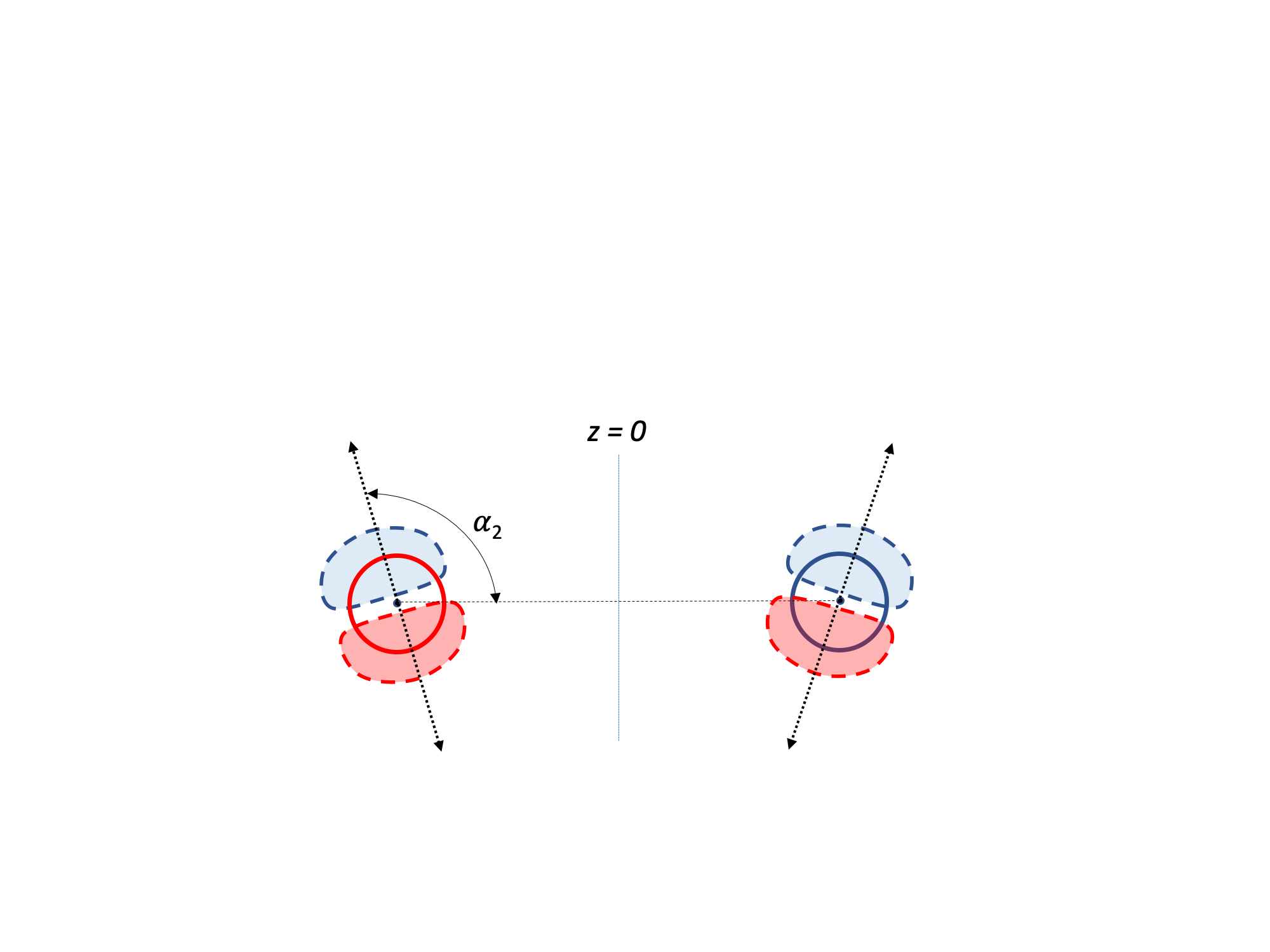}\caption{POD Mode 2}\end{subfigure}\\
 \caption{Sketch depicting spatial displacement of vortex cores due to the leading POD modes.} 
  \label{fig:podmode3}
\end{figure}

The movement of each vortex core in the plane depends on the alignment of the vortex dipoles compared to the mean vortex as illustrated in figure ~\ref{fig:podmode3}.
The inclination angles of vortex dipoles in the first and second POD modes are defined as $\alpha_1$ and $\alpha_2$, as shown in the figure, and double-sided arrows depict the oscillation axis of vortex core due to these individual dipoles.
The dipole increases the vorticity in one half of the time-averaged vortex while decreasing it in the other half, thus causing the displacement, which is governed by the inclination of the dipole.
In the simulation,  $\alpha_1$  and $\alpha_2$  are approximately \ang{16} and \ang{106} for the left vortex, while for the right vortex, they are   \ang{-16} and \ang{-106} suggesting that the coherent structures are also symmetric to each other.
A \ang{90} difference between the inclination angles, $|\alpha_1-\alpha_2|$, for both vortices, confirm that these modes are orthogonal to each other.


The temporal dynamics of these modes are examined in figures~\ref{fig:podmode}(e) and (f), by considering the power spectral density (PSD) of the \textbf{L} and \textbf{R} vortices respectively.
Modes for both the vortices show spectral energies largely distributed in $St_D \in [0.3 ~2.0]$ as also observed in the analysis of point probe data (figure~\ref{fig:pdf}(b)).  
The first POD mode for both the vortices show a peak at $St_D \simeq 1.0$, while the second POD mode has more spectral energy at low-frequencies. 
Because of the difference in energies ($\Lambda$) of the two modes in flow, the overall motion of the vortex core is seemingly random as shown in Fig.~\ref{fig:meandering1}.  
However, the high energy content of leading modes in the flow can be exploited to reconstruct meandering phenomena using a low-rank model as presented in section~\ref{sec:lods}. 
\section{Linear Stability Analysis}\label{sec:stability}
Linear stability analysis is now employed to provide insights into the physical mechanism of meandering by extracting the underlying modal structures, frequencies and wavenumbers that form the basis for the observed phenomena. 
\citet{jacquin2001properties} have observed that the cooperative interactions between the vortices affect the similar frequency range as meandering. 
Two typical instabilities of interest due to the mutual induction between counter-rotating vortex pairs are those designated elliptic and Crow. 
The former is of short-wavelength and incurs an internal deformation of the vortex cores with a wavy displacement of the vortex center \citep{widnall1971theoretical,tsai1976stability,sipp2003widnall}. 
This instability thus can be viewed as a resonance between two normal modes of a vortex and an external strain field induced by the neighboring vortex.
The Crow instability \citep{crow1970stability}, on the other hand, is three-dimensional, in which perturbations displace the vortices locally as a whole without any change in their core structure; the observed wavelengths are large compared to the core radius.
This instability can be observed in aircraft tip vortices \citep{leweke1998cooperative}. 


A rough estimate of the features of these instabilities, if present,  can be obtained by considering the characteristic length scales in the vortex pair, namely the core radius $\delta$ and the inter-vortical separation $b$; their ratio, $\delta/b$, indicates the relative distance between the vortices.
Values obtained in the current simulations, reported in Table~\ref{tab:vortex},  are consistently between $0.1$ and $0.2$ at all streamwise locations.
According to \citet{leweke2016dynamics}, these lie in the range that can trigger both long and short wave instabilities, though Reynolds number considerations are also important.

The instabilities existing in the flowfield are now investigated in detail through a linear stability analysis. 
The mean vortex pair is first matched to a theoretical vortex model, and both temporal and spatial stability analyses are performed.
Analyzing a fitted model vortex is advantageous because it offers the flexibility to extend the domain arbitrarily without compromising effects due to boundaries.
This also facilitates the examination of the long-wavelength Crow instability, if present.
The possibility of a global mode is not examined as it is beyond the scope of the present study.
Parallel flow is assumed in the streamwise direction; this is a reasonable assumption since the streamwise variation of the mean flow is relatively small at distances sufficiently downstream of the body such as those considered here.
As such, stability characteristics of the vortices predicted with parallel flow assumption provide a reasonable description of the meandering phenomena~\citep{edstrand2018parallel}.
\subsection{Theoretical Vortex Model}\label{sec:vortexfit}
\begin{figure}
\centering
 \begin{subfigure}[b]{0.6\textwidth}
    \includegraphics[width=\textwidth, trim={1.5cm 9cm 14.5cm 9cm},clip]{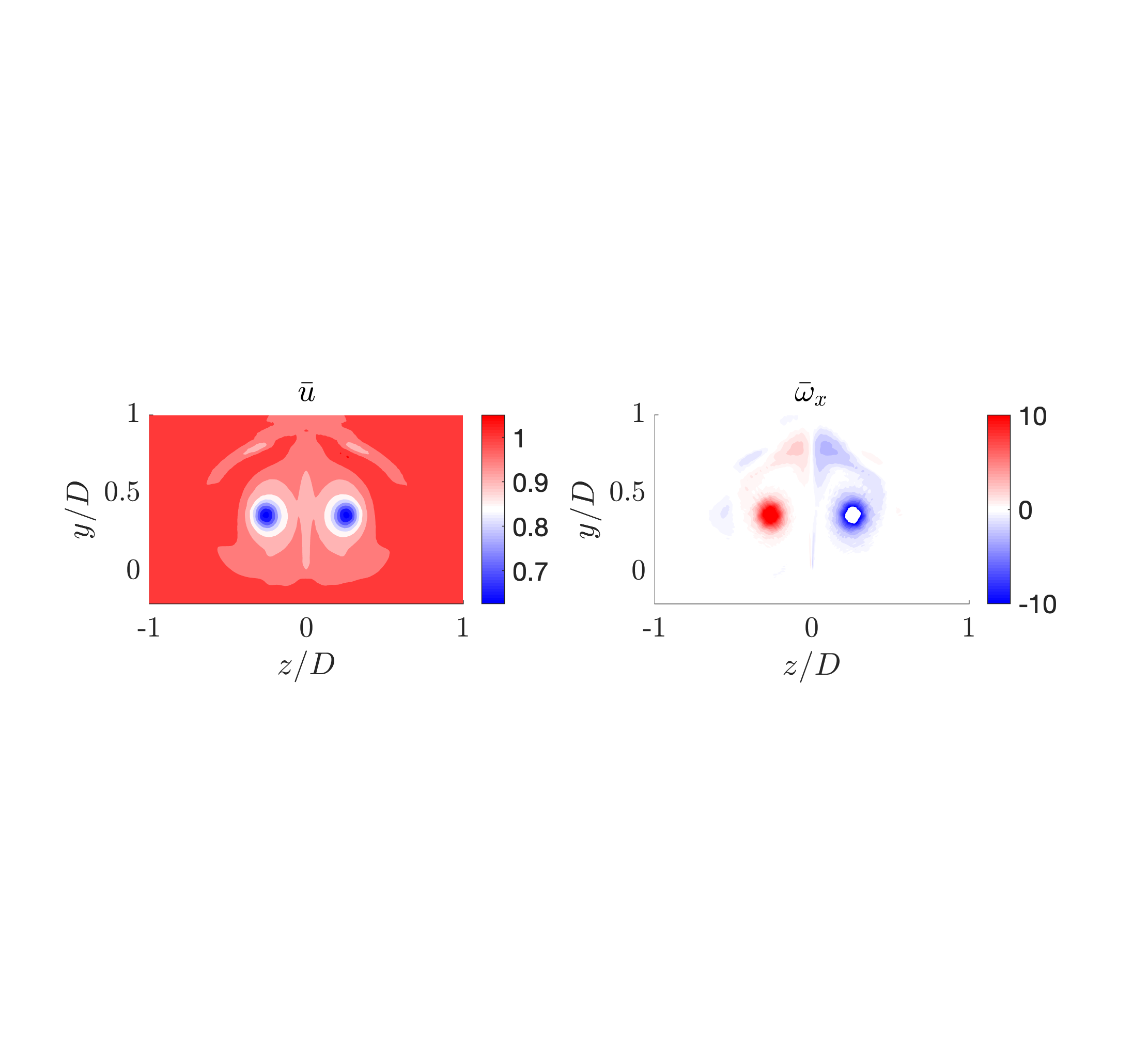}
  \caption{}
  \end{subfigure}\\
 \begin{subfigure}[b]{0.46\textwidth}
 \includegraphics[width=\textwidth, trim={0.1cm 0cm 1cm 0cm},clip]{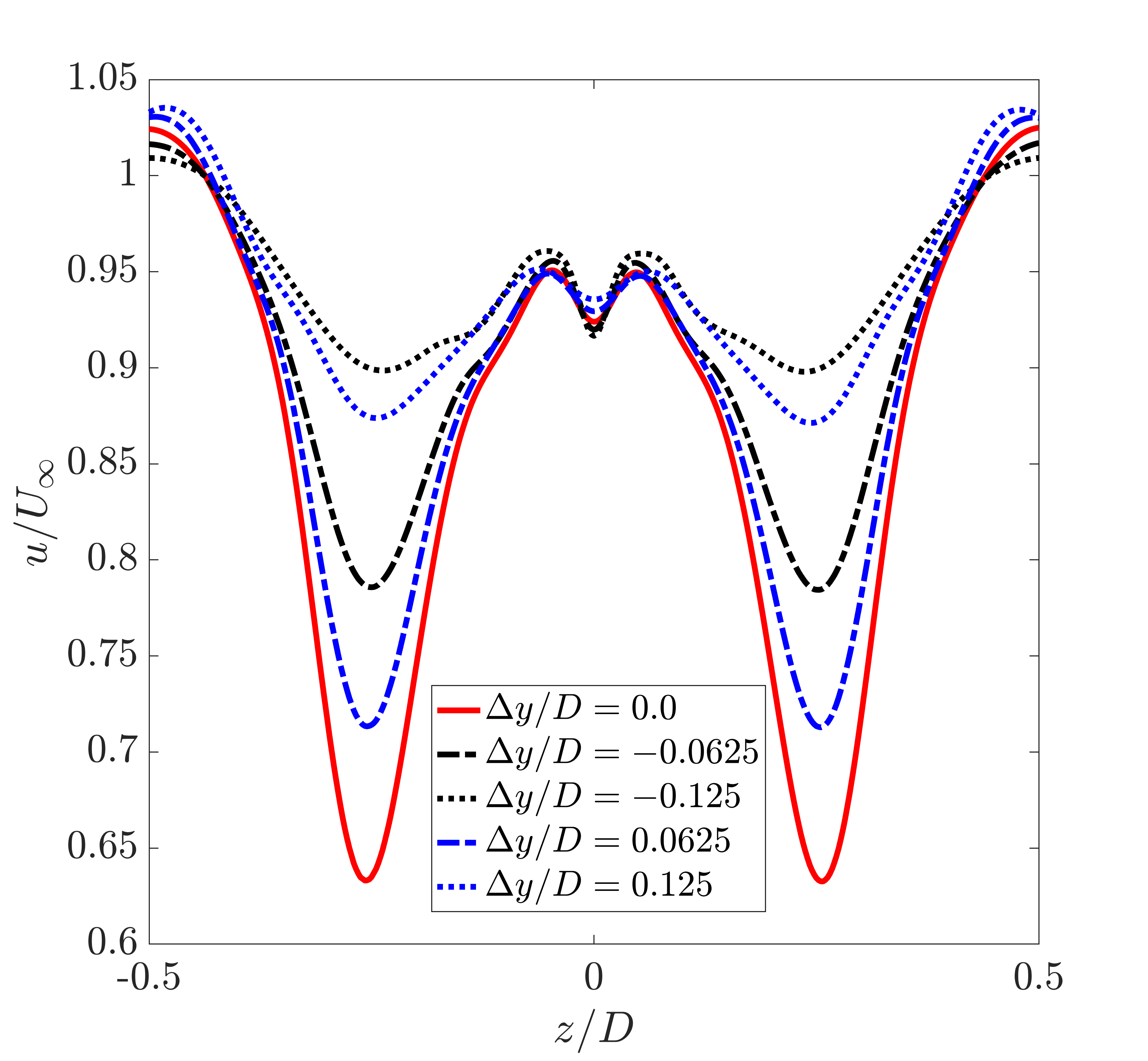}
\caption{}
\end{subfigure}\quad
 \begin{subfigure}[b]{0.46\textwidth}
 \includegraphics[width=1.0\textwidth, trim={0.1cm 0cm 1cm 0cm},clip]{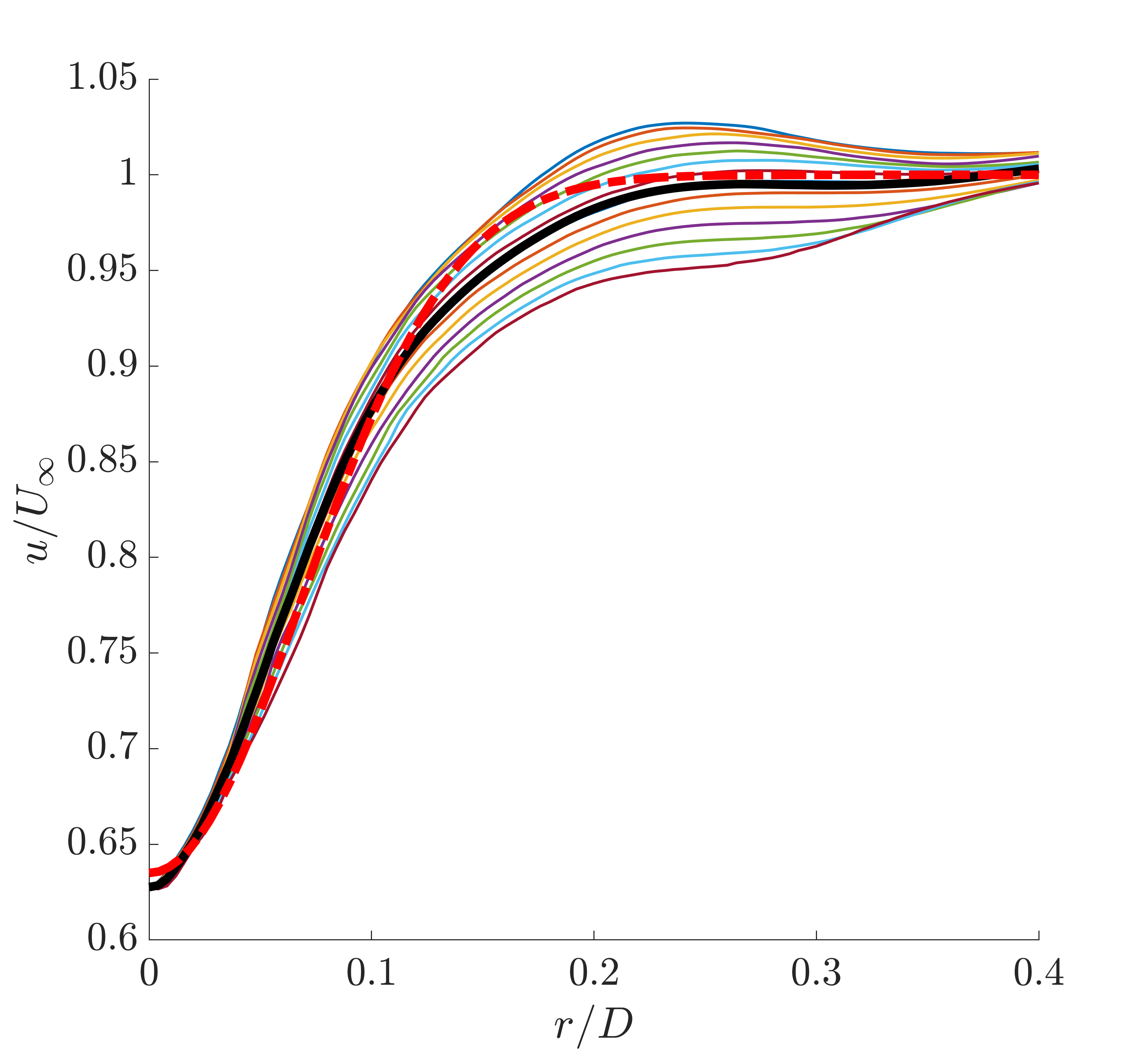}
 \caption{}
\end{subfigure}
\caption{Axial velocity distribution at a streamwise location $X/L=2.0$. (a)$\bar{u}$ contours. (b) Wake-like behavior shown by velocity distributions at different vertical locations measured from the vortex center. (c) Fitting of axial velocity from simulations to the Batchelor vortex. The thick black line indicates the azimuthally averaged axial velocity, and the red dashed line indicates the axial velocity in the fitted Batchelor vortex. Note the vortex in the simulation is recentered at (0,0) after interpolating to the cylindrical coordinate system.}
\label{fig:vortex_sketch2}
\end{figure}

Afterbody vortices, such as the ones of interest, are characterized by two components: swirl and strain. 
While the swirl effect is accounted through the radially varying in-plane  velocity field, the axial velocity provides the strain field.
The mean vortical field at location $X/L=2.0$ was shown earlier in figure~\ref{fig:2dsections}, while 
the mean axial velocity distribution in the vortex is shown in  figure~\ref{fig:vortex_sketch2}(a). 
As expected, the velocity is minimum at the center of the vortex and increases along the radial direction, eventually reaching free-stream values about $10$ core radii ($\delta$) away from the center. 
The quantitative profiles in figure~\ref{fig:vortex_sketch2}(b) at different vertical locations show typical wake-like behavior. 
An important point to note is that the magnitude of $\bar{u}$ never reaches the free-stream value between the vortices because of their relatively close proximity to each other. 
This motivates the use of a vortex pair as opposed to an isolated \textbf{L} or \textbf{R} vortex for the present analysis.

Wake vortices have been modeled with different theoretical models that contain their characteristic information.
Vortex models used in the literature for analyses of wingtip or similar vortices include the Batchelor or $q$-vortex~\citep{edstrand2016mechanism}, Lamb-Oseen vortex~\citep{antkowiak2004transient}, Lamb–Chaplygin vortex~\citep{jugier2020linear}, Rankine vortex~\citep{lacaze2005elliptic} and Moore-Saffman vortex model~\citep{feys2016elliptical}.
Among these, the Batchelor and Moore-Saffman models account for axial flow and hence are more realistic in providing the dynamics of interest in this work. 
We employ the Batchelor model to match the individual vortex in the current streamwise vortex pair as the role of the axial velocity in the stability mechanism may be crucial.
Although this model vortex is typically expressed in cylindrical coordinates, we analyze them in the Cartesian system to account for the two vortices in the pair.
In this system, the velocity components for the Batchelor vortex can be written as~\citep{paredes2011pse}:
\begin{subequations}
\begin{equation}
\displaystyle u = 1 - \gamma(x) \operatorname{exp}\left(\frac{-r^2}{\delta(x)^2}\right)
\end{equation}    
\begin{equation}
\displaystyle v = -\frac{\kappa}{r^2}(z-z_0)\left(1 -  \operatorname{exp}\left(\frac{-r^2}{\delta(x)^2}\right)\right)
\end{equation}
\begin{equation}
\displaystyle w = \frac{\kappa}{r^2}(y-y_0)\left(1 -  \operatorname{exp}\left(\frac{-r^2}{\delta(x)^2}\right)\right)\end{equation}
\end{subequations}
where $(y_0,z_0)$ is the center of the vortex and $r^2 = (y-y_0)^2 + (z-z_0)^2.$ The quantity $\delta(x)$ corresponds physically to the dimensionless local vortex core radius. 
$\gamma(x)$ is the axial velocity defect and $\kappa$ is a swirl strength parameter.

For the current analysis, one vortex is first matched to the model, and the other anti-symmetric vortex is then placed at the separation distance $b$.
The simulation results are first interpolated on a cylindrical grid and the axial velocity $u$ is then azimuthally averaged to obtain the parameters for the fitted Batchelor vortex.
\citet{edstrand2016mechanism} have found that stability results are insensitive to the averaging process.
Figure~\ref{fig:vortex_sketch2}(c) shows the radial variation of axial velocity at different azimuthal locations. 
The azimuthally averaged vortex velocity as well as the fitted Batchelor vortex model are also shown. 

\begin{figure}
\centering
 \includegraphics[width=\textwidth, trim={1cm 1.5cm 1cm 2cm},clip]{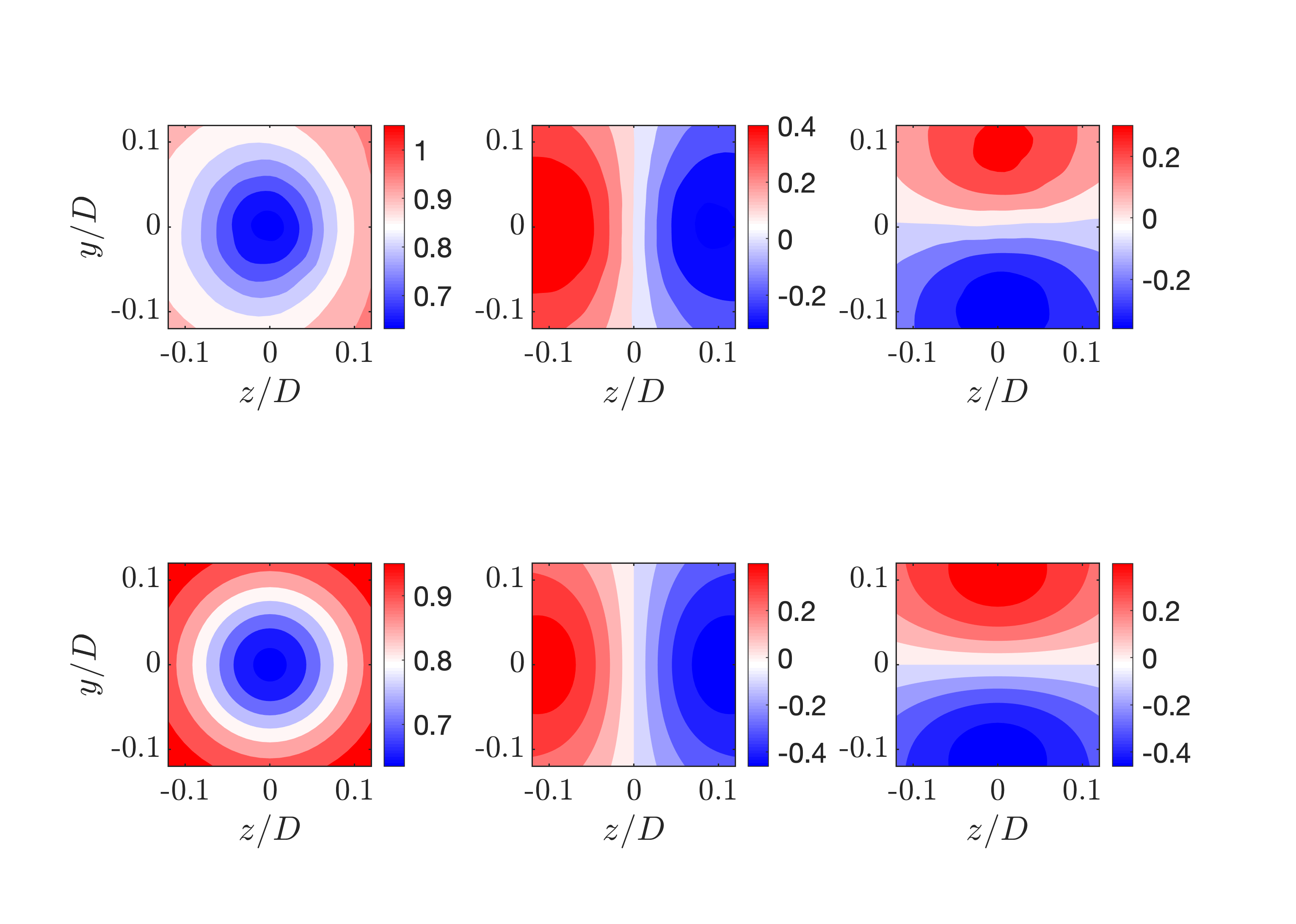}
\caption{Top: \textbf{R}-vortex from the simulation. Bottom: Theoretical Model (Batchelor vortex). $U, V, W$ velocity components are shown from left to right. Note the \textbf{R}-vortex in simulation is centered to (0,0) to compare with the fitted Batchelor vortex. }
\label{fig:Model}
\end{figure}

Figure~\ref{fig:Model} compares the velocity fields in mean \textbf{R}-vortex obtained directly from simulations with those of the fitted vortex used for stability analysis. 
All the three velocity components considered in the model show good match with the LES data. 
The parameters of this matched Batchelor vortex are $\delta=0.0972, \gamma=0.365, \kappa=0.07$, to obtain a non-dimensional swirl parameter $q=\kappa/(\delta \gamma) = 1.9731$. 
This value of $q$ is much smaller than the $q=4.41$ obtained for trailing edge vortex in \citet{edstrand2016mechanism}. 
The present parameters correspond to a Reynolds number based on the core radius as $Re_{\delta} \equiv U_{\infty} \delta/ \nu \simeq 2500$, which is much lower compared to $Re_{\delta}=27,200$ used in the stability study of wingtip vortices~\citep{edstrand2016mechanism}. 

    




\subsection{Stability Approach} \label{sec:stability-approach}
Most stability studies in the literature consider an isolated vortex and efforts on vortex pairs separated by a relatively small distance,  and in the presence of an axial flow, are relatively rare.
The latter problem presents substantial computational challenges as the domain becomes significantly larger~\citep{hein2004instability}.
To ease the computational burden, temporal stability analyses are performed for a given wavenumber range while spatial analysis at the most unstable frequency is used to further confirm the results. 
Further, incompressible linearized Navier-Stokes equations are considered for the eigenvalue analysis instead of compressible equations as the latter increases the linear operator size significantly due to an extra variable. 
This is acceptable because of the low Mach regime considered and a comparison with the fully compressible formulation is given below to ensure confidence.
Consistent with the simulations, a spanwise coordinate system is used in which two inhomogeneous spatial directions, $y$ and $z$, are resolved simultaneously, while the axial direction, $x$, is considered to be locally homogeneous. 
The resulting eigenvalue problems along with boundary conditions for both temporal and spatial analyses are summarized in appendix~\ref{app:LSE}.

\begin{figure}
\centering
  \begin{subfigure}[b]{0.47\textwidth}
    \includegraphics[width=\textwidth, trim={1cm 0cm 1cm 0cm},clip]{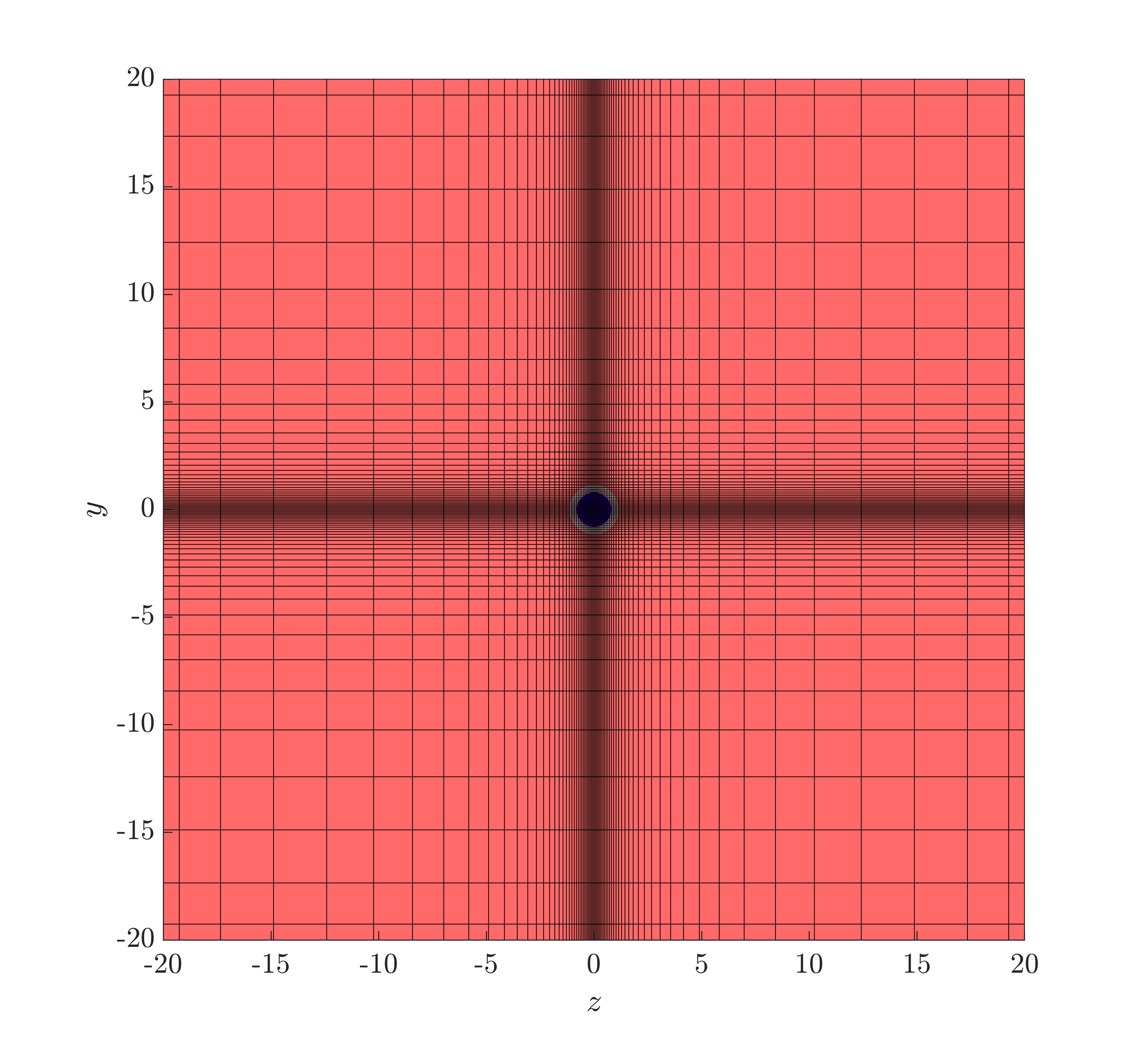}
    \caption{Isolated vortex $(72\times 72, c=0.975)$.}
  \end{subfigure}
  \enspace
    \begin{subfigure}[b]{0.49\textwidth}
     \includegraphics[width=\textwidth, trim={1cm 0cm 1cm 0cm},clip]{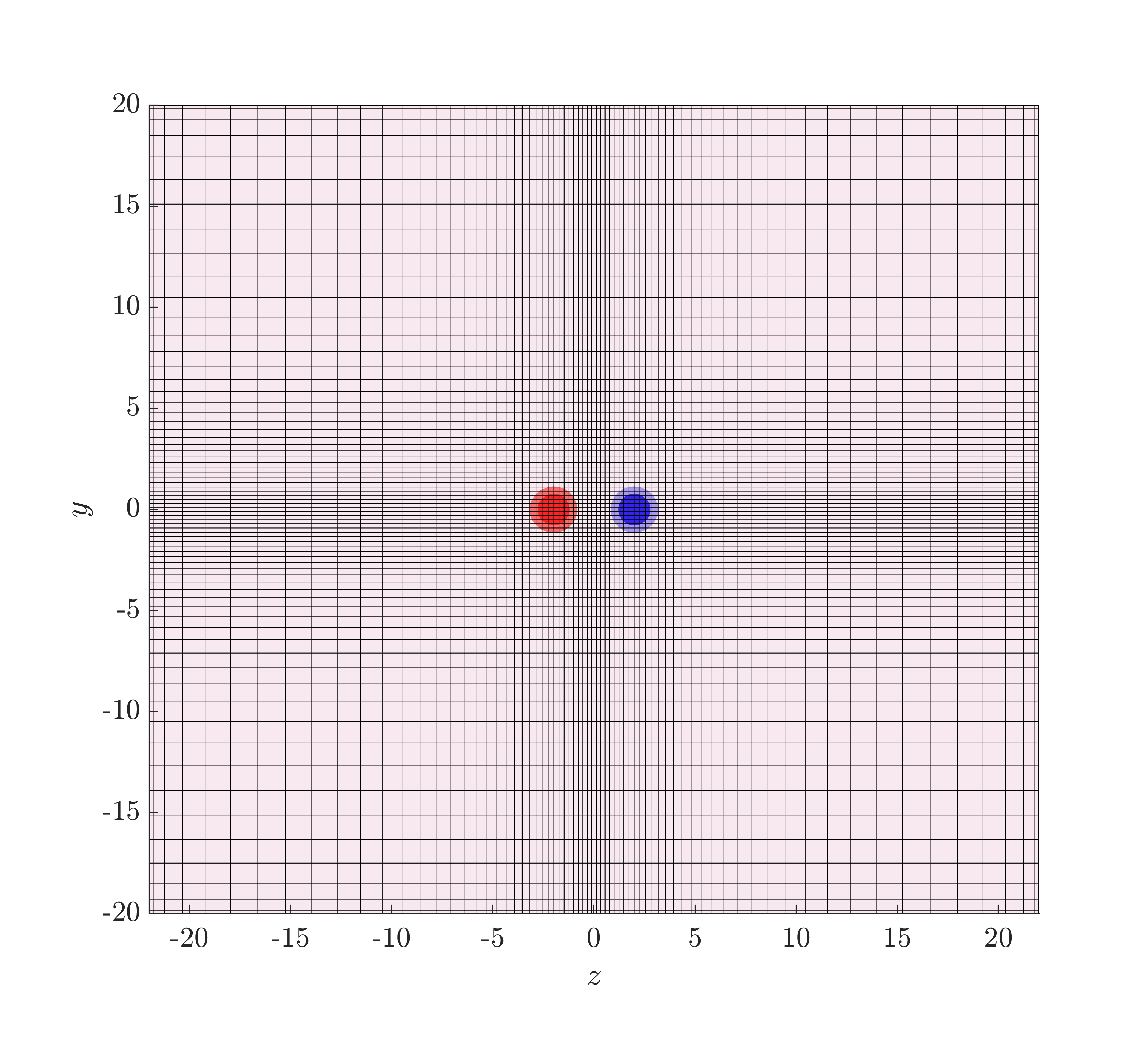}
    \caption{Vortex pair ($100\times 100, c=0.75$).}
  \end{subfigure}
\caption{Spatial discretization for stability analysis.}
\label{fig:grid}
\end{figure}

\cite{hein2004instability, gonzalez2008eigenmodes} have described the importance of choosing a sufficiently large domain for solving the eigenvalue problem, and a suitable grid discretization scheme to resolve the vortex core and wake regions adequately, while ensuring proper treatment of the relatively innocuous farfield.
Because of the large domain, whose size scales as $y_{\infty}/\delta \approx 20$, $z_{\infty}/\delta \approx 20$ \citep{edstrand2018parallel, hein2004instability, paredes2011pse}, finite difference schemes are not generally a preferred choice because of the large number of grid points required in the farfield region. 
\cite{hein2004instability} use a mapping $\zeta$ of Chebyshev collocation points $\eta_j \in [-1; ~1]$ such that:
\begin{equation}\label{eq:tanh}
    \zeta_j = \zeta_{\infty} \frac{\operatorname{tan}\frac{c\pi}{2}\eta_j}{\operatorname{tan}\frac{c\pi}{2}}
\end{equation}
where $\zeta_{\infty}$ is the farfield boundary. 
The stretching parameter $c \in (0, ~ 1]$ defines the clustering of grid points near the vortex core. 
Higher $c$ provides more clustering near the vortex core while imposing a relatively coarser grid away from the core.
This distribution of points will be referred to as the CHEB-TANH grid.
Figure \ref{fig:grid}(a) shows a typical grid clustering for $\delta=1,y/\delta \in [-20 ~ 20], z/\delta \in [-20 ~ 20], N_y=N_z=72, c=0.975$, ensuring that the vortex at the center is adequately resolved.

For a vortex pair, the choice of discretization scheme requires more care than for a single vortex, because of their separation along the $z$-axis connecting their centers. 
The domain used for the two vortex system is $y/\delta \in [-20; ~ 20], z/\delta \in [-20-b/2; ~ 20+b/2]$, where $b$ is the separation between vortices as defined earlier. 
Naturally, for similar accuracy, the number of grid points required is much higher than the single vortex case, as the stretching parameter, $c$, in the CHEB-TANH needs to be relaxed to encompass both the vortices. Figure \ref{fig:grid}(b) shows a representative grid clustering for $c=0.75$, resolving both the vortices as well as the spacing between them.

Table~\ref{tab:qvortex} shows validations for both temporal and spatial stability codes used in the study. 
For the temporal approach, the classical $q$-vortex at $q = 0.475, Re = 100, k = 0.418$ is chosen; this problem was first examined by \citet{mayer1992viscous} and subsequently used for validation in several studies~\citep{hein2004instability, paredes2011pse, edstrand2018parallel}. 
The results for the unstable mode are reported along with the reference results from \citet{paredes2011pse}. 
Excellent convergence is obtained with the mapped CHEB-TANH discretization, even for a slightly smaller domain in $[-15 ~ 15]$ and moderate grid size.
The stretching parameter in this discretization is fixed at $c=0.975$, ensuring that the vortex at the center is adequately resolved.
A good match is also obtained with the sixth-order finite difference (FD6) approach albeit at a higher resolution. 
The table also includes a result when a fully compressible stability solver is used for the same $q$-vortex analysis at a low Mach number ($M=0.1$). 
Comparing the results between incompressible and compressible solvers with similar grid and numerical schemes, we note only a minor difference of about 1\% in the frequency and 0.44\% in the growth rate. 
\begin{table}
 \begin{center}
\def~{\hphantom{0}}
 \begin{tabular}{lcccc}
 \toprule
        Domain &  Diff. Scheme & Resolution  & Temporal & Spatial  \\[3pt]
      \midrule
      Paredes & FD10 & & 0.028353+0.009617i & 0.543-0.185i \\
  $[-15 ~ 15]$ & CHEB-TANH  & $40 \times 40$ & 0.028337 + 0.009607i & 0.5534-0.1839i \\
    $[-20 ~ 20]$ & CHEB-TANH  & $60 \times 60$ & 0.0283321 + 0.0096043i & 0.549-0.18i \\
    $[-15 ~ 15]$ & FD6 & $101 \times 101$ & 0.028533 + 0.009980 i &  \\
    $[-15 ~ 15]$ & FD6(C) & $101 \times 101$ & 0.028225 + 0.010024 i &  \\
            \bottomrule
 \end{tabular}
 \caption{Validation of stability analyses techniques. 
 The temporal analysis is performed for a $q$-vortex with parameters $q=0.475, Re=100, k=0.418$. 
 The spatial analysis is performed for the Batchelor vortex ($\kappa=0.8,\gamma=0.8, \delta=1$) at $\omega=0.86$ and $Re=3000$. 
 Temporal analysis using compressible formulation, FD6(C), at low Mach number ($M=0.1$) yields leading  modes close to those obtained with fully incompressible code.}
 \label{tab:qvortex}
 \end{center}
\end{table}

For the spatial code validation, the Batchelor vortex study by \citet{paredes2014advances} at $\kappa=0.8, \gamma =0.8, \delta=1, Re=3000$ is used for reference.  
In table~\ref{tab:qvortex}, results for $m=1, \omega = 0.86$) are compared against the values reported in ~\cite{paredes2014advances}.
Even with a moderate $40\times40$ grid, the frequency obtained in the current study is within 2\% of the reported value; the comparison improves with increase in grid size. For the subsequent results for current afterbody vortices, CHEB-TANH grid is employed.

\subsection{Stability Analysis of Afterbody Vortices}
\label{sec:stab-analys-afterb}
\begin{figure}
\centering
    \begin{subfigure}[b]{0.63\textwidth}
    \includegraphics[width=\textwidth, trim={0cm 0cm 1cm 0cm},clip]{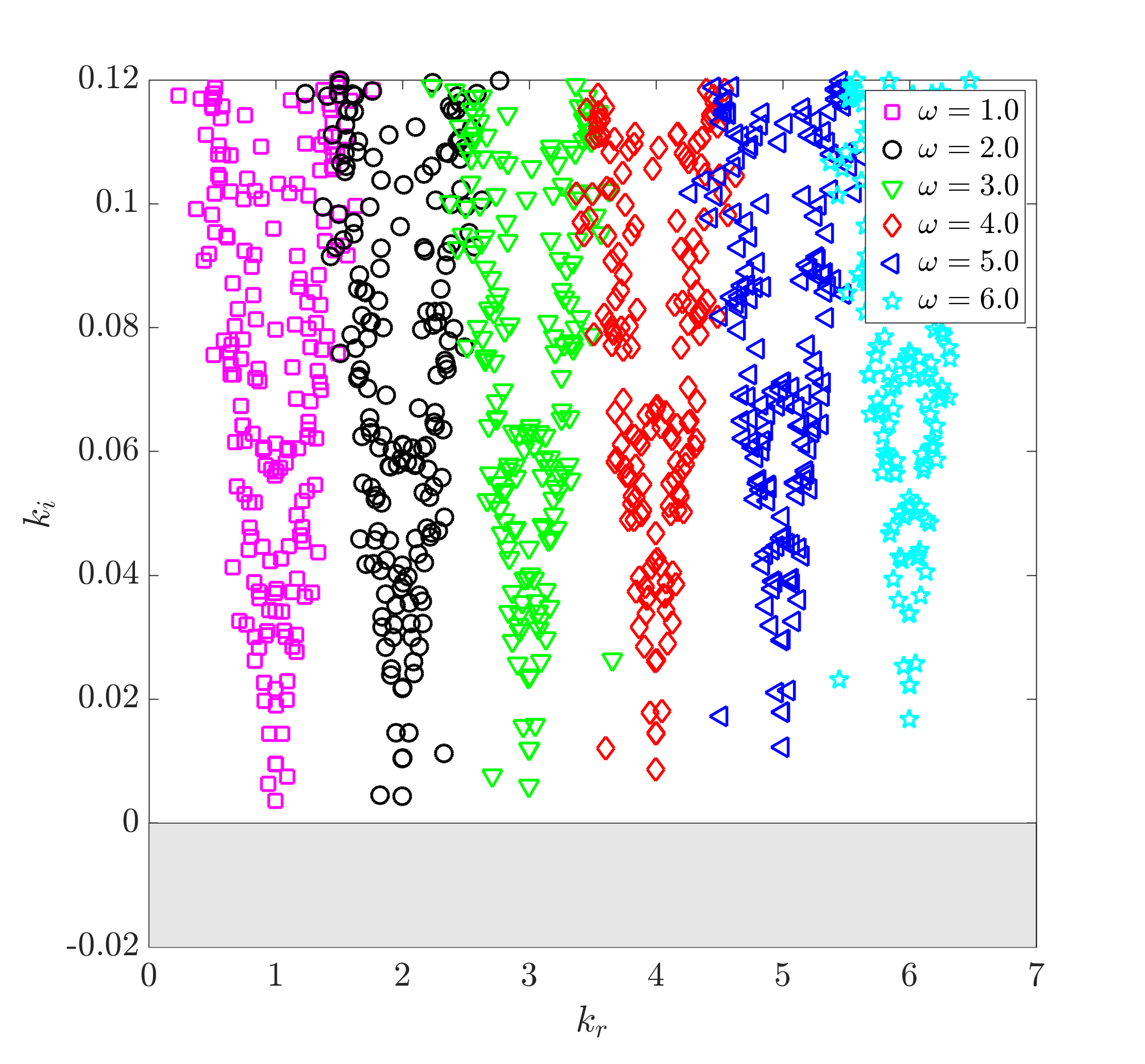}
    \caption{Spatial analysis for $\omega \in [1.0~ 6.0]$.}
  \end{subfigure}
    \enspace
  \begin{subfigure}[b]{0.32\textwidth}
    \includegraphics[width=\textwidth, trim={7cm 0cm 7cm 0cm},clip]{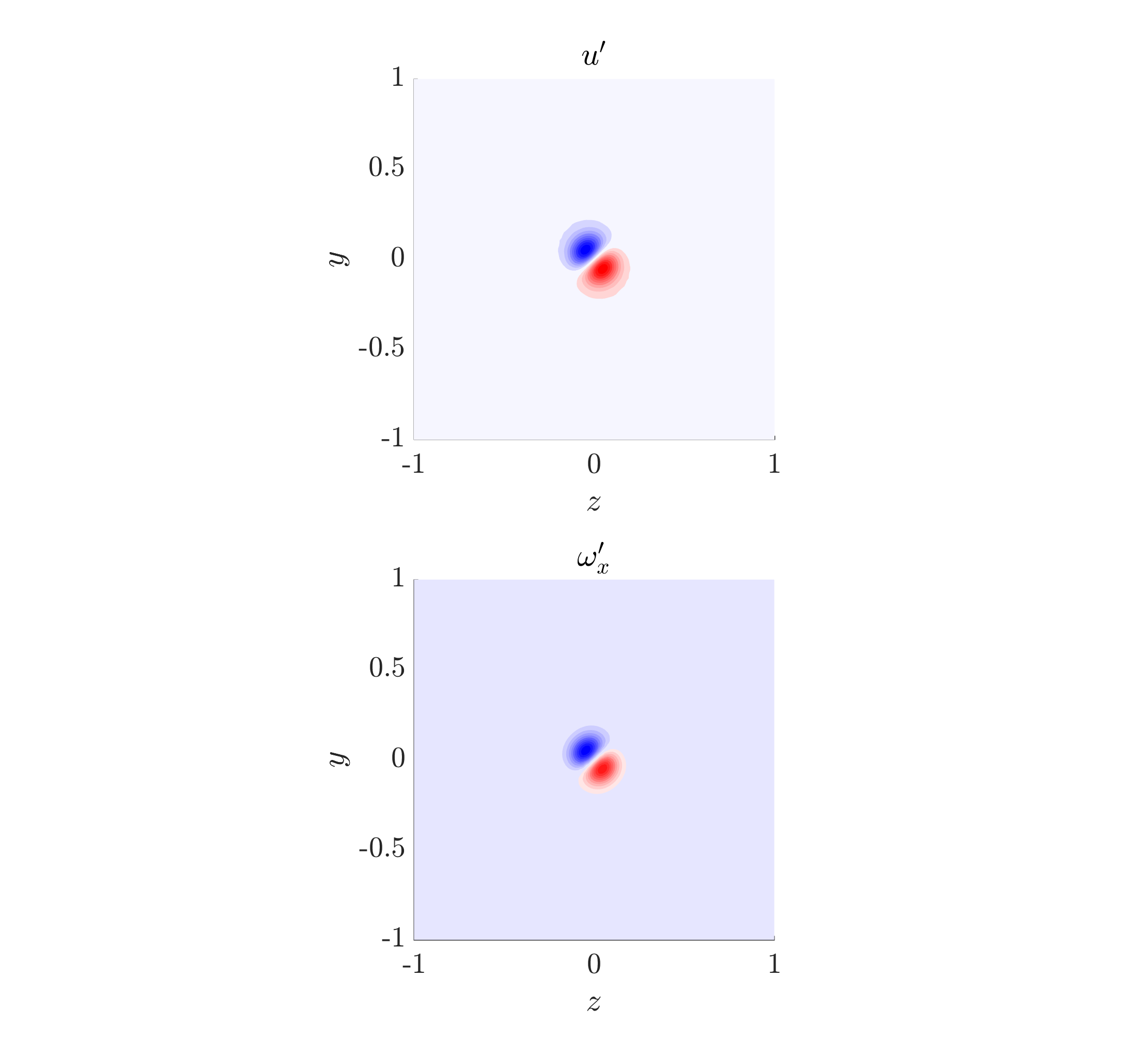}
    \caption{Least stable mode at $\omega=1$ ($k = 0.9346 + 0.0025i$).}
  \end{subfigure}
\caption{Eigenspectrum from spatial analyses of an isolated vortex for a range of frequencies(a). Shaded portion indicates the region of growing modes ($k_i < 0$). The least stable mode at $\omega=1$ shows $m=1$ elliptic instability (b).}
\label{fig:stab1}
\end{figure}

The stability dynamics of the afterbody vortices are now delineated. 
Before showing the dynamics of the vortex pair, we first discuss the stability characteristics of an isolated vortex.
The fitted Batchelor vortex described in section \ref{sec:vortexfit} is used for this study.
Figure~\ref{fig:stab1}(a) shows results from spatial analyses of this vortex for input frequency varying from $\omega=1$ to $\omega=6$. 
A $60 \times 60$ grid with $c=0.975$ is used for these analyses. 
As seen from the figure, the vortex is stable for the entire frequency range $\omega \in [1.0~6.0]$. 
The vortex becomes increasingly more stable as the frequency is increased. 
For input frequency $\omega =1$,  the least stable mode is obtained at $k \equiv k_r + i k_i = 0.9346 + 0.0025i$. 
Note that the positive value of $k_i$ in the spatial analysis indicates a decaying mode (see appendix~\ref{app:LSE}). 
The spatial structure of this mode is shown in Fig.~\ref{fig:stab1}(b). 
Both the axial velocity and vorticity fluctuations show dipole structures indicating the $|m|=1$ elliptic mode discussed earlier. 
This mode is typically observed to be the dominant instability mechanism  in wake vortices~\citep{edstrand2016mechanism}.

\begin{figure}
\centering
    \begin{subfigure}[b]{0.55\textwidth}
        \includegraphics[width=\textwidth, trim={0cm 0cm 0cm 0cm},clip]{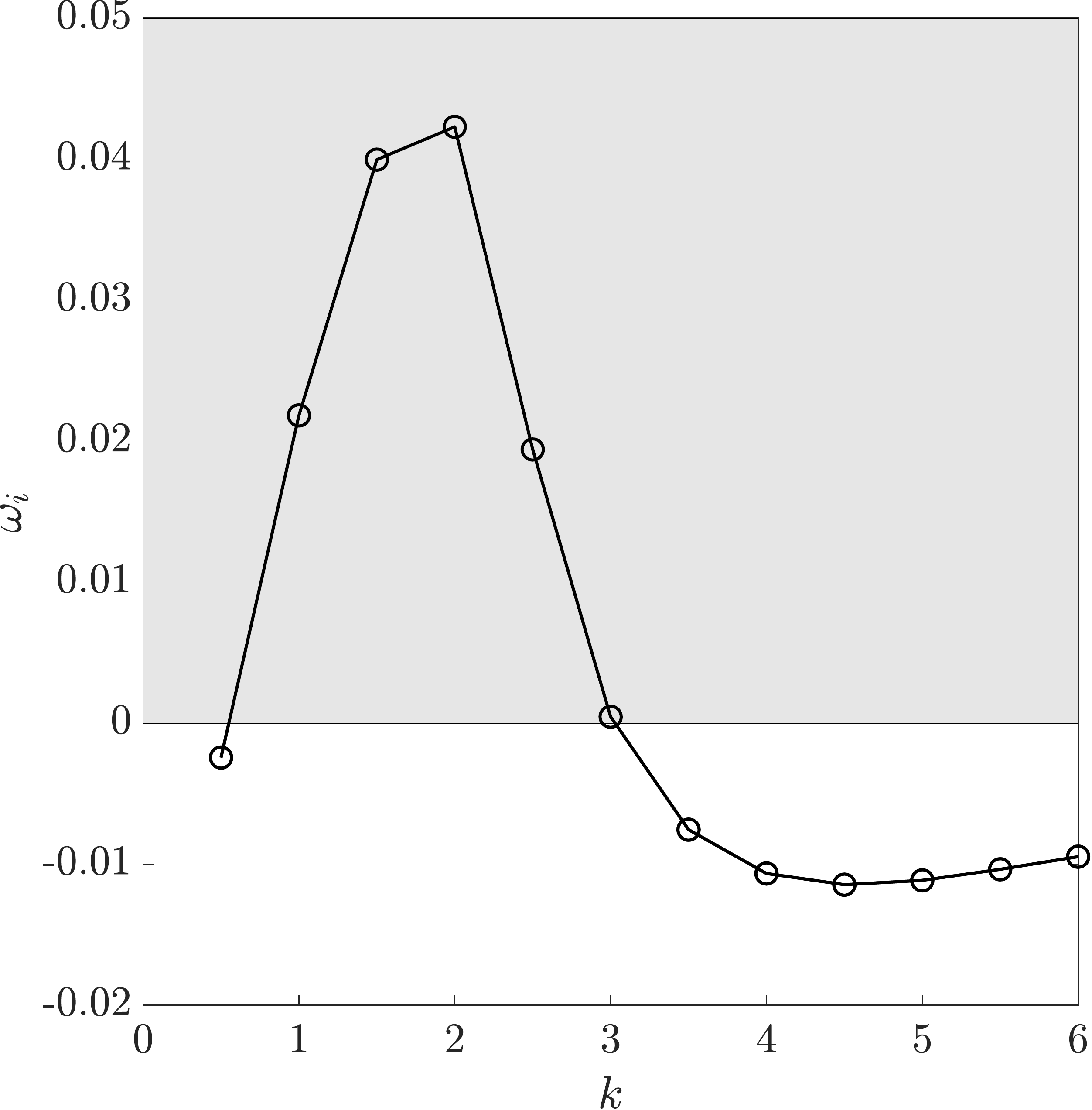}
    \caption{Temporal analysis for $k \in [0.0~ 6.0]$.}
  \end{subfigure}
    \enspace
  \begin{subfigure}[b]{0.32\textwidth}
    \includegraphics[width=\textwidth, trim={7cm 0cm 7cm 0cm},clip]{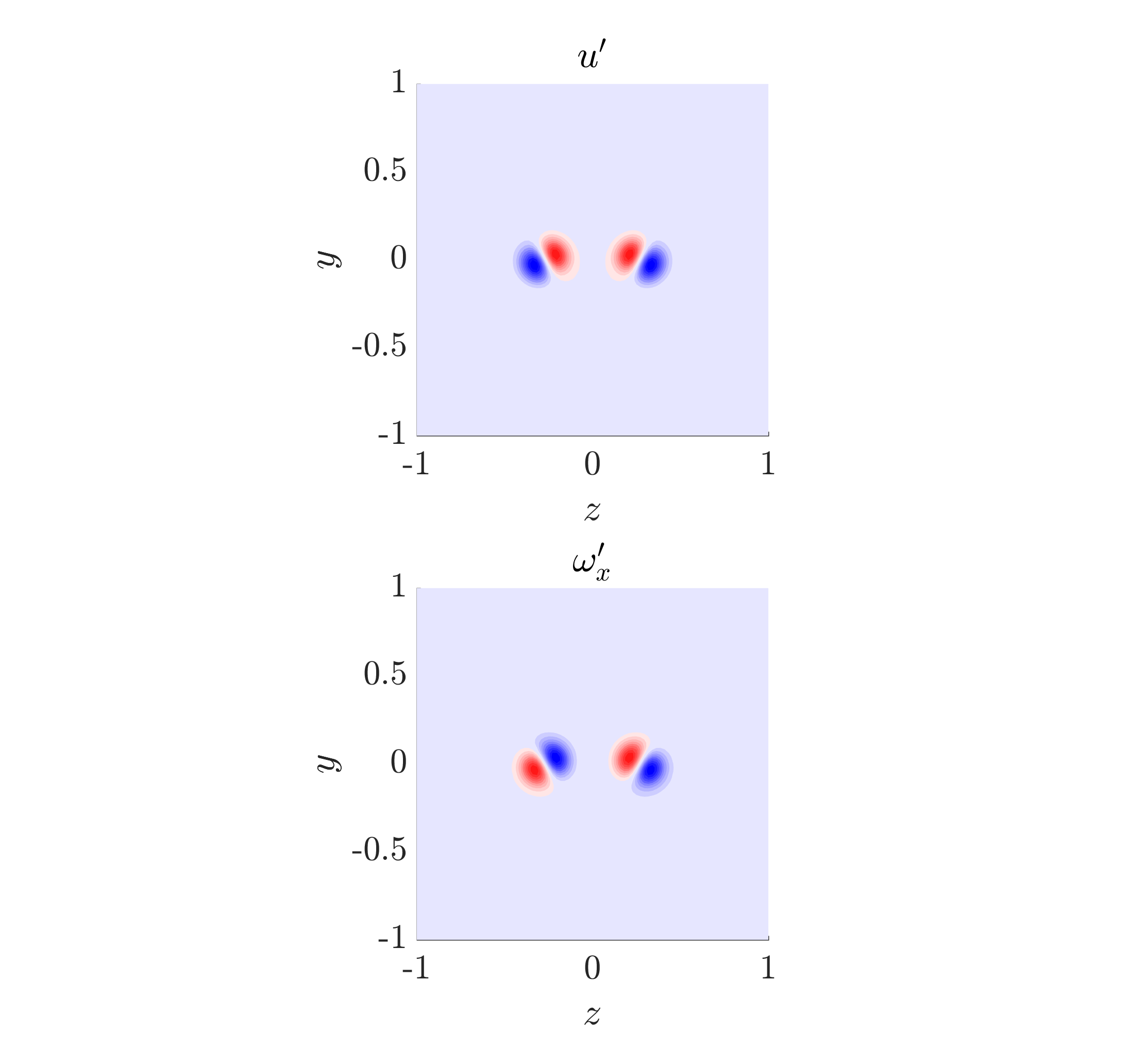}
    \caption{Unstable mode at $k=2$.}
  \end{subfigure}
\caption{Eigenspectra from temporal analyses of the vortex pair for a range of wavenumbers (a). Shaded portion indicate the region of growing modes ($\omega_i > 0$).  Structure of the fastest growing mode at all wavenumbers is shown in (b).}
\label{fig:stab2}
\end{figure}

Now we consider the change in stability characteristics due to the presence of a counter-rotating vortex next to the above stable vortex.  
Following the results of Section~\ref{sec:meanvortex}, the counter-rotating vortices are separated by $b/\delta = 5.6132$. 
As alluded to earlier, this relatively large separation between the vortices increases the computational requirement enormously because the stretching parameter needs to be relaxed to as low as $0.75$, with a corresponding increase in the grid size to ensure sufficient accuracy. 
The parallel spatial analysis, performed above for the isolated vortex, is natural for the study of the vortex pair also, since it directly provides the wavenumber at which the flow can become unstable. 
However, since the grid size required for vortex pair is significantly larger (up to $150 \times 150$) than that for an isolated vortex, we use a temporal approach with wavenumber variation to extract the least stable wavenumber.

\begin{figure}
\centering
    \begin{subfigure}[b]{0.48\textwidth}
    \includegraphics[width=\textwidth, trim={0cm 0cm 1cm 0cm},clip]{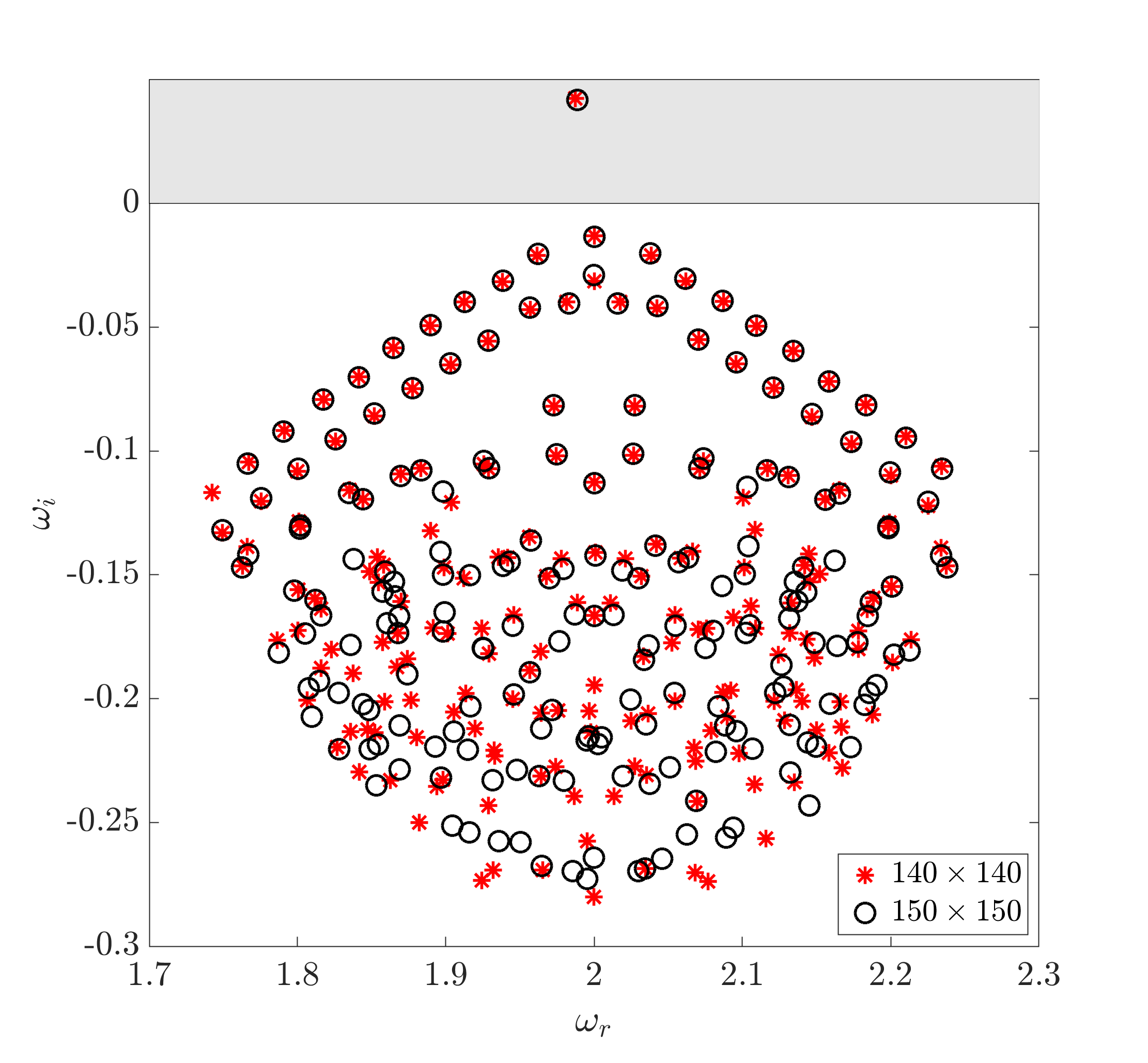}
    \caption{Temporal ($k=2$).}
  \end{subfigure}
    \enspace
  \begin{subfigure}[b]{0.48\textwidth}
    \includegraphics[width=\textwidth, trim={0cm 0cm 1cm 0cm},clip]{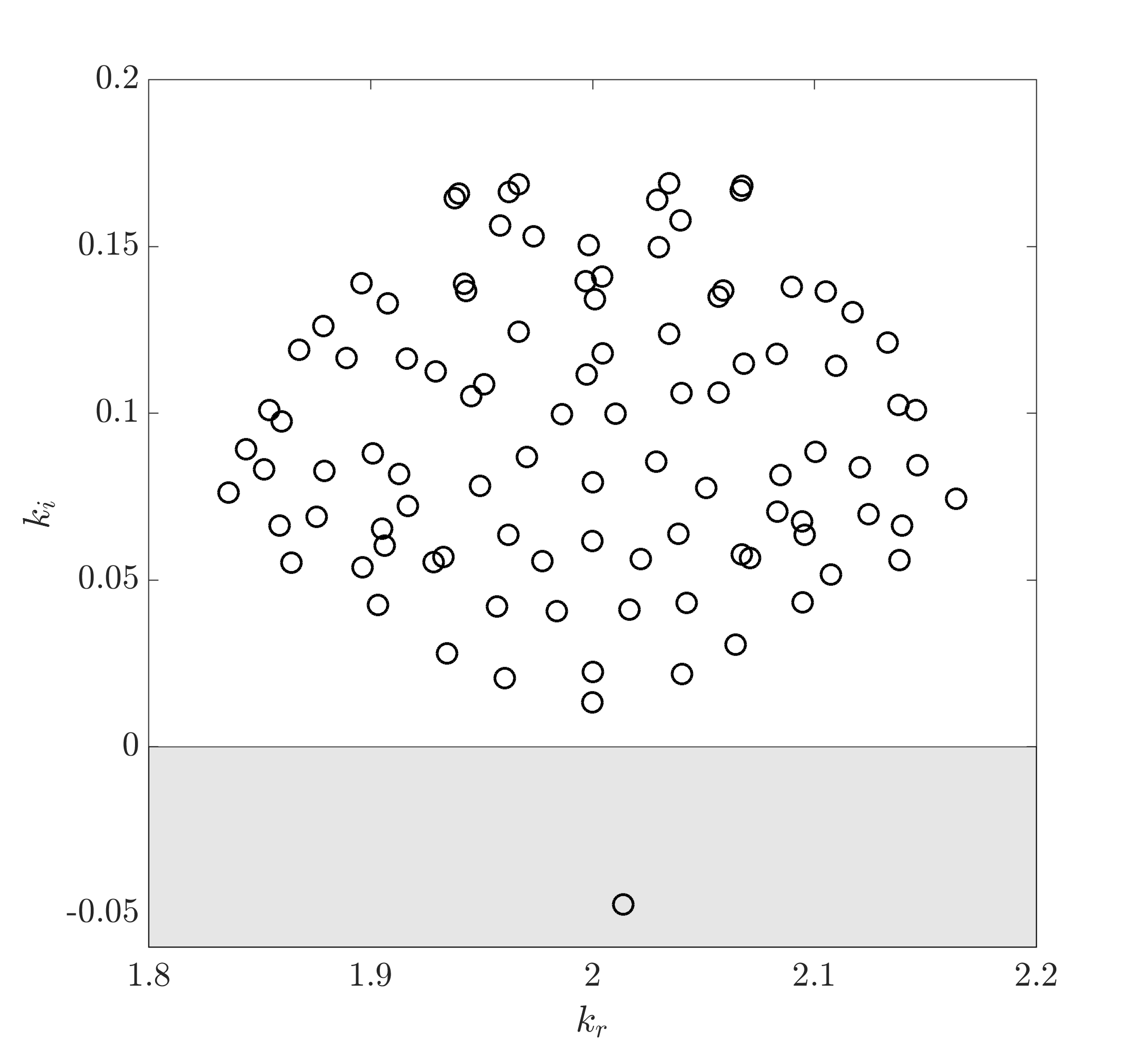}
    \caption{Spatial ($\omega=2$).}
  \end{subfigure}
\caption{Eigenspectra from  temporal(a) and spatial(b) analyses of the afterbody vortex pair. Both approaches show analogous instabilities at a phase speed of about $0.99$.}
\label{fig:stab3}
\end{figure}

The temporal analysis results for a streamwise wavenumber sweep of $k \in [0.5 ~6.0]$ is shown in Fig.~\ref{fig:stab2}(a). 
These results are obtained by using a $140\times140$ CHEB-TANH grid that gives a dense matrix of $78{,}400 \times 78{,}400$. 
The vortex pair displays unstable eigenmodes at wavenumbers $1.0 \le k \le 3.0$, with growth rate increasing between $k=1.0$ and $k=2.0$. As the input wavenumber increases beyond $k=2.0$, the modes become less unstable and only stable modes are recovered for $k > 3.0$.
For the subsequent analyses, we focus on the wavenumber $k=2.0$ at which the fastest growing modes are recovered. Figure~\ref{fig:stab3}(a) shows the convergence of dominant modes in the eigenspectrum at this wavenumber as obtained by performing the analysis on a finer $150 \times 150$ grid. 
Both grids recover most unstable modes at $\omega \equiv \omega_r + i \omega_i  = \pm 1.988 + 0.0417i$ (note only the positive frequency $\omega_r$ mode is shown). 
The phase speed of this mode is $c_r \equiv \omega_r/k  = 0.9940$. 
In order to confirm that spatial analysis also recovers similar instabilities, we show the eigenspectrum from this approach at input frequency $\omega = 2.0$ in figure~\ref{fig:stab3}(b). 
Note that the matrix size for the spatial problem is very large ($137{,}200 \times 137{,}200$) for a similar grid size of $140 \times 140$ and solving this eigenvalue problem with a Krylov size of $100$, requires about $140$ gigabytes of the physical system memory.
The spatial analysis recovers an unstable mode at $k \equiv k_r + i k_i = \pm 2.0138 - 0.0473i$ as shown in the unstable region ($k_i < 0$) in figure~\ref{fig:stab3}(b). 
The phase speed of this mode, $c_r \equiv \omega/k_r  = 0.9932$, is very similar to that obtained from the temporal analysis earlier. 
The recovery of analogous instabilities in both spatial and temporal approaches further validates present observations.

The structure of this mode as obtained from both temporal and spatial approaches is very similar as shown in figure~\ref{fig:stab2}(b).
The shape exhibits a dipole form representing the $|m|=1$ elliptic mode  for both vortices in the pair, as observed in POD analysis. 
This dipole is extracted in both the axial velocity and vorticity fluctuations, but it is symmetric about the mid-plane ($z=0$) in the former, but anti-symmetric in the latter. 
This is consistent with the symmetry of the mean flow. 
The inclination of the dipole with respect to the horizontal axis, $\alpha_1$ (see sketch~\ref{fig:podmode3}), is about \ang{30}, which is higher than that observed in POD mode $1$. 
This suggests that the interaction of the downstream shear layer with the vortices,  as well as non-linearities present in the flow, modifies these structures and introduces high-frequency content as evident in the spectral signatures of the POD modes (figure~\ref{fig:podmode}(e,f)).

The frequency of the most unstable mode as obtained from the temporal analysis is $\omega \simeq 1.988$, or equivalently $St_D \simeq 0.32$. 
This low-frequency was observed in the vortex flowfield (figure~\ref{fig:pdf}(b)) as well as in the POD modes (figure~\ref{fig:podmode}(e,f)). 
As mentioned earlier, in the experimental work of~\citet{zigunov2020dynamics} this low-frequency was observed and was conjectured to be related to fluctuations due to interactions of cores. 
The present stability analysis confirms this hypothesis and establishes meandering as a low-frequency phenomenon in afterbody flows in similarity with vortices in aircraft wakes~\citep{jacquin2001properties}.

\begin{figure}
\centering
    \begin{subfigure}[b]{0.48\textwidth}
    \includegraphics[width=\textwidth, trim={0cm 0cm 1cm 0cm},clip]{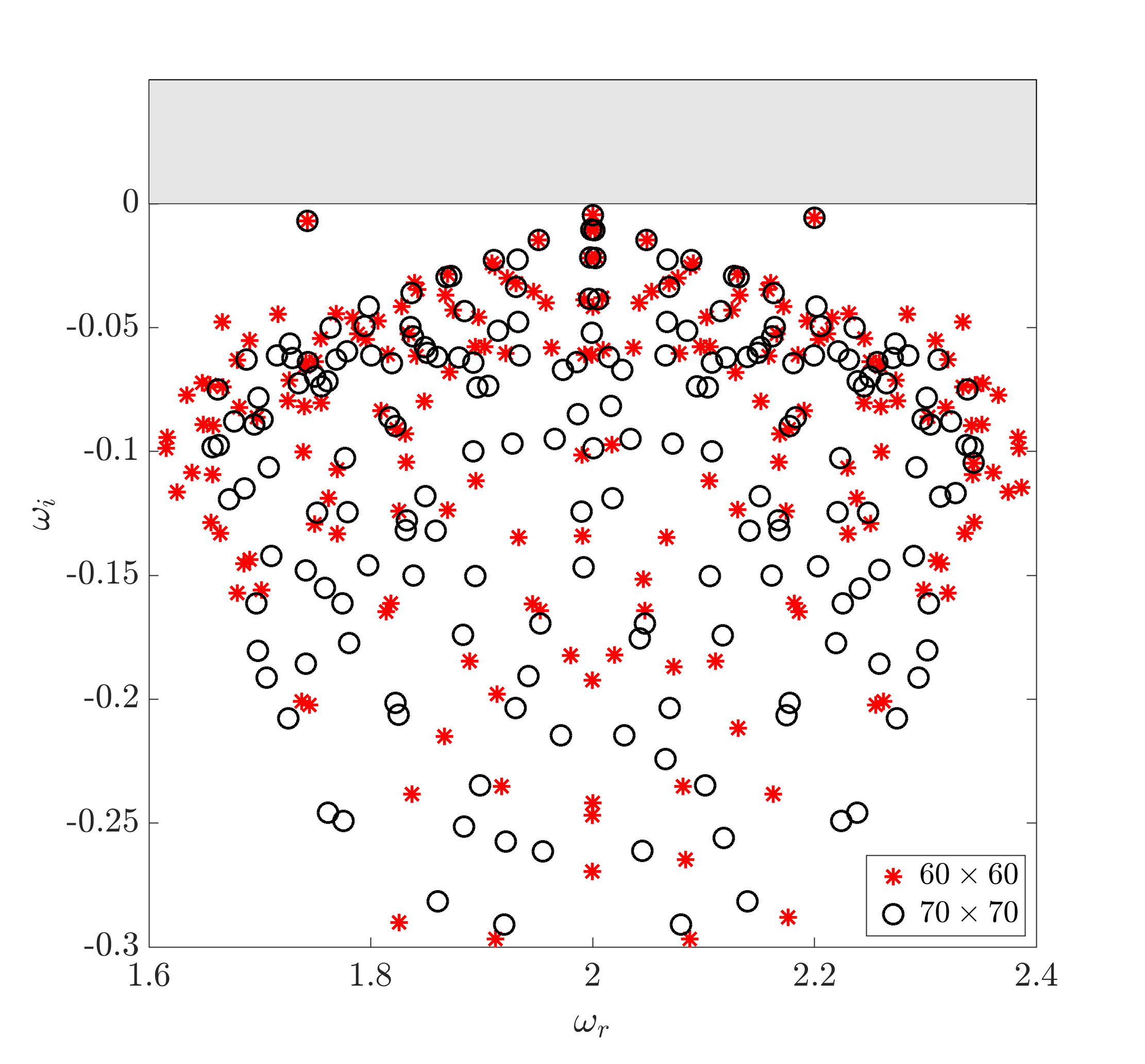}
    \caption{Isolated vortex.}
  \end{subfigure}
    \enspace
  \begin{subfigure}[b]{0.48\textwidth}
    \includegraphics[width=\textwidth, trim={0cm 0cm 1cm 0cm},clip]{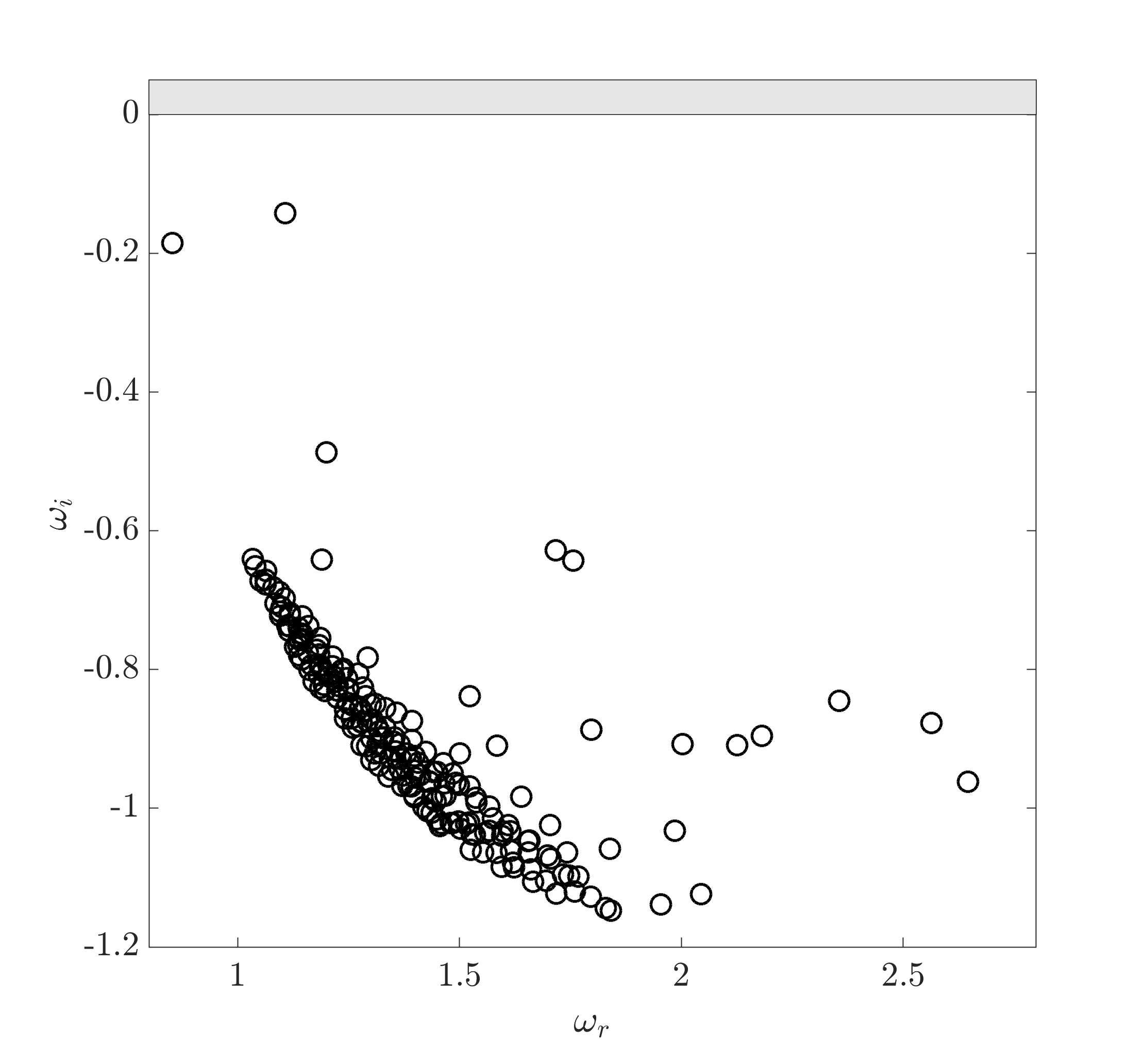}
    \caption{Vortex pair $(u=0)$.}
  \end{subfigure}
\caption{Eigenspectra from temporal analysis at $k = 2.0$. (a) Isolated vortex with axial velocity. (b) Vortex pair with no axial velocity.}
\label{fig:stab4}
\end{figure}

\begin{figure}
\centering
                \includegraphics[width=0.9\textwidth, trim={0cm 0cm 0cm 0cm},clip]{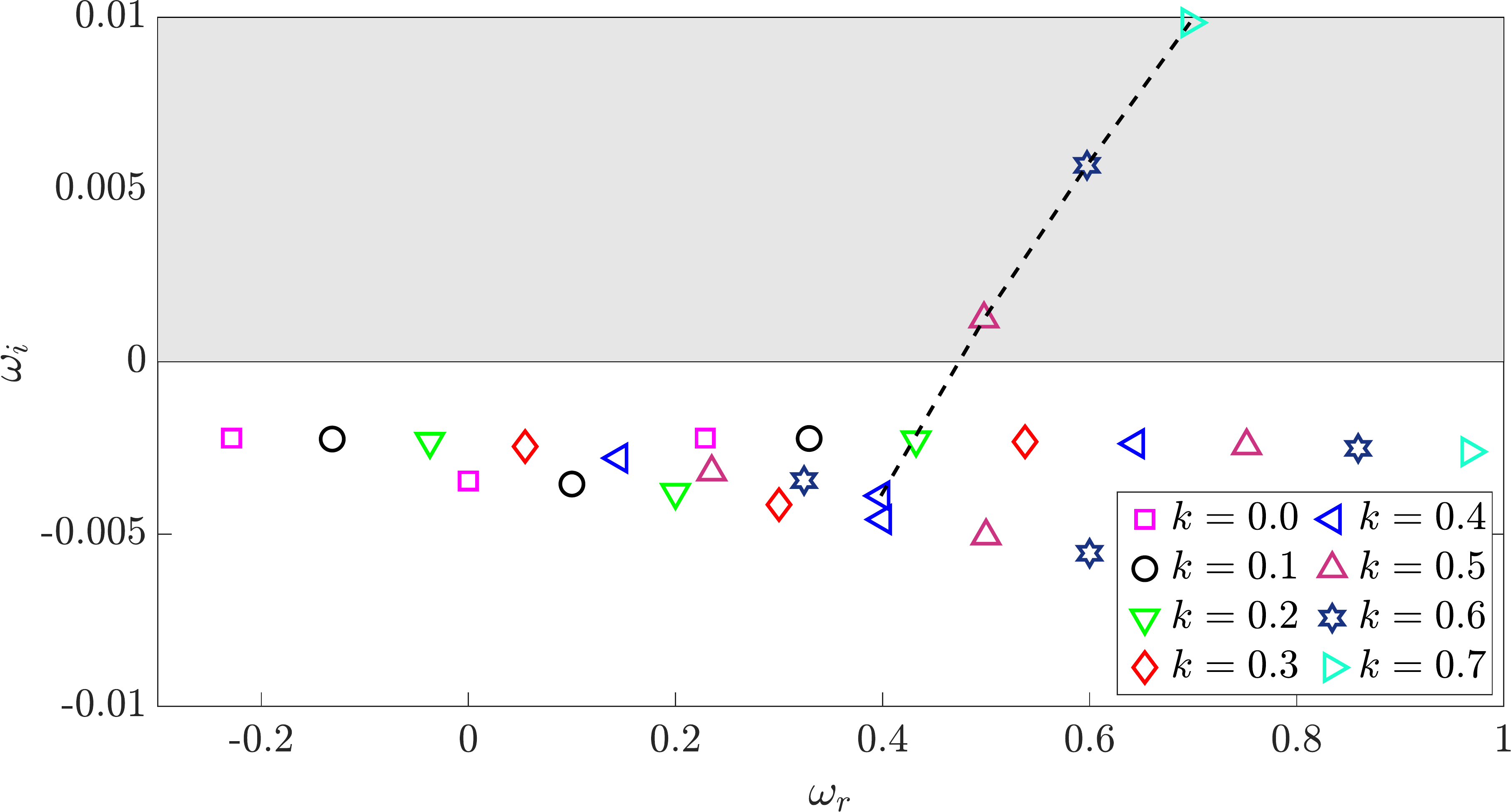}
\caption{Eigenspectra from temporal analyses of the afterbody vortex pair at low wavenumbers. No unstable modes are recovered below $k=0.5$.}
\label{fig:stab5}
\end{figure}

To further confirm the role of the interaction between the vortices as a necessary mechanism for the recovery of unstable modes, we perform the temporal stability analysis for an isolated vortex with the same swirl and strain as in each individual vortex in the pair. 
Figure~\ref{fig:stab4}(a) shows the result from this analysis along with grid convergence for the dominant modes as obtained with two grids. 
Unlike in the two-vortex system, all modes recovered in this analysis are decaying. 
The least stable modes are found at $\omega = \pm 2.0001 - 0.0047i$ exhibiting dipole structures for both vorticity and axial velocity, which is similar to that shown in figure~\ref{fig:stab1}(b).

The external strain field is influenced by the neighboring vortex, and is accounted for in the stability calculations through the axial velocity component.
The correlation analysis of section~\ref{sec:phenomenology}, based on the phenomenological description of instantaneous vortex cores on a sectional plane, indicates only weak influence of the vortices on one another.
That analysis, however, did not include the axial velocity component, whose role in destabilizing the vortices is now clarified.
A stable vortex can become unstable even without the presence of axial flow (as in the Kelvin vortex) due to an elliptic instability of the core; however, the presence of an axial flow can make an otherwise stable vortex unstable \citep{lacaze2005elliptic, jacquin2005unsteadiness}.
In order to examine the effects of this strain,  the temporal stability analysis was repeated on the afterbody vortex pair by setting the axial velocity to zero. 
The problem then becomes essentially similar to Rankine vortex models used in the literature to study elliptic instabilities~\citep{kerswell2002elliptical}.
The eigenspectrum thus obtained is shown in figure~\ref{fig:stab4}(b).
A key observation is that unlike the analysis with the axial velocity (figure~\ref{fig:stab3}(a)), no unstable mode is recovered in this calculation.
Further, the modes are far removed from the zero axis.
Although not shown, the shapes of dominant modes continue to show helical structures. 
The present analysis thus shows that it is necessary to consider a vortex model that includes the strain field in order to recover the correct instability behavior in afterbody vortices. 

Finally, the possible presence of long-wavelength (small $k$) Crow instability in the flow is briefly investigated.
The least stable mode at $k=2$ in figure~\ref{fig:stab2}(b) is inclined at \ang{30} to the horizontal.
The Crow instability on the other hand typically manifests a \ang{45} inclination in the structure~\citep{brionoptimal}. 
To confirm this further at low wavenumbers, the temporal analysis was performed using a $k$-parameter sweep between $0.0$ and $1.0$ at increments of $0.1$.  
Figure~\ref{fig:stab5} shows the resulting eigenspectra for all these wavenumbers for the current vortex pair.
No unstable mode is observed in the $k$-range of $[0.0 ~ 0.4]$, and the first unstable mode occurs at $k=0.5$.
The growth rate of modes continues to increase with wavenumber until $k=2.0$ as discussed earlier.
This indicates the absence of the Crow instability in this flow, although simulations with a longer domain may be necessary to establish this fact numerically. 
Further, the flow conditions that lead to a change in the non-dimensional swirl parameter or the Reynolds number may affect this observation, as discussed in \citet{hein2004instability}.
Another key parameter in the analysis that determines the interaction between the two vortices is the separation distance, $b/\delta$.
If this distance varies substantially with the upsweep angle $\phi$ of the afterbody, the analysis needs to be repeated to examine the stability mechanism. 

\section{Low-rank Representation}\label{sec:lods}

\begin{figure}
\centering
\begin{subfigure}[b]{0.33\textwidth}\includegraphics[width=1.0\textwidth, trim={0cm 0cm 0cm 0cm},clip]{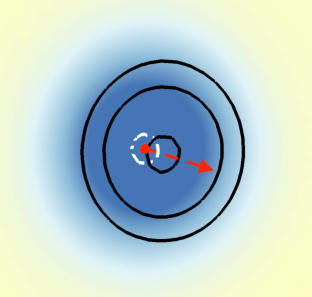}
\caption{$\vect{\tilde{q}_{1,1}}(\vect{x},t)$}\end{subfigure}  \enspace
\begin{subfigure}[b]{0.32\textwidth}\includegraphics[width=1.0\textwidth, trim={0cm 0cm 0cm 0cm},clip]{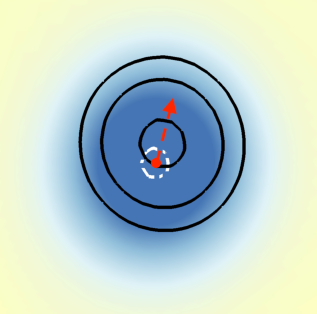}\caption{$\vect{\tilde{q}_{2,2}}(\vect{x},t)$}\end{subfigure}
  \enspace
\begin{subfigure}[b]{0.31\textwidth}\includegraphics[width=1.0\textwidth, trim={0cm 0cm 0cm 0cm},clip]{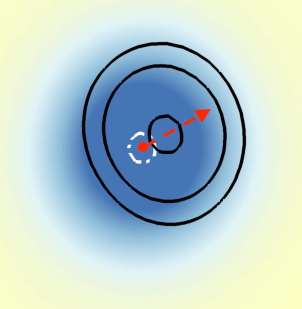} \caption{$\vect{\tilde{q}_{1,2}}(\vect{x},t)$} \end{subfigure}
 \caption{Displacement of mean vortex core due to most energetic POD modes. 
 Mean \textbf{R}-vortex ($\vect{\overline{q}}(\vect{x})$) is shown with flooded contours using the streamwise vortex at contour levels between $-10$ and $10$. 
 Broken white lines indicate the core.  
 Reconstructed vortex cores ($\vect{\tilde{q}_{m_1,m_2}}(\vect{x},t)$) are shown in solid black lines. 
The red arrow shows the direction of movement of the mean vortex center due to superposition of first mode(a), second mode (b) as well as both first and second modes (c). In (a,b), the arrow directions are consistent with the alignment of modes, $\alpha_1$ and $\alpha_2$ in figure~\ref{fig:podmode3}. } 
  \label{fig:podmode2}
\end{figure}

The characteristics of the POD and flow stability modes discussed above indicate the potential for a low-rank unsteady vortex meandering model using POD-based coherent structures. 
The reconstruction using a range of modes between $m_1$ and $m_2$, $\vect{\tilde{q}_{m_1,m_2}}(\vect{x},t)$, may be performed using the definition of POD in equation~\eqref{eq:pod} as:
\begin{eqnarray} \label{eq:pod2}
  \displaystyle \vect{\tilde{q}_{m_1,m_2}}(\vect{x},t) = \vect{\bar{q}}(\vect{x}) + \sum\limits_{i=m_1}^{m_2} \sqrt{\Lambda_i} 
  \vect{\Phi}_i (\vect{x}) \vect{a}_i(t) 
\end{eqnarray}

The influence of POD modes $1$ and $2$, as described in section~\ref{sec:pod}, is examined first by adding them individually to the mean flow and reconstructing the solution.
Subsequently, the combined effect of both helical modes is examined in reproducing meandering effects.
Figures~\ref{fig:podmode2}(a,b) show the displacement of the \textbf{R}-vortex core at $X/L=2$ when the first two POD modes are added individually to the mean flow at a representative time.
The reconstructed vortices display slightly non-axisymmetric structure due to the addition of these modes.
Furthermore, the core, as estimated by the innermost black contour, is displaced from its mean position shown with a broken white contour.
The displacement of the core is in the direction of the inclination of the dipoles shown earlier in figure~\ref{fig:podmode3}.  
When the two modes are added, the effective displacement of the core from the mean position is shown in Figure \ref{fig:podmode2}(c).

\begin{figure}
\centering
  \begin{subfigure}[b]{0.47\textwidth}
    \includegraphics[width=\textwidth, trim={0.1cm 0.1cm 1cm 1cm},clip]{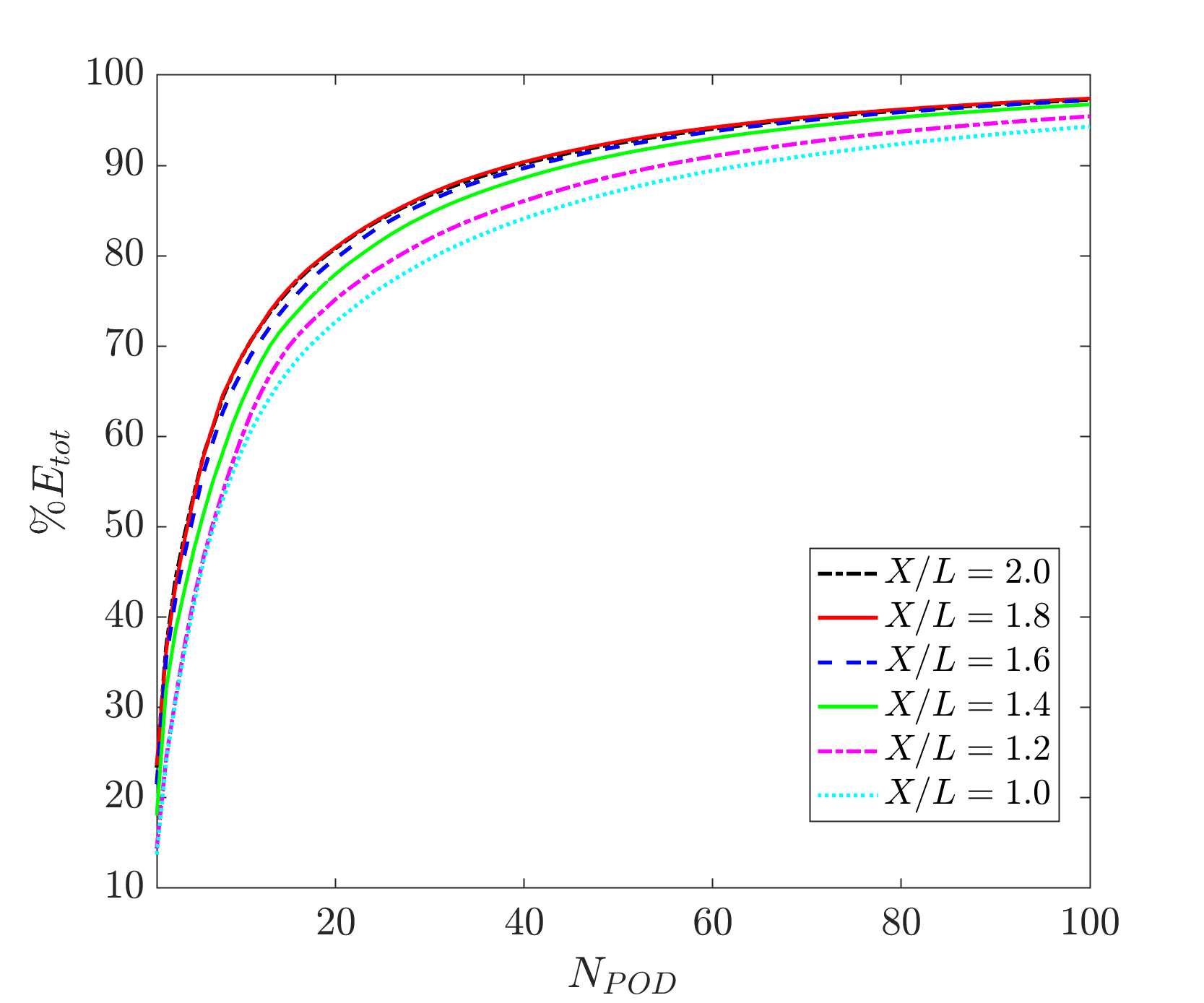}
    \caption{}
  \end{subfigure}
  \enspace
    \begin{subfigure}[b]{0.47\textwidth}
    \includegraphics[width=\textwidth, trim={0.1cm 0.1cm 1cm 1cm},clip]{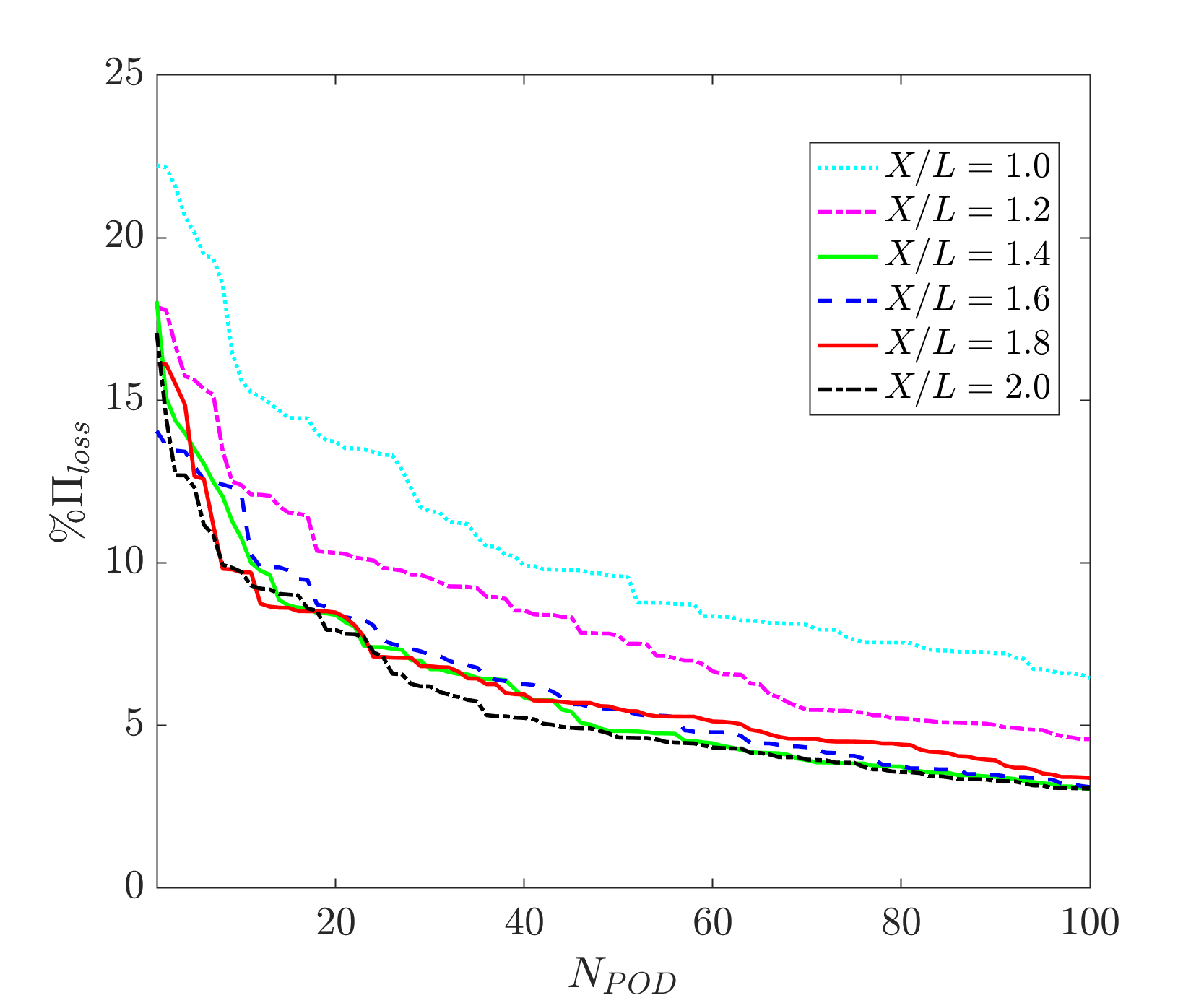}
    \caption{}
  \end{subfigure}
\caption{Comparison of energy of POD modes (a), and performance losses in reconstruction (b) at different streamwise locations.}
\label{fig:locations_compare}
\end{figure}

The efficacy of a low-rank approximation may be evaluated by comparing the LES result with reconstructed spatio-temporal evolution of the vortex cores due to a small subset of modes, taking into account the different energy content of each mode. 
The rank-order of the approximation is a function of the streamwise location chosen since the intensity of meandering as well as the proximity of the separating shear layer can alter the properties of the POD modes (\textit{e.g.,} compare leading modes at $X/L=1.0$ and $2.0$ in figures~\ref{fig:pod20} and~\ref{fig:podmode} respectively). 
The cumulative energy content,
\begin{equation}
E_{tot} = \sum\limits_{i=1}^{N_{POD}} \Lambda_i    
\end{equation}
due to $N_{POD}$ modes is shown in figure~\ref{fig:locations_compare}(a) for six locations downstream of the body at an interval of $\frac{\Delta x}{L}=0.2$. 
The total energy contribution due to dominant modes gradually increases with the downstream distance, with nearly overlapping mode contributions for $X/L=1.8$ and $2.0$. 
Comparing the situation at $X/L=1.0$ and $2.0$,  the first two modes contribute to about $24\%$ and $37\%$ of the total energy of flow respectively, while the first $100$ modes respectively account for  $97.3\%$ and $94.3\%$.

\begin{figure}
\centering
    \begin{subfigure}[b]{0.23\textwidth}
    \includegraphics[width=\textwidth, trim={5cm 1cm 7cm 3cm},clip]{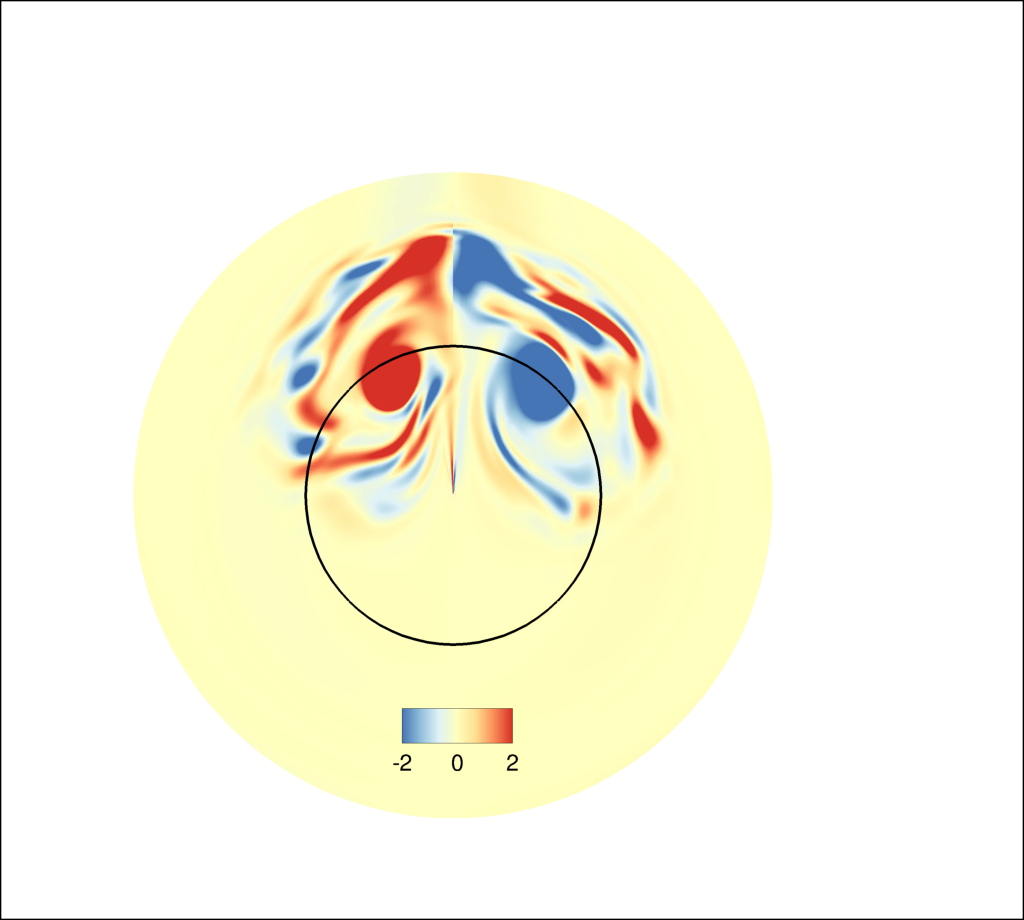}
  \end{subfigure}
  \enspace
    \begin{subfigure}[b]{0.23\textwidth}
    \includegraphics[width=\textwidth, trim={5cm 1cm 7cm 3cm},clip]{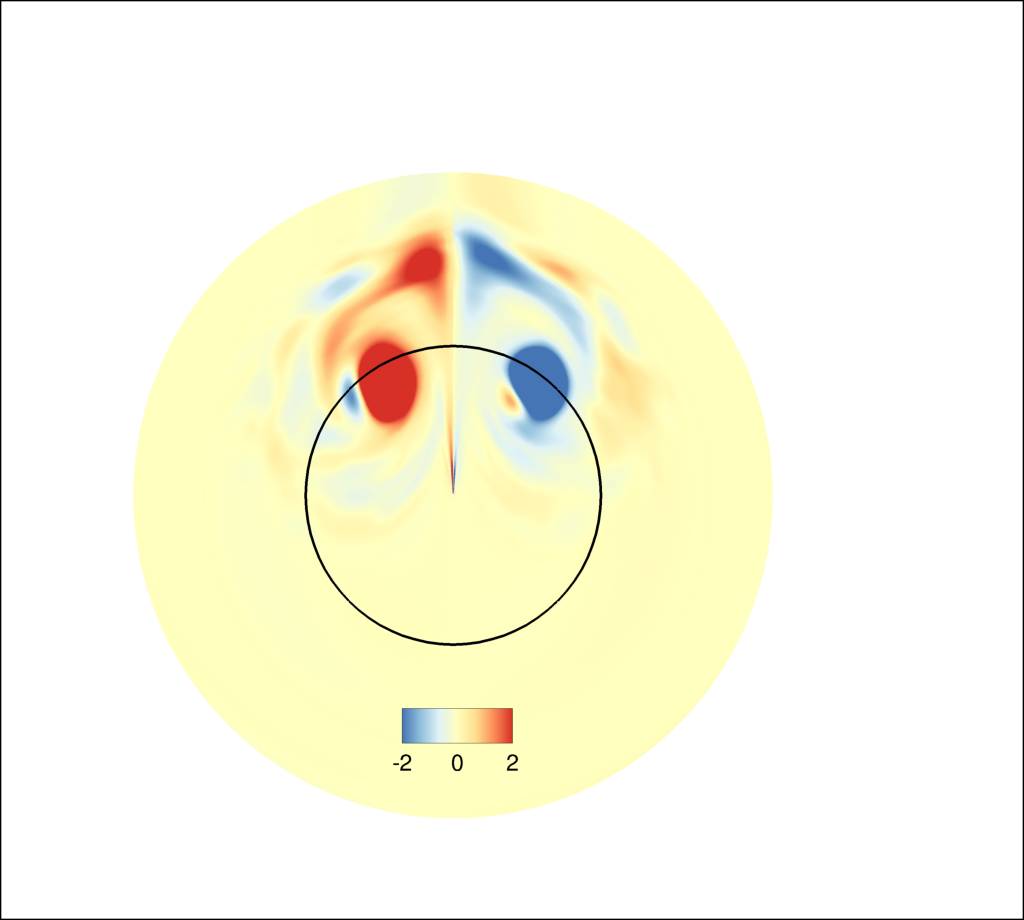}
  \end{subfigure}
 \enspace
    \begin{subfigure}[b]{0.23\textwidth}
    \includegraphics[width=\textwidth, trim={5cm 1cm 7cm 3cm},clip]{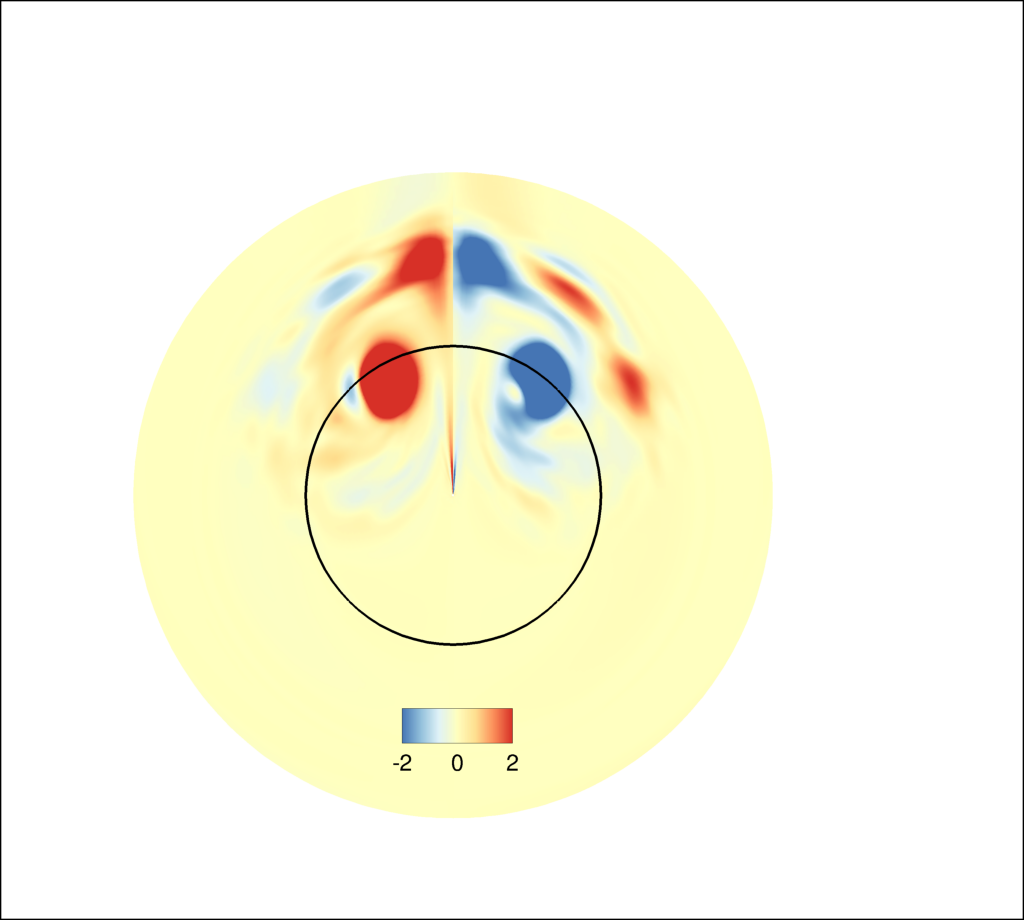}
  \end{subfigure}
   \enspace
    \begin{subfigure}[b]{0.23\textwidth}
    \includegraphics[width=\textwidth, trim={5cm 1cm 7cm 3cm},clip]{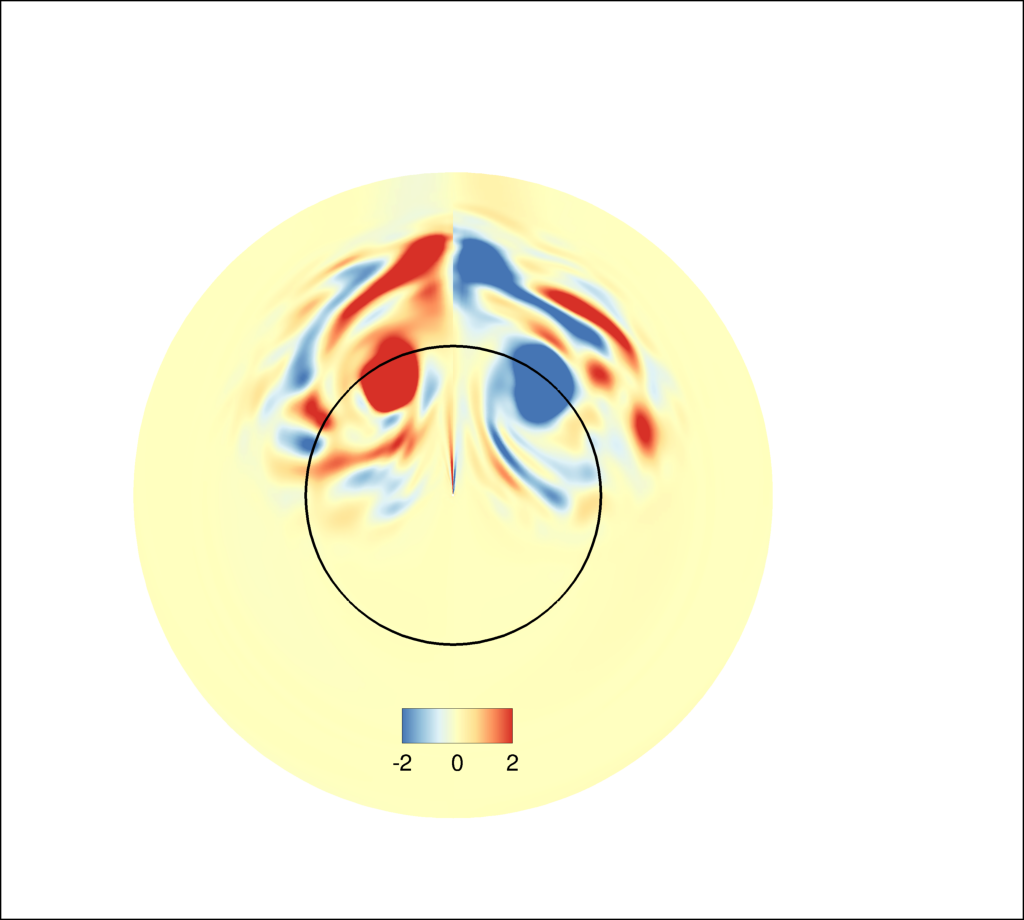}
  \end{subfigure}
\\
     \begin{subfigure}[b]{0.9\textwidth}
    \caption{$X/L=2.0$}
  \end{subfigure}
  \\
    \begin{subfigure}[b]{0.23\textwidth}
    \includegraphics[width=\textwidth, trim={5cm 1cm 7cm 3cm},clip]{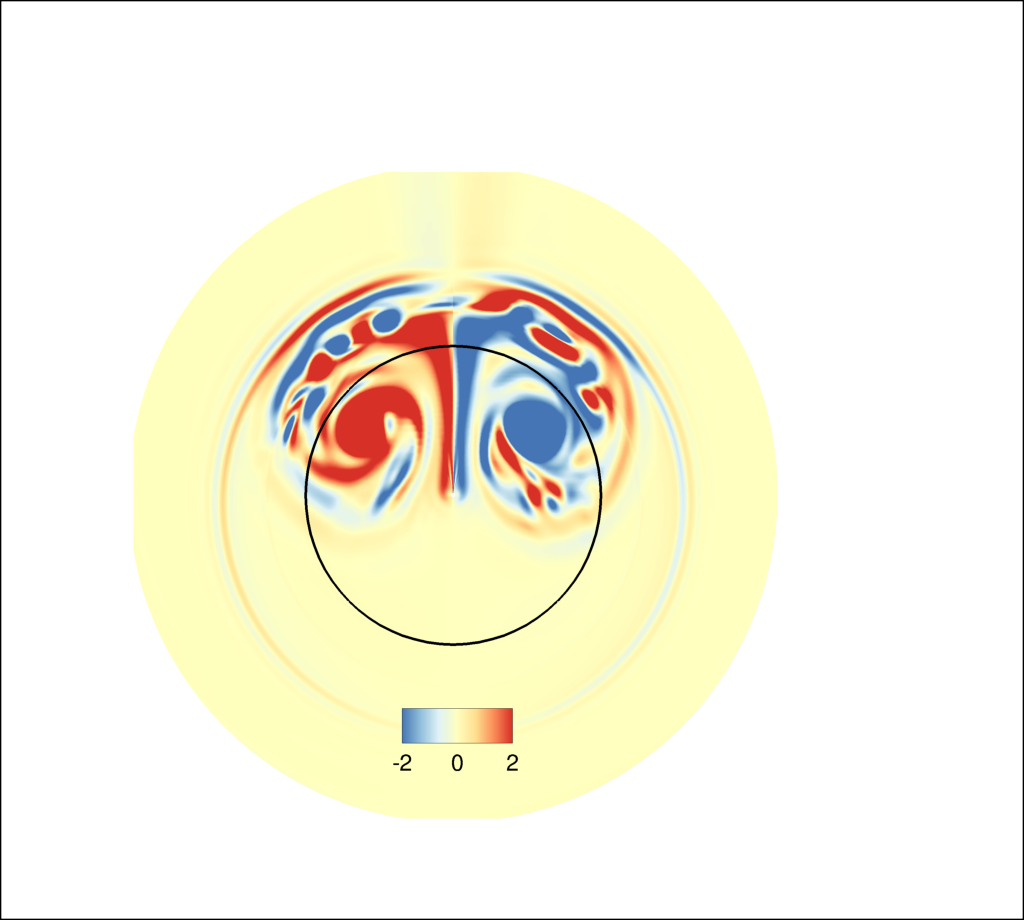}
  \end{subfigure}
  \enspace
    \begin{subfigure}[b]{0.23\textwidth}
    \includegraphics[width=\textwidth, trim={5cm 1cm 7cm 3cm},clip]{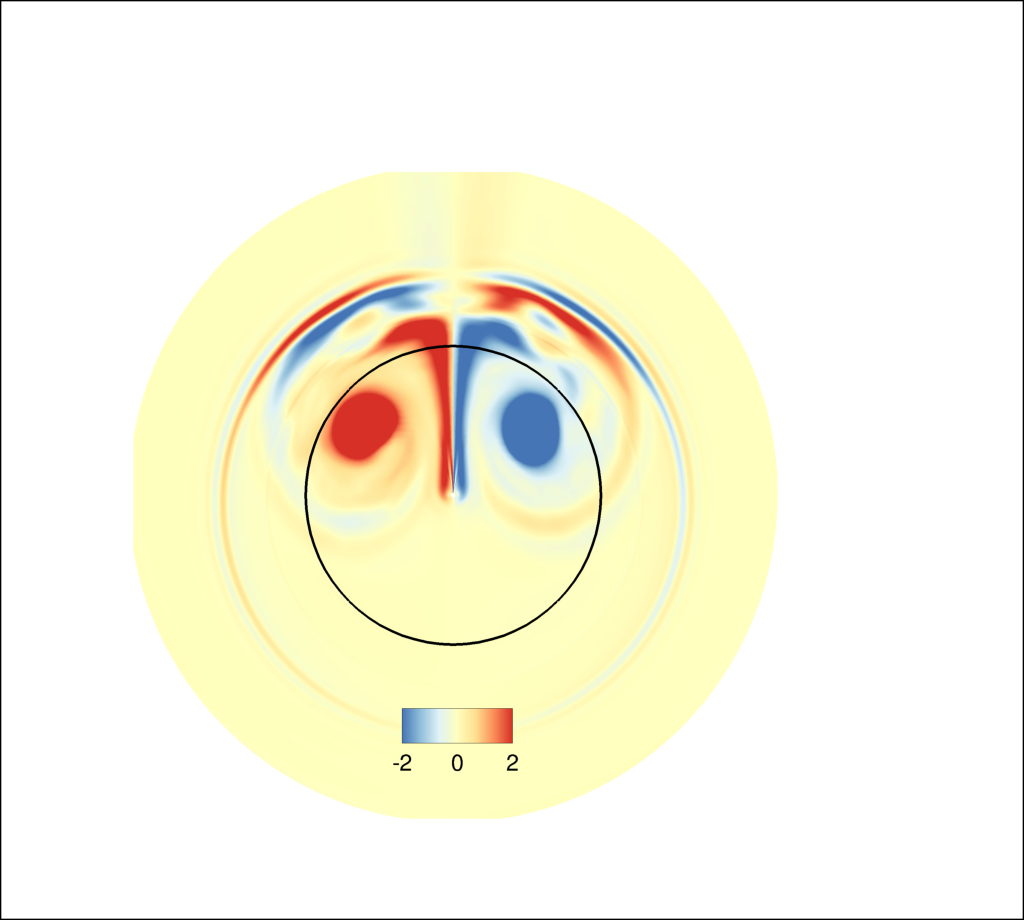}
  \end{subfigure}
 \enspace
    \begin{subfigure}[b]{0.23\textwidth}
    \includegraphics[width=\textwidth, trim={5cm 1cm 7cm 3cm},clip]{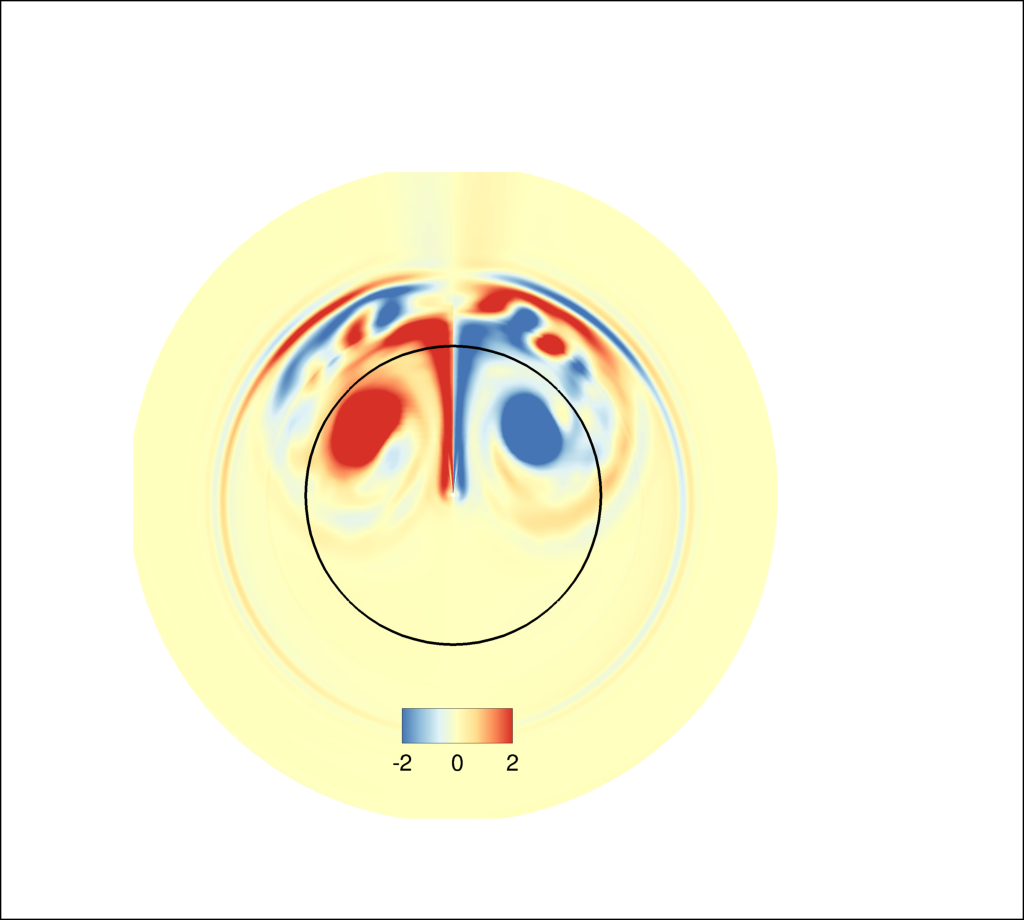}
  \end{subfigure}
   \enspace
    \begin{subfigure}[b]{0.23\textwidth}
    \includegraphics[width=\textwidth, trim={5cm 1cm 7cm 3cm},clip]{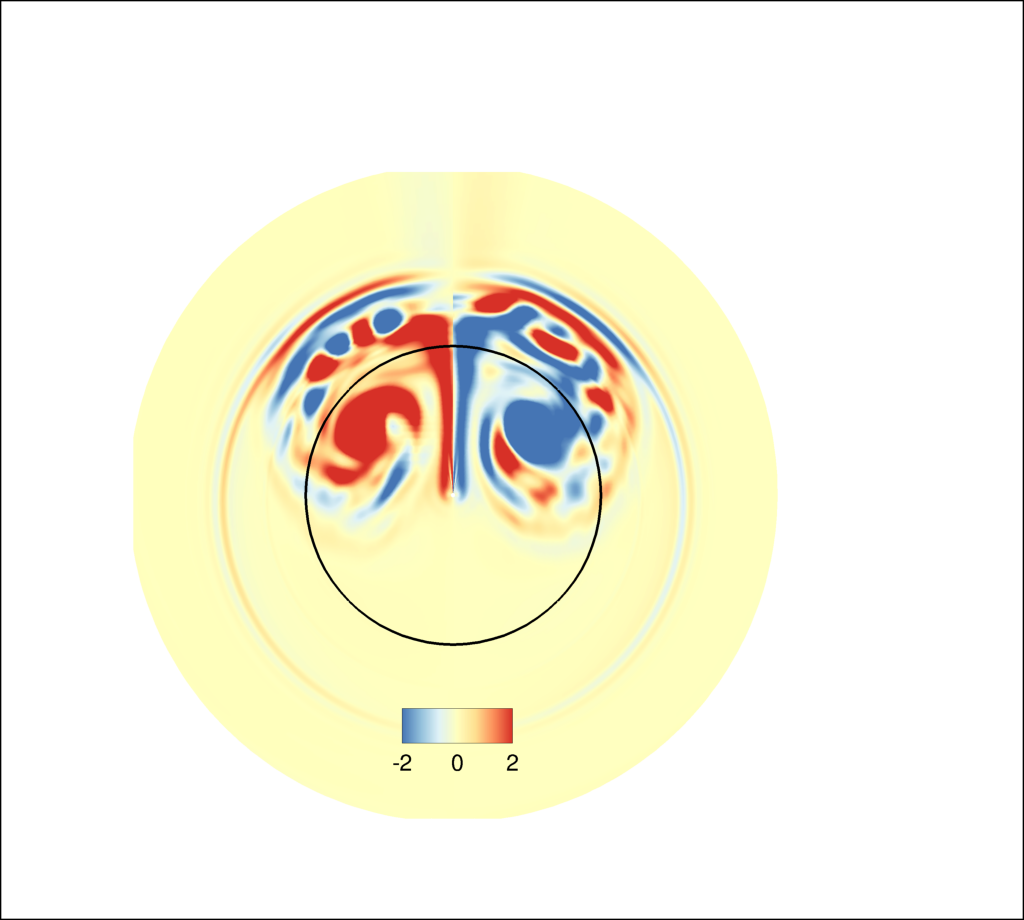}
  \end{subfigure}\\
       \begin{subfigure}[b]{0.9\textwidth}
    \caption{$X/L=1.4$}
  \end{subfigure}
  \\
   \begin{subfigure}[b]{0.23\textwidth}
    \includegraphics[width=\textwidth, trim={5cm 1cm 7cm 3cm},clip]{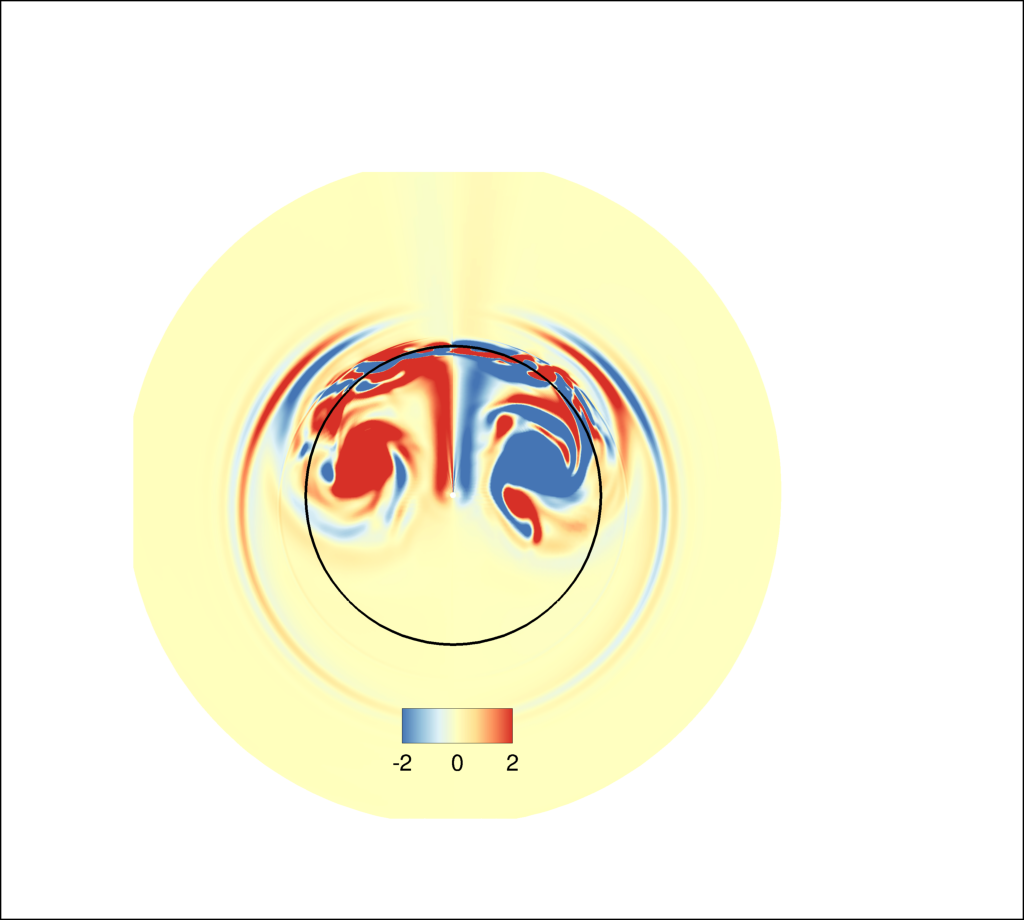}
        \caption*{$\vect{q}$}
  \end{subfigure}
  \enspace
    \begin{subfigure}[b]{0.23\textwidth}
    \includegraphics[width=\textwidth, trim={5cm 1cm 7cm 3cm},clip]{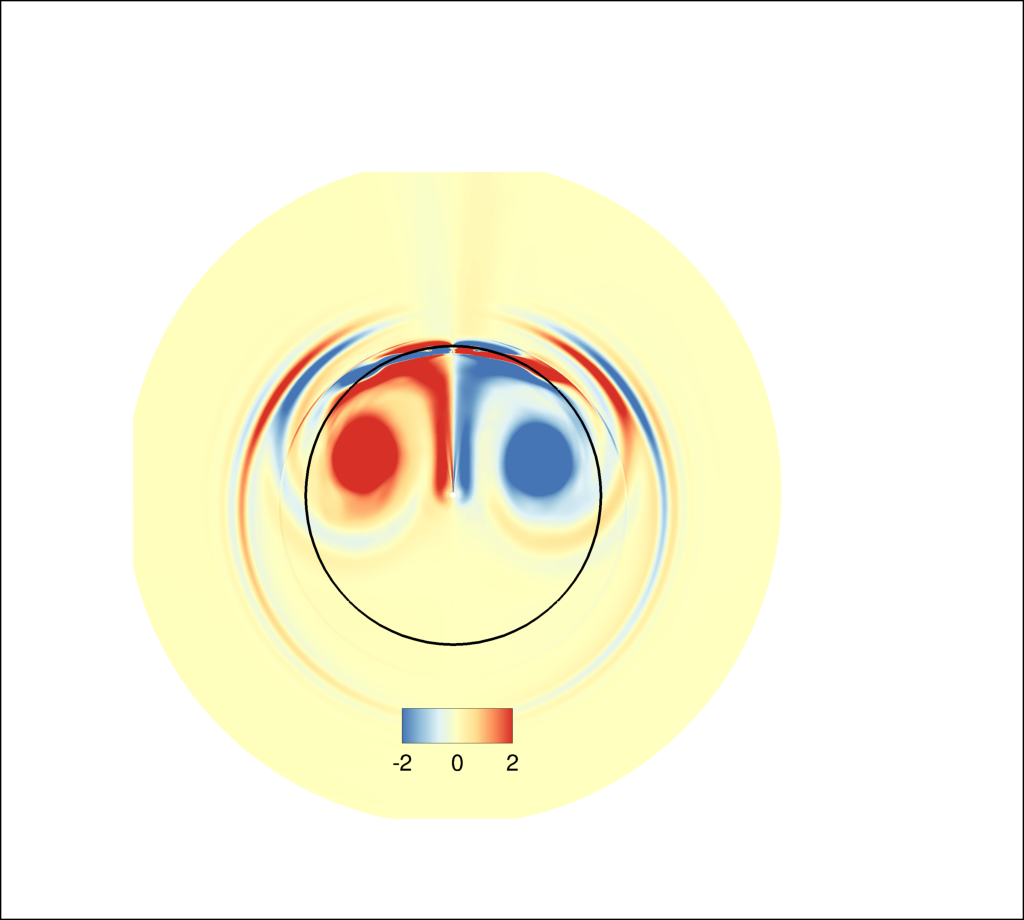}
        \caption*{$\vect{\tilde{q}_{1,2}}$}
  \end{subfigure}
 \enspace
    \begin{subfigure}[b]{0.23\textwidth}
    \includegraphics[width=\textwidth, trim={5cm 1cm 7cm 3cm},clip]{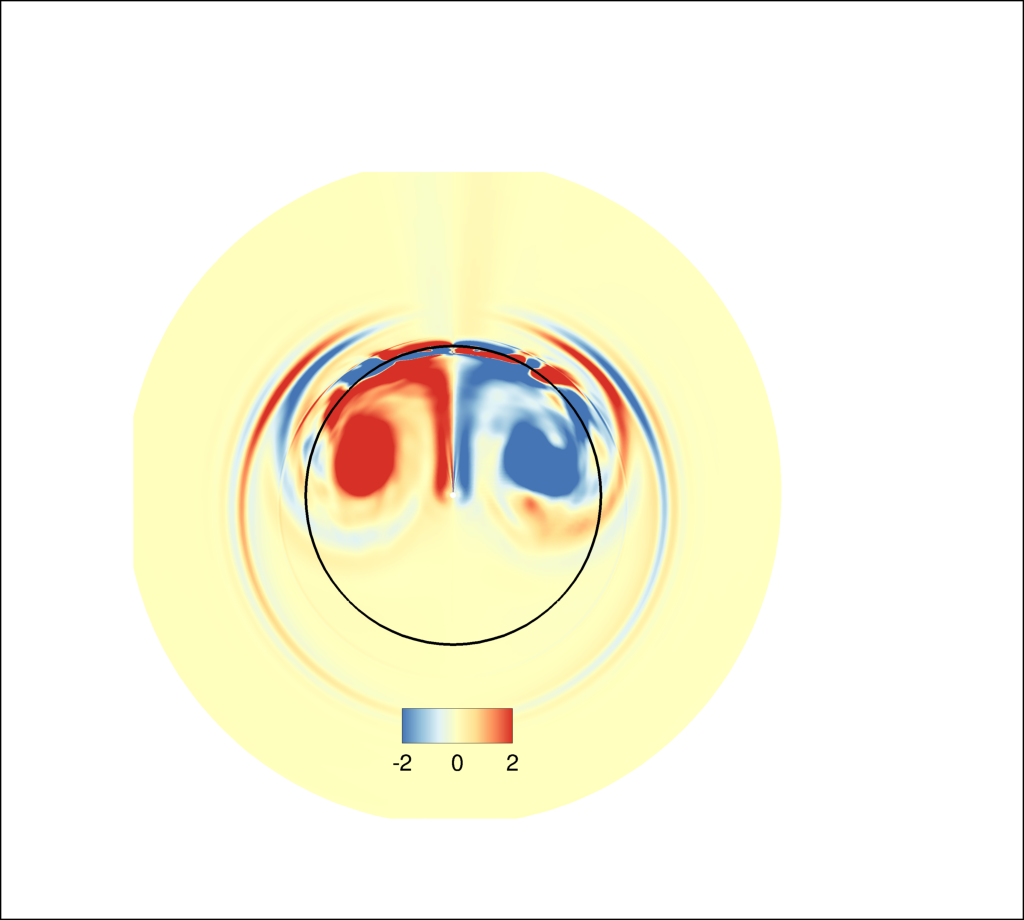}
        \caption*{$\vect{\tilde{q}_{1,10}}$}
  \end{subfigure}
   \enspace
    \begin{subfigure}[b]{0.23\textwidth}
    \includegraphics[width=\textwidth, trim={5cm 1cm 7cm 3cm},clip]{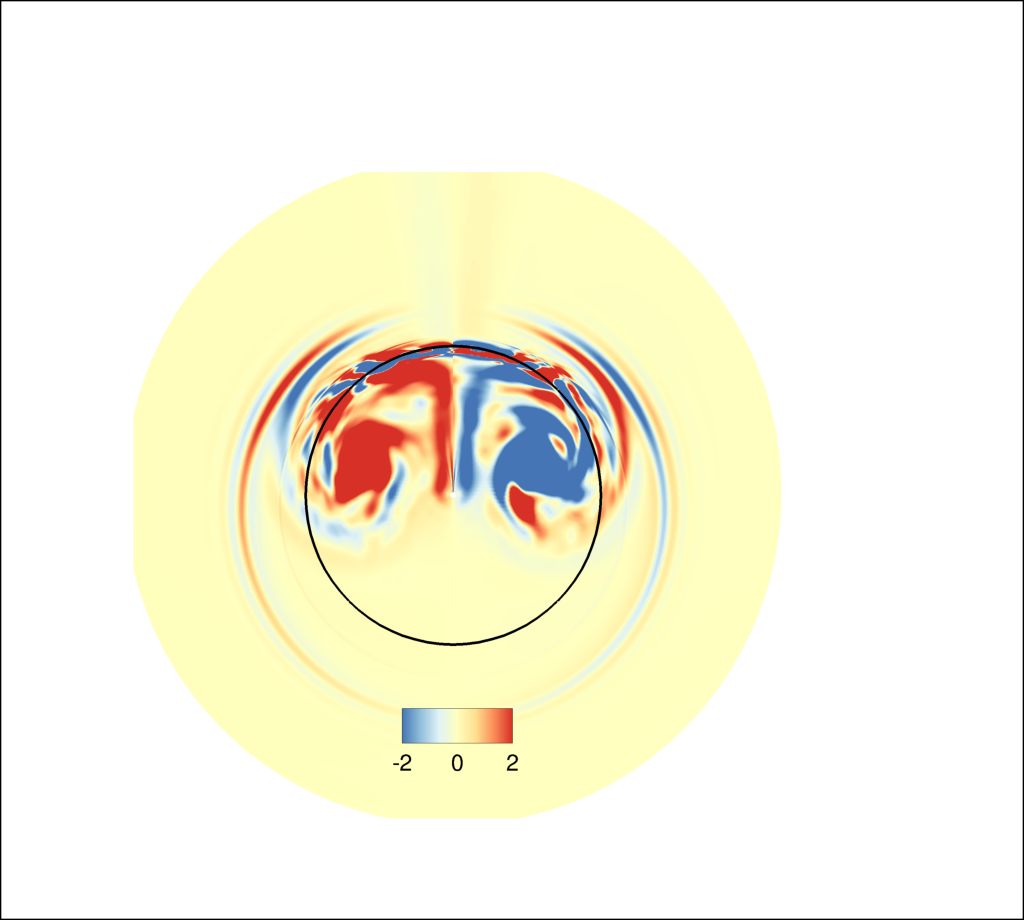}
        \caption*{$\vect{\tilde{q}_{1,100}}$}
  \end{subfigure}\\
     \begin{subfigure}[b]{0.9\textwidth}
    \caption{$X/L=1.0$}
  \end{subfigure}
  \caption{Reconstruction of flow fields using POD modes. The first two modes are major contributors to the coherent structures of the streamwise vortices. 
  At locations closer to the body e.g., at $X/L=1.0$ and $X/L=1.4$, the meandering intensities are higher and therefore reconstructed vortices with just two modes differ significantly from the LES. 
  In all three cases, $100$ of the total $3{,}000$ modes reconstruct the complete flowfield to reasonable accuracy.}
\label{fig:modes_recon}
\end{figure}

The disparity in energy content of dominant modes at different locations is expected to affect the reconstruction of flowfields with these modes.
Figure~\ref{fig:modes_recon} compares the actual and reconstructed flowfields at three locations when $2$, $10$ and $100$ modes are considered respectively. 
At the primary location of interest, $X/L=2.0$ in Fig.~\ref{fig:modes_recon}(a), a reasonable description of the streamwise vortices is obtained with only the first two modes. 
These modes also contribute to the large-scale structures in the top shear layer away from the vortices.  
The next $8$ modes primarily contribute to small-scale structures surrounding the vortices and the top shear-layer. 
When $100$ of the total $3{,}000$ modes are used for reconstruction, the structure is very similar, qualitatively and quantitatively, to the LES result.

At locations closer to the body, $X/L=1.4$ and $1.0$, shown in Fig.~\ref{fig:modes_recon}(b) and (c) respectively, the first two modes contribute to the coherent structures of the vortices. 
However, due to high meandering intensity at these locations, the structures of the reconstructed vortices are significantly different from the LES snapshots. 
The addition of more modes improves the result as expected, with about $100$ modes nearly reconstructing the complete flowfield with reasonable accuracy. 

A performance loss parameter $\Pi_{loss}$ is used to quantify the error in reconstruction, similar to that used in~\citet{jovanovic2014sparsity}:
\begin{equation}
    \Pi_{loss} = \frac{||\vect{q} - \vect{\tilde{q}}||_2}{||\vect{q}||_2}
\end{equation}
Figure~\ref{fig:locations_compare}(b) displays these losses at six locations between $X/L=1.0$ and $2.0$. 
Even with only one POD mode, the maximum performance error is only 23\% across all locations, further confirming the low-rank behavior of the flow.  
As expected, the losses are higher at locations near the body, with a difference of about 5\% between $X/L=1.0$ and $1.2$. 
Away from the body, for example at $X/L=2.0$, the loss is about 3\% with $100$ modes. 
At $X/L=1.0$, this loss is slightly higher ($8\%$).    
Further, as described above in the context of Fig.~\ref{fig:modes_recon}(a), the major factor contributing to error observations at the farthest location ($X/L=2.0$) is the absence of finer structures in the first few modes; two modes by themselves recreate the streamwise vortices reasonably well.

\begin{figure}
\centering
    \includegraphics[width=1.0\textwidth, trim={1.5cm 0.5cm 1cm 0.25cm},clip]{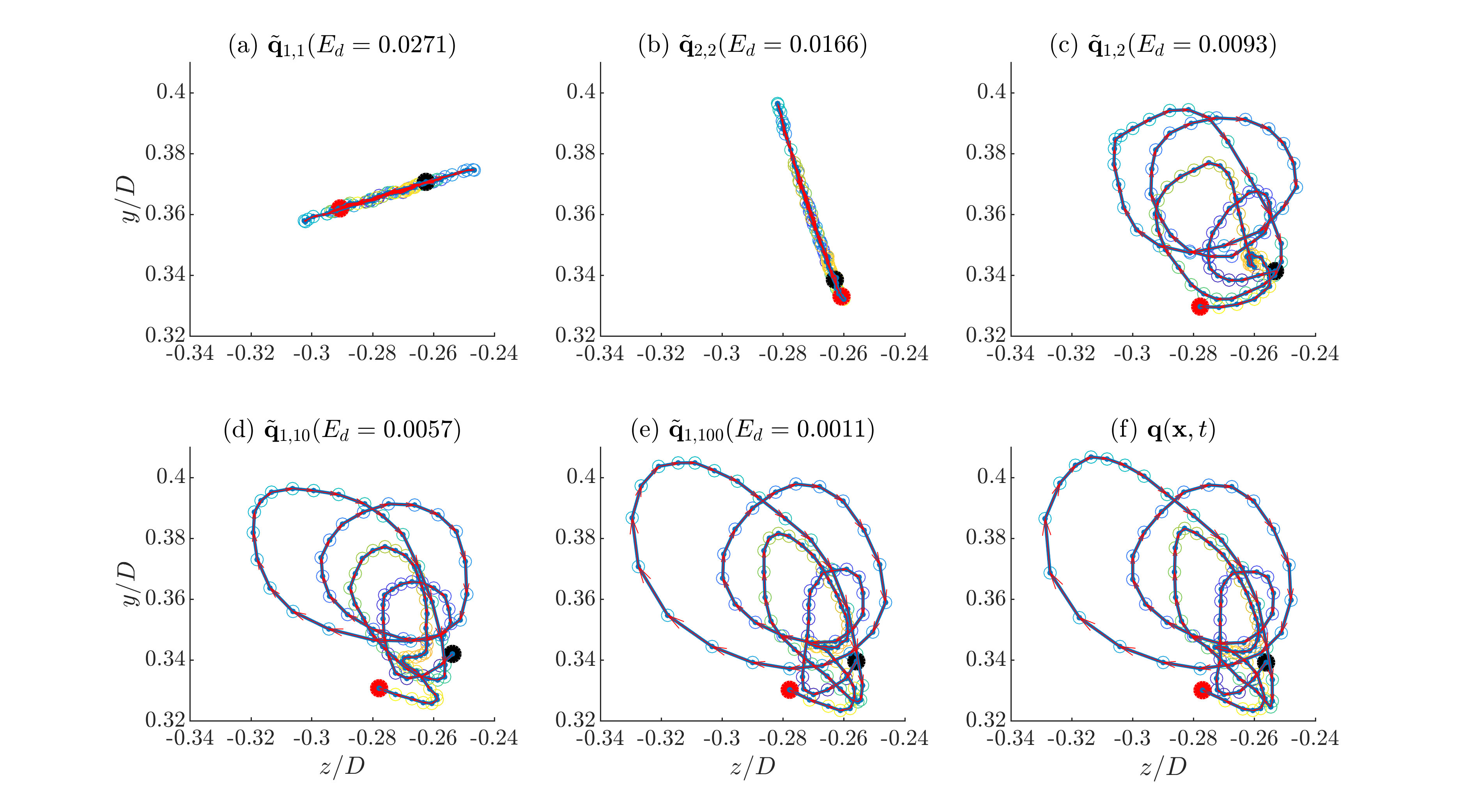}
 \caption{Meandering of the \textbf{L}-vortex using reconstructed flow fields. Black and red circles indicate the beginning and the end of the meandering trajectory.} 
  \label{fig:meandering3}
\end{figure}

The effects of low-rank approximation on the temporal dynamics at this location are now compared to the meandering observed in the LES field. 
As explained earlier, the reconstruction results in vortex motion that does not follow a preferred displacement direction but rather moves around in a manner that represents the meandering observed in the LES.  
The temporal displacement of the vortex cores at $X/L=2.0$, when the effects of the leading modes are added are shown in figure~\ref{fig:meandering3} for the \textbf{L} vortex. 
The individual displacement due to the first and second modes considered separately are depicted in figures~\ref{fig:meandering3}(a) and (b), respectively for $100$ time-steps.  
These figures confirm that the motions of the core induced by mode $1$ and $2$, by themselves, lie along straight lines. 
The slopes of these lines,  $\alpha_1$ and $\alpha_2$, are consistent with the structures of the vortex dipoles as described earlier in \S~\ref{sec:pod}.

Although the displacements due to the individual modes are relatively simple, their combined effect is not.
The trajectory of vortex cores due to both modes, shown in figure \ref{fig:meandering3}(c), are  clearly very complex and are established due to the combination of two nearly straight lines in an effect similar to Lissajous curves \citep{fahy1952geometrical} generated by a system of parametric equations. 
The combined displacement due to both these modes may be compared with the actual meandering motion shown in figure~\ref{fig:meandering3}(f).
Even when only the two dominant POD modes are considered, the result qualitatively reproduces the complex clockwise movement of the core, although the maximum displacement of core in the reconstructed flowfield is smaller than the original flowfield. 
The error in reconstruction may be characterized using the displacement from the vortex location in the root-mean-square sense as:  
\begin{equation}
\displaystyle E_{d} = \sqrt{\frac{\sum\limits_{i=1}^{N} ((y_i-\tilde{y}_i)^2 + (z_i-\tilde{z}_i)^2)}{N}} 
\end{equation}
where $y_i,z_i$ are the coordinates of the vortex center from the LES flowfield for snapshots $i=1,2...,N$. 
The tilde quantities ($\tilde{y}_i,\tilde{z}_i$) represent the vortex centers in the reconstructed flowfield. 
These errors are also shown in Fig.~\ref{fig:meandering3}. 
This error is $0.0271$ and $0.0166$ considering only the first and second mode respectively, and decreases to $0.0093$ when both modes are added. 
When $10$ modes are considered (Fig.~\ref{fig:meandering3}(d)), the  trajectory approaches the actual meandering path even further, and the error decreases to $E_d=0.0057$. 
Further, when the vortex core movement is computed on a flowfield reconstructed with $100$ modes (Fig.~\ref{fig:meandering3}(e)), the core trajectory nearly mimics the complete meandering path. 
The addition of high-order modes mainly contributes to correcting the maximum displacement from the mean vortex, while the direction of the core movement is mainly dictated by only two modes.  

\begin{figure}
\centering
      \includegraphics[width=1.0\textwidth, trim={1.5cm 0.5cm 1cm 0.25cm},clip]{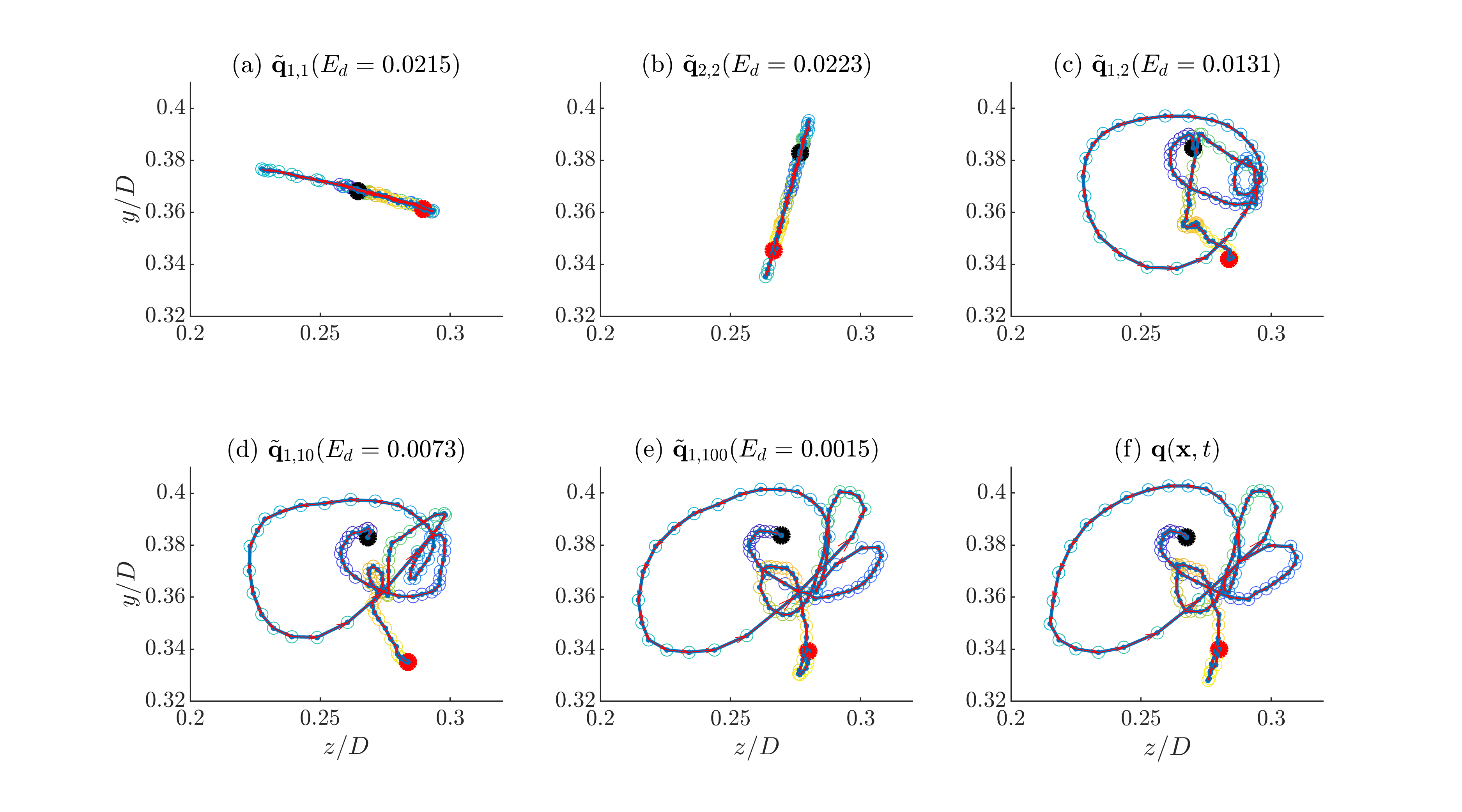}
 \caption{Meandering of \textbf{R}-vortex using reconstructed flow fields.} 
  \label{fig:meandering4}
\end{figure}

In order to confirm that this observation also holds true for the \textbf{R} vortex, its reconstructed meandering motions are shown in figure~\ref{fig:meandering4}(a-e).
Consistent with the observations for the \textbf{L} vortex, linear displacements are observed due to the first and second POD modes in this vortex also, as shown in figure~\ref{fig:meandering4}(a) and (b) respectively.
The inclinations of the linear paths for this vortex are different from those obtained for \textbf{L} case of course, since they are aligned with displacements due to the individual modes of the \textbf{R} vortex. 
The displacement errors $E_d$ due to these modes are comparable to each other. 
The combined displacement effect due to both modes, shown in figure~\ref{fig:meandering4}(c), again mimics the actual meandering path (figure~\ref{fig:meandering4}(f)), except for lower maximum displacements. 
The addition of $8$ more modes brings the trajectory closer to the actual meandering path (figure~\ref{fig:meandering4}(d)), and with a total $100$ modes, the paths  nearly overlap with each other (figure~\ref{fig:meandering4}(e)). 
Although not shown, for locations closer to the body, the reconstruction error of the meandering path using these modes is slightly higher, as expected from their performance in reconstructing spatial structure (figure~\ref{fig:modes_recon}(b)).
However, the key observation that they continue to exhibit low-rank behavior remains valid.    

%
The reconstruction of the meandering path with only a few leading modes confirms the lower-dimensional dynamics and a smaller number of Lyapunov exponents in the dynamical system that  characterize the chaos present in the system.  
\citet{singler2008approximate} show that POD can be used in the manner above to compute approximate low-rank solutions of the Lyapunov equations.
The solution of these equations can be used with standard methods to compute reduced-order models of linear systems either in a Galerkin-based~\citep{couplet2005calibrated} or a Galerkin-free approach~\citep{shinde2016galerkin}.
This would constitute a significant step in ultimately determining control laws to modulate the meandering of vortices.

\section{Conclusions}\label{sec:conclusion}
High fidelity simulations are employed to examine the dynamics of the flow behind an axisymmetric cylinder with a slanted base.
The base upsweep angle considered, $\phi=\ang{20}$, is representative of a cargo fuselage afterbody geometry. 
The primary feature is a horseshoe-shaped structure that develops around the periphery of the base and ultimately lifts away from the surface to manifest a three-dimensional generally streamwise-oriented vortex pair.
Relatively far downstream from the base, these wake vortices, while axisymmetric in the mean sense, exhibit low-frequency meandering in space and time. 
This motion is characterized by seemingly random displacement of the vortex cores, although the time trajectory of the displacement of the core shows a rotation sense commensurate with the swirl in the vortices. 
These complex phenomena are first analyzed through proper orthogonal decomposition (POD) at several locations on the ramp as well as downstream of the body.
Near the base, the dominant structures show fluctuations in both the vortex and the shear layer.
Downstream, however, the coherent structures are characterized by a pair of orthogonal helical modes with azimuthal wave number ($|m|=1$).
These two most energetic modes provide a first-order approximation of the displacement of the vortex cores, confirming that the meandering is primarily associated with changes in large-scale coherent structures. 
The instability mechanism underlying meandering is analyzed using a linear stability approach on the time-averaged streamwise vortex pair.
The mean vortex obtained from the simulations is first matched to a suitable Batchelor vortex and then a counter-rotating model vortex is placed at an appropriate separation distance from the first vortex. 
Both spatial and temporal analyses confirm the presence of an elliptic instability with the structures of the unstable modes being similar to those obtained from POD.
The unstable modes are obtained at relatively low frequency with the presence of axial velocity as primary mechanism for instability.  

Low-rank representation behavior is examined by considering different numbers of POD modes.
Consideration of only the first or second mode generates temporal displacement of each core along a relatively straight line, which is very different from the actual meandering path.
However, when both the modes are considered together, the effective trajectory recreated recovers the broad meandering motion. 
In the spatial reconstruction, these two modes alone capture the vortex cores reasonably well, while high-order modes further correct both spatial and temporal reconstructions to capture small scale features. 
Both spatial and temporal reconstruction errors diminish rapidly with the superposition of additional modes, especially in downstream regions.
These low-rank observations of the complex meandering phenomena should facilitate the construction of a reduced-order model as a surrogate for the full dynamical system in flow control analyses.

\section*{Acknowledgements}
This material is based upon work supported by the Air Force Office of Scientific Research under award number FA9550-17-1-0228 (monitor: Gregg Abate). 
The authors thank F. Alvi, F. Zigunov, and P. Sellappan for several fruitful communications.
 Simulations were carried out using resources provided by the U.S. Department of Defense High-Performance Computing Modernization Program and the Ohio Supercomputer Center.

\appendix
\section{Governing linear stability equations}\label{app:LSE}
The incompressible Navier-Stokes equations in non-dimensional form are given by:
\begin{eqnarray}\label{eqn:ns}
\nabla. \vect{u} &=& 0 \\
\frac{\partial \vect{u}}{\partial t} +  (\vect{u}.\nabla) \vect{u} &=& \nabla p +  Re^{-1}\nabla^2 \vect{u}
\end{eqnarray}
where $\vect{u} = {\{u,v,w\}}^T$. 
The flow field is decomposed into a basic state ($\vect{\bar{u}},\bar{p}$) and perturbations ($\vect{u'},p'$), with the former satisfying equations~\eqref{eqn:ns}. The equations governing linear perturbations are then:
\begin{eqnarray}\label{eqn:pert}
\nabla. \vect{u'} &=& 0 \\
\frac{\partial \vect{u'}}{\partial t} +  (\vect{u'}.\nabla) \vect{\bar{u}} +  (\vect{\bar{u}}.\nabla) \vect{u'} &=& \nabla p' +  Re^{-1}\nabla^2 \vect{u'}
\end{eqnarray}
where the small non-linear terms, $(\vect{u'}.\nabla) \vect{u'}$, are ignored.
For stability analysis, a modal form of the perturbations is assumed, with homogeneity in the $x$-direction
\begin{eqnarray}
\{u',v',w',p'\} (x,y,z,t) = \{\tilde{u},\tilde{v},\tilde{w},\tilde{p}\}(y,z)e^{i(kx - \omega t)}
\end{eqnarray}
where the quantities with tilde($\tilde{\bullet}$) are two-dimensional amplitude functions. 
Substituting this in the perturbation equations~\eqref{eqn:pert}, the stability equations are obtained:
 \begin{align}\label{eqn:stability}
 \Bigg( \underbrace{\begin{bmatrix}
    \mathcal{L}   & \overline{U}_y & \overline{U}_z & 0 \\[0.3em]
        0 & \mathcal{L} + \overline{V}_y   & \overline{V}_z & D_y \\[0.3em]
    0 & \overline{W}_y & \mathcal{L}+\overline{W}_z  & D_z \\[0.3em]
        0 & D_y & D_z & 0 
  \end{bmatrix}}_{\mathbb{A}_0} 
  + k
    \underbrace{\begin{bmatrix}
      i\overline{U} & 0 & 0 & i \\[0.3em]
      0 & i\overline{U}  & 0 & 0 \\[0.3em]
    0& 0 & i\overline{U}  & 0 \\[0.3em]
        i & 0 & 0 & 0 
  \end{bmatrix}}_{\mathbb{A}_1} 
    + k^2
    \underbrace{\begin{bmatrix}
      I Re^{-1} & 0 & 0 & 0 \\[0.3em]
      0 & I Re^{-1}  & 0 & 0 \\[0.3em]
    0& 0 & I Re^{-1}  & 0 \\[0.3em]
        0 & 0 & 0 & 0 
  \end{bmatrix}}_{\mathbb{A}_2}  \notag \\
  - \omega \underbrace{\begin{bmatrix}
    i & 0 & 0 & 0 \\[0.3em]
        0 &i  & 0 & 0 \\[0.3em]
    0 & 0 &i & 0 \\[0.3em] 
        0 & 0 & 0 & 0 
  \end{bmatrix}}_{\mathbb{B}} \Bigg)
  \begin{bmatrix}
    \tilde{u} \\[0.3em]
    \tilde{v} \\[0.3em]
      \tilde{w}\\[0.3em]
    \tilde{p} 
  \end{bmatrix}
  = 0
  \end{align}
  where
\begin{eqnarray}
\mathcal{L} &=& \overline{V}D_y + \overline{W}D_z - \frac{D_{yy} + D_{zz}} {Re}. 
\end{eqnarray}
$D_y, D_{yy}$ etc. represent first and second-order differentiation matrices. $I$ is the identity matrix.

For the temporal analysis, the real wavenumber $k$ is specified, while  $\omega = \omega_r + i \omega_i$ is a complex quantity denoting the circular frequency, $\omega_r$, and growth rate, $\omega_i$. 
The generalized eigenvalue problem, following equations~\eqref{eqn:stability}, is then:  
\begin{eqnarray}
\mathbb{A} \tilde{\vect{q}} = \omega \mathbb{B} \tilde{\vect{q}}. 
\end{eqnarray}
with $\tilde{\vect{q}} = \{\tilde{u},\tilde{v},\tilde{w},\tilde{p}\}^T$ and $\mathbb{A} = \mathbb{A}_0 + k\mathbb{A}_1 + k^2\mathbb{A}_2$.

For spatial stability analysis, the streamwise
wavenumber $k = k_r + i k_i$ represents the complex eigenvalue, with  $k_i < 0$ indicating spatial exponential growth, and $\omega$ is a real specified circular frequency.
In this case, equations~\eqref{eqn:stability} present a quadratic eigenvalue problem: 
\begin{eqnarray}
(\mathbb{A}'_{0} + k \mathbb{A}_1 +  k^2 \mathbb{A}_2 ) \tilde{\vect{q}} = 0
\end{eqnarray}
with $\mathbb{A}'_{0} = \mathbb{A}_0 - \omega \mathbb{B}$.

Although the problem can be solved directly, the quadratic eigenvalue problem can become very expensive for large matrices. 
Therefore, it is converted into a linear eigenvalue problem by using the companion matrix technique \citep{tisseur2001quadratic}. 
The matrix size of this spatial eigenvalue problem is therefore 75\% higher than the temporal approach due to the padding of three auxiliary variables at every grid point.
For both temporal and spatial analysis, distant farfield boundaries are used to allow for homogeneous Dirichlet boundary conditions for all perturbation variables except for pressure, for which the Neumann boundary condition is appropriate.
The eigenvalue problems are solved using the iterative Arnoldi~\citep{arnoldi1951principle} algorithm. Further, the shift-and-invert approach~\citep{lehoucq1998arpack} in the Arnoldi algorithm is used to obtain faster convergence to desired eigenvalues.
A large Krylov dimension ($k=100$) is chosen to ensure that the relevant unstable eigenspectrum around the shift value is captured.

\bibliographystyle{jfm}
\bibliography{loftbase_references,stability_wake_vortices}


\end{document}